\documentclass[12pt]{article}

\usepackage{amsmath,amssymb,graphicx,hyperref}
\usepackage{authblk}
\usepackage[utf8]{inputenc}
\usepackage[separate-uncertainty=true,list-units=repeat,multi-part-units=brackets]{siunitx}
\usepackage{subcaption}
\usepackage{placeins}
\usepackage[export]{adjustbox}
\usepackage{cprotect}

\usepackage[style = phys]{biblatex}
\graphicspath{ {./figures/} }

\newcommand{\aSn}{$\alpha$-Sn}
\DeclareSIUnit{\angstrom}{\textup{\AA}}
\newcommand{\kp}{\textbf{\textit{k}}$\cdot$\textbf{\textit{p}}}
\DeclareUnicodeCharacter{2212}{-}

\begin{document}

\title{3D Topological Semimetal Phases of Strained $\alpha$-Sn on Insulating Substrate}

\author[1]{Jakub Polaczy{\'n}ski}
\author[2]{Gauthier Krizman\thanks{\href{mailto:gauthier.krizman@jku.at}{gauthier.krizman@jku.at}}}
\author[1]{Alexandr Kazakov\thanks{\href{mailto:kazakov@MagTop.ifpan.edu.pl}{kazakov@MagTop.ifpan.edu.pl}}}
\author[1]{Bart{\l}omiej Turowski}
\author[3]{Joaqu{\'i}n Bermejo~Ortiz}
\author[1]{Rafa{\l} Rudniewski}
\author[1]{Tomasz Wojciechowski}
\author[4]{Piotr D{\l}u{\.z}ewski}
\author[4]{Marta Aleszkiewicz}
\author[1]{Wojciech Zaleszczyk}
\author[4]{Bogus{\l}awa Kurowska}
\author[1]{Zahir Muhammad}
\author[5]{Marcin Rosmus}
\author[5]{Natalia Olszowska}
\author[3]{Louis-Anne de Vaulchier}
\author[3]{Yves Guldner}
\author[1]{Tomasz Wojtowicz}
\author[1,6]{Valentine V. Volobuev\thanks{\href{mailto:Volobuiev@magtop.ifpan.edu.pl}{Volobuiev@magtop.ifpan.edu.pl}}}

\affil[1]{International Research Centre MagTop, Institute of Physics, Polish Academy of Sciences, Aleja Lotnik{\'o}w 32/46, PL-02668 Warsaw, Poland}
\affil[2]{Institut f{\"u}r Halbleiter und Festk{\"o}perphysik, Johannes Kepler Universit{\"a}t, Altenberger Stra{\ss}e 69, 4040 Linz, Austria}
\affil[3]{Laboratoire de Physique de l’{\'E}cole normale sup{\'e}rieure, ENS, Universit{\'e} PSL, CNRS, Sorbonne Universit{\'e}, 24 rue Lhomond 75005 Paris, France}
\affil[4]{Institute of Physics, Polish Academy of Sciences, Aleja~Lotnik{\'o}w 32/46, PL-02668 Warsaw, Poland}
\affil[5]{Narodowe Centrum Promieniowania Synchrotronowego SOLARIS, Uniwersytet Jagiello{\'n}ski, ul. Czerwone Maki 98, 30-392 Krak{\'o}w, Poland}
\affil[6]{National Technical University "KhPI", Kyrpychova Str. 2, 61002 Kharkiv, Ukraine}

\maketitle

\begin{abstract}
{\aSn} is an elemental topological material, whose topological phases can be tuned by strain and magnetic field. Such tunability offers a substantial potential for topological electronics. However, InSb substrates, commonly used to stabilize {\aSn} allotrope, suffer from parallel conduction, restricting transport investigations and potential applications. Here, the successful MBE growth of high-quality {\aSn} layers on insulating, hybrid CdTe/GaAs(001) substrates, with bulk electron mobility approaching 20000~cm$^2$V$^{-1}$s$^{-1}$ is reported. The electronic properties of the samples are systematically investigated by independent complementary techniques, enabling thorough characterization of the 3D Dirac (DSM) and Weyl (WSM) semimetal phases induced by the strains and magnetic field, respectively. Magneto-optical experiments, corroborated with band structure modeling, provide an exhaustive description of the bulk states in the DSM phase. The modeled electronic structure is directly observed in angle-resolved photoemission spectroscopy, which reveals linearly dispersing bands near the Fermi level. The first detailed study of negative longitudinal magnetoresistance relates this effect to the chiral anomaly and, consequently, to the presence of WSM. Observation of the $\pi$ Berry phase in Shubnikov-de Haas oscillations agrees with the topologically non-trivial nature of the investigated samples. Our findings establish {\aSn} as an attractive topological material for exploring relativistic physics and future applications.
\end{abstract}

\pagebreak

\section{Introduction} \label{section:Introduction}

Tin is a well-known functional material that has two allotropes stable in near-ambient conditions. Above \qty{13.2}{\celsius}, it exits in metallic form with a tetragonal crystal structure known as $\beta$-Sn or white tin. It is an industrially important material widely used in soldering alloys \cite{Plumbridge_JMatSciMatElec2007}, which has recently been proposed as a promising superconductor for fault-tolerant quantum computing \cite{Pendharkar_Science2021, Khan_Science2023, JArdine_ACSMatInt2023}. Below \qty{13.2}{\celsius}, $\beta$-Sn transforms into diamond-cubic form known as {\aSn} or grey tin. This low-temperature allotrope is attracting increasing attention of the scientific community after it was predicted to be a rare example of elemental topological material \cite{Fu_PRB2007, HuangLiu_PRB2017, Zhang_PRB2018}, with possible applications in spin-to-charge conversion \cite{RojasSanchez_PRL2016,Ding_AdvFunMat2021}, spin-orbit torque devices \cite{Ding_AdvMat2021, Binda_PRB2021} and as an anode for high efficient Na-based ion batteries \cite{zhu2022microsizedSn}. The promise of lower disorder, higher purity and lack of defects related to non-perfect stoichiometry compared to compounds and alloys makes grey tin an attractive material for topological electronics \cite{wang2020topotronics, gilbert2021topological}. Even both allotropes can be combined in a single electronic device showing a giant superconducting diode effect promising for superconducting spintronics \cite{ishihara2023giant}.

\begin{figure}[htbp]
  \centering
  \includegraphics[width = \textwidth]{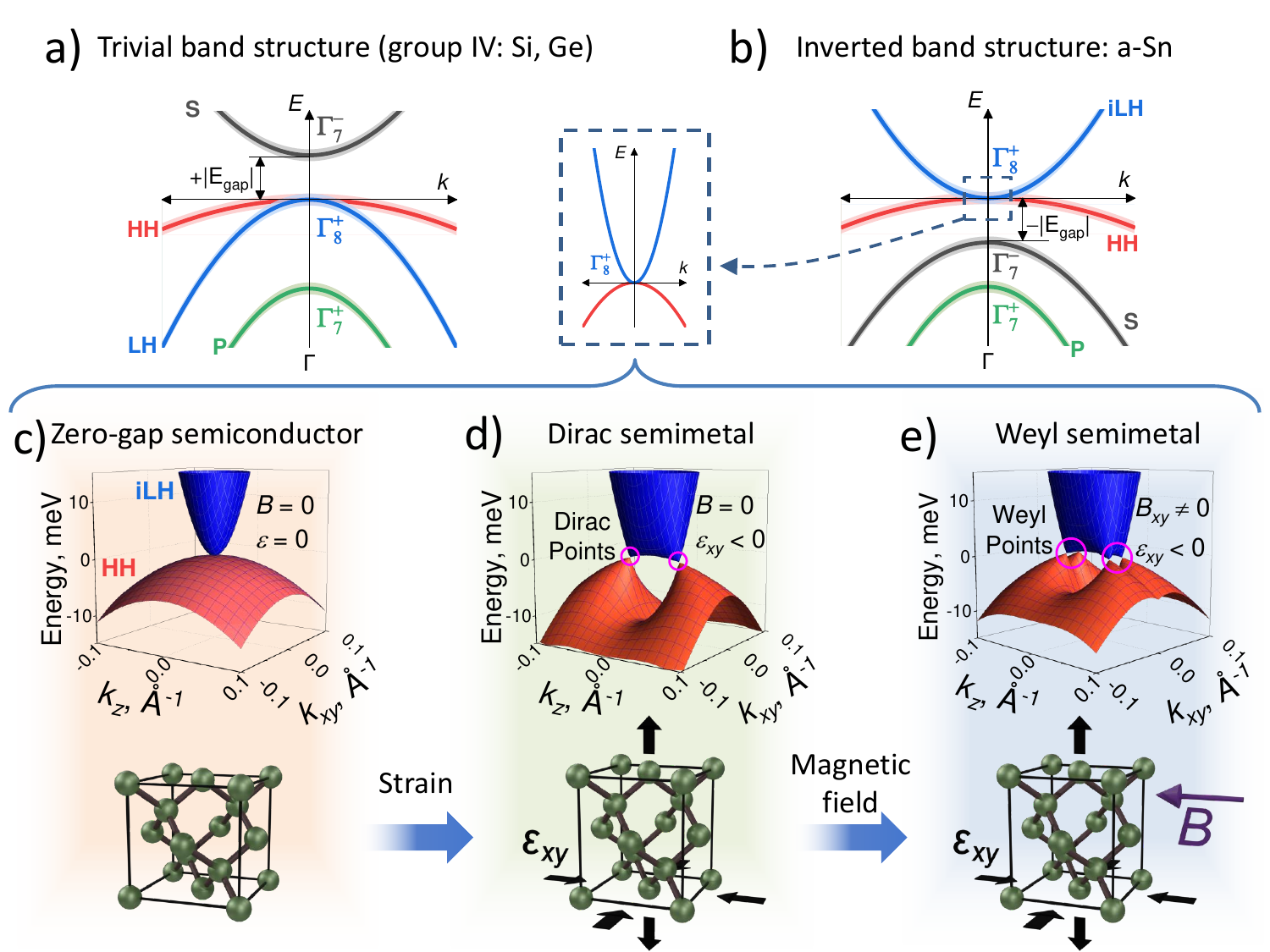}
  \caption{a) Schematic band structure of group-IV semiconductors with diamond crystal structure (Si, Ge) close to the $\Gamma$ point of the Brillouin zone. The conduction band is formed by S-states, the top valence band is formed by degenerate, P-type light hole (LH) and heavy hole (HH) states. b) In {\aSn}, which has the same crystal structure, {\kp} interaction switches the curvature of LH and S states and inverts their ordering. The inverted LH (iLH) states now form conduction band, while the S states - the lower valence band, forming "negative" gap with respect to Si and Ge. c-e) The evolution of LH and HH states from zero-gap semiconductor to Dirac semimetal under compressive biaxial strain to Weyl semimetal under additional external magnetic field. Lower panels show application of perturbations to the {\aSn} crystal. Dirac and Weyl points are marked by pink circles.}
  \label{fig:BandStr}
\end{figure}

In its bulk form, {\aSn} is a zero-gap semiconductor with a diamond crystal structure \cite{Groves_PRL1963, Pollak_PRB1970}. Unlike other group-IV elements like Si and Ge (\autoref{fig:BandStr}a), its conduction band and the valence band are degenerate at \textit{k} = 0, and originate from the light-hole (LH) and heavy-hole (HH) parts of the \textit{p}-type $\Gamma_{8}^{+}$ states, respectively (\autoref{fig:BandStr}b,c). The LH states acquire positive curvature, becoming electron-like or inverted LH (iLH), due to the {\kp} interaction with the lower $\Gamma_{7}^{-}$ valence band  \cite{Groves_PRL1963,averous1979symmetry}. Since this valence band is \textit{s}-type, the band structure and the gap of {\aSn} is inverted, which is a prerequisite for a non-trivial topology \cite{Fu_PRB2007, HuangLiu_PRB2017, Zhang_PRB2018}. The realization of topological phases (Figure 1d,e) additionally requires the lifting of the degeneracy of the $\Gamma_{8}^{+}$ conduction and valence bands. This can be achieved, for example, by applying external strains which break crystal symmetry protecting the zero gap \cite{Roman_PRB1972, HuangLiu_PRB2017, Xu_PRL2017}. In particular, as presented schematically in \autoref{fig:BandStr}d, a tetragonal distortion (compressive biaxial strain associated with tensile uniaxial strain) transforms {\aSn} into DSM, with two Dirac cones formed along parallel to the tensile strain axis. The application of an external magnetic field breaks the time reversal symmetry (TRS) and leads to the splitting of each of the Dirac cones into a pair of Weyl cones -- a Weyl semimetal (WSM) is formed, as shown in \autoref{fig:BandStr}e. Control over electronic properties of {\aSn} can also be realized by reduction of the sample's thickness: below some critical value, the TI phase is formed regardless of strain \cite{deCoster_PRB2018}. Further reduction of thickness to several monolayers leads to the formation of the buckled phase of {\aSn} called stanene \cite{Yong_PRL2013} known by its superconducting \cite{Liao_NatPhys2018, Falson_Science2020} and other intriguing characteristics \cite{Sahoo_AdvMatInt_2019}. In addition to the potential spintronic applications, such a multitude of possible phases makes {\aSn} a convenient topological playground for studying relativistic physics in a simple solid state system due to the so-called relativistic analogy \cite{Zawadzki_JourPhysCondMat2017}.

Experimental studies of {\aSn} are hindered by its low stability at room temperature \cite{Hillman_JourElecMat2022}. This limitation can be overcome by the fabrication of thin layers, e.g. by molecular beam epitaxy (MBE). The lattice-matched substrate can stabilize the crystal structure of grey tin \cite{Farrow_JourCrystGrowth1981}, enabling room-temperature sample characterization and processing. The common choices for the substrates are nearly lattice-matched CdTe and InSb \cite{Farrow_JourCrystGrowth1981}. Both introduce a small biaxial compressive strain in {\aSn} layers, paving the way for the engineering of the topological phases in this material. The topologically non-trivial character of grey tin grown on InSb was confirmed by extensive angle-resolved photoemission spectroscopy (ARPES) \cite{Barfuss_PRL2013, Ohtsubo_PRL2013, RojasSanchez_PRL2016, Xu_PRL2017, Rogalev_PRB2017, Scholz_PRB2018, Barbedienne_PRB2018, Rogalev_PRB2019, Madarevic_APLMat2020, Chen_PRB2022}. The main focus of those reports was on the characterization of 2D topological surface states, leaving bulk bands of DSM phase weekly explored \cite{Xu_PRL2017, Chen_PRB2022}. Additionally, intrinsically doped InSb creates a parallel conduction channel and makes it difficult to extract properties of {\aSn} from magneto-transport measurements \cite{Barbedienne_PRB2018, LeDucAnh_AdvMat2021, Y-Ding_PhysRevAppl2022, Madarevic_PSS2022, Ding_APL2022}. The recent attempts to grow on an insulating CdTe suffered from the lower surface quality of the substrate and resulted in reduced carrier mobility of {\aSn} \cite{Vail_PSSB2020, Y-Ding_JourVacSciTech2021}. Further improvement was received by inserting a thin conducting InSb interlayer between {\aSn} and an insulating CdTe substrate \cite{Ding_APL2022}. To the best of our knowledge, no systematic experimental investigation of the band structure and magneto-transport properties of {\aSn} grown directly on an insulating substrate has been conducted so far.

In this report, we demonstrate the reproducible MBE growth of {\aSn} on hybrid CdTe/GaAs(001) substrates. The fabrication of CdTe on GaAs is well-established \cite{Bicknell_APL1984, Karczewski_JCrystGrowth1998, Wichrowska_ActPhysPolA2014} and allows to obtain high-quality heterostructures exhibiting quantum Hall effect \cite{Karczewski_JCrystGrowth1998}. Therefore, we can simultaneously take advantage of the insulating character of the substrates and of the enhanced surface quality of the CdTe/GaAs structure. By expanding the thickness range of the films up to \qty{200}{\nano\metre}, well above the critical value for the transition to the TI phase, we were able to provide the first comprehensive characterization of the bulk Dirac fermions in this material. The diversity of applied experimental techniques allows us to extract a number of band- and material parameters. It results in a self-consistent description of {\aSn} in agreement with the predicted DSM and WSM phases. The high quality of the grown samples is confirmed by transmission electron microscopy (TEM) and X-ray diffraction (XRD), which also provide evidence of the presence of the in-plane compressive strain (\autoref{section:struc_char}). Magneto-optical studies yield a set of band structure parameters for the 4-band Hamiltonian (\autoref{section:magnetooptics}), which, together with the value of the strain determined by XRD, describes the bulk bands in agreement with the expected DSM phase. Modelled band structure, accompanied by a set of surface states, is observed in ARPES (\autoref{section:arpes}). The insulating character of the hybrid substrates used for epitaxy allowed us to perform the first systematic magneto-transport studies of {\aSn} and, particularly, to observe negative longitudinal magnetoresistance (NLMR). By ruling out alternative sources, we are able to relate this feature to the chiral anomaly - a signature of the WSM \cite{Nielsen_PhysLett1983, Son_PRB2013, Burkov_PRL2014}, in agreement with earlier prediction \cite{HuangLiu_PRB2017} (\autoref{section:transport}). Analysis of SdH oscillations clearly shows a $\pi$-Berry phase and a 3D nature of the carriers (\autoref{section:transport}), in consistency with the rest of the obtained results. 

\section{Results and discussion} \label{section:results}

\subsection{MBE growth and structural characterization}\label{section:struc_char}

\begin{figure}[htbp]
\centering
    \includegraphics[width = \textwidth]{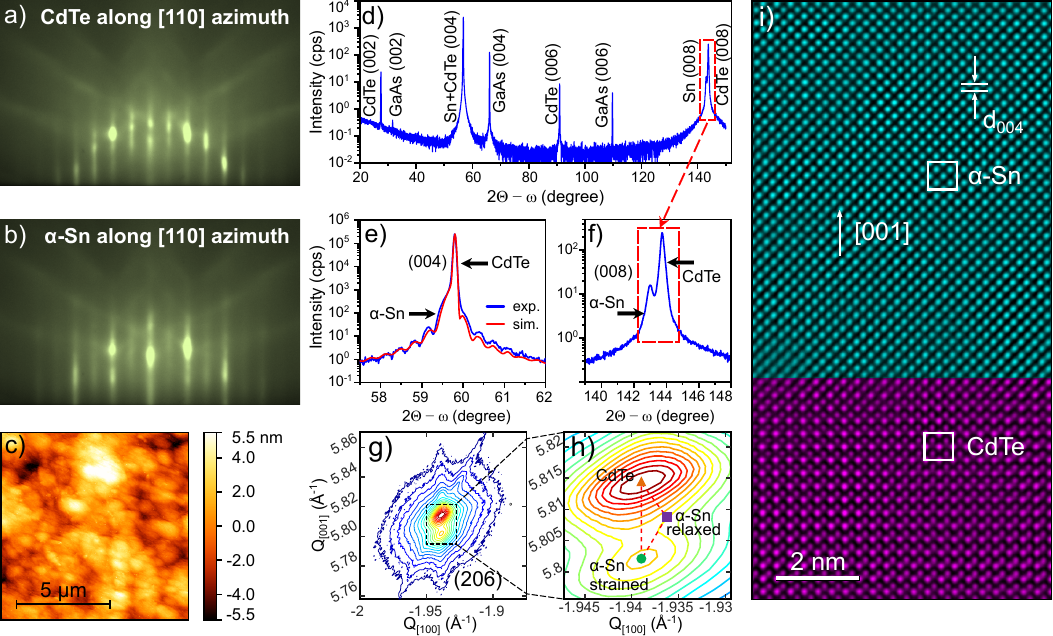}
    \caption{Structural and morphological characterization of {\aSn} films grown on CdTe/GaAs(001) substrates. a) and b) RHEED patterns along [110] azimuth of CdTe substrate and 30 nm {\aSn} epilayer, respectively. c) AFM \qty{10}{\micro\meter} $\times$ \qty{10}{\micro\meter} image of 50 nm thick {\aSn} epilayer with RMS $\sim$ \qty{2}{\nano\meter}. d) Wide-range 2$\theta$-$\omega$ scan of the Sn(150 nm)/CdTe/GaAs (001) film. e) (004) reflection HR-XRD pattern of 30 nm Sn film demonstrating thickness fringes, experimental and simulated curves are marked with blue and red lines respectively. f) Magnified (008) XRD pattern for the same film as in d) showing resolved grey tin and CdTe peaks. g), h) (206) asymmetric RSM for \qty{150}{\nano\meter} film from panel d; in h) the relaxation triangle calculated with Poisson's ratio $\upsilon$ = 0.298 is marked with a red dashed line, and the positions of CdTe, fully relaxed and fully strained {\aSn} are represented with an orange triangle, a violet square and a green circle, respectively. i) False-color, Fourier filtered high-resolution TEM image of the {\aSn}/CdTe interface of \qty{200}{\nano\meter} film (sample D) for which magneto-optical and magneto-transport studies were performed. The arrow indicates the growth direction. The squares represent the projections of the Sn and CdTe unit cells, note the difference in the intensities of the Cd and Te atomic columns.}
    \label{fig:structure}
\end{figure}

Layers of {\aSn} in the thickness range of \qtyrange{30}{200}{\nano\meter} were grown by MBE on (001) GaAs substrates with CdTe buffer ($\approx \qty{4}\mu m$ thick); details of growth are given in the \autoref{section:experimental}. The quality of the CdTe buffer obtained in the present work was monitored in-situ by reflection high-energy electron diffraction (RHEED), which showed sharp, streaky reflections on Laue semicircle and intense Kikuchi lines - evidence of a smooth single-crystalline surface (see \autoref{fig:structure}a). A similar RHEED pattern was obtained for the Sn epilayers subsequently grown on fresh CdTe (\autoref{fig:structure}b). Atomic force microscopy (AFM) of the Sn films also shows a smooth surface with root mean square (RMS) roughness on the order of \qty{2}{\nano\meter} (\autoref{fig:structure}c), resulting from the roughness of the underlying CdTe buffer. XRD $2\theta-\omega$ scans confirm the presence of the $\alpha$ phase of Sn. Only \{001\} maxima are present, with (004) and (008) reflections of Sn clearly visible and (002) and (006) reflections fully suppressed, as expected for the structural factor of diamond crystal structure (\autoref{fig:structure}d-f). The smoothness of the {\aSn}/CdTe interface and of the film surface is further confirmed by the presence of a large number of thickness fringes in the high-resolution XRD (HR-XRD) pattern obtained for thin (\qtyrange{30}{50}{\nano\meter}) layers (\autoref{fig:structure}e).

The small lattice mismatch between {\aSn} and CdTe ($\approx 0.1\%$) is expected to introduce biaxial in-plane compressive strain in the MBE-grown layers. Its presence is confirmed by the X-ray reciprocal space maps (RSM) of asymmetric (206) (\autoref{fig:structure}g, h) and (115) reflections. The epilayers were found to be fully strained, regardless of their thickness, with an in-plane lattice constant equal to that of the CdTe buffer, $a_{\parallel} = \qty{6.481}{\text{\AA}}$. Such in-plane bi-axial compressive strains induce uniaxial out-of-plane strain ($a_{\perp} = \qty{6.497}{\text{\AA}}$). Therefore, our layers are subject to tetragonal distortion, described previously in the introduction. The values of the unstained lattice constant $a_{0}$, as well as the in-plane and out-of-plane strains ($\epsilon_{\parallel}$ and $\epsilon_{\perp}$, respectively) were calculated using elastic constants of the bulk {\aSn} \cite{Price_PRB1971} (see \autoref{tab:structure}). The obtained value of $a_{0}$ agrees with the previous reports, and the amount of strains is similar to that in {\aSn} grown on InSb \cite{Farrow_JourCrystGrowth1981, Xu_PRL2017, Song_AdvEngMat2019, Asom_APL1989, Thewlis_Nature1954, Chen_PRB2022, Carrasco_APL2018}. We note that the DSM phase should also emerge if the uniaxial tensile strain is applied along [111] axis \cite{HuangLiu_PRB2017}. Therefore, we do not expect any qualitative difference between epitaxial layers grown on (001) and (111) substrates. We suggest that the difference between topological phases observed in earlier ARPES studies (TI or DSM) was due to different layer thicknesses rather than different substrate orientations.

The structural quality of the obtained samples is further confirmed by high-resolution scanning transmission electron microscopy (HR-STEM) (see \autoref{fig:structure}i). It showed a homogeneous epitaxial {\aSn} layer with a smooth {\aSn}/CdTe interface, characterized by the absence of mismatch dislocations. The absence of cracks in the volume of the layer suggests uniform strain distribution, even in 200 nm thick films. Importantly, neither the XRD nor the TEM studies revealed the presence of $\beta$-Sn inclusions. Observed tetragonal distortion in {\aSn} induced by compressive in-plane strains is a prerequisite for the formation of the DSM and WSM phases \cite{HuangLiu_PRB2017, Zhang_PRB2018}. We note that despite the room-temperature characterization, we expect the same amount of strains at lower temperatures due to similar thermal expansion of the film, buffer, and substrate \cite{Thewlis_Nature1954, Jochym_PhysRevMat2022, SpringerHandbookOfMaterials}.

\begin{table}
    \caption{Lattice constants and strain in 150 nm {\aSn} and CdTe calculated from the RSM asymmetric (206) and (115) reflections.}
    \centering
    \begin{tabular}{c c c c c c c}
        \hline
        Material & Reflection & $a_{\parallel}$ [\AA] & $a_{\perp}$ [\AA] & $a_{0}$ [\AA] & $\epsilon_{\parallel}$ [\%] & $\epsilon_{\perp}$ [\%] \\
        \hline
        Sn & (206) & 6.481 & 6.497 & 6.490 & -0.14 & 0.12 \\
        CdTe & (206) & 6.481 & 6.484 & 6.4822 & -0.02 & 0.02 \\
        Sn & (115) & 6.481 & 6.498 & 6.490 & -0.14 & 0.12 \\
        CdTe & (115) & 6.481 & 6.484 & 6.4823 & -0.02 & 0.03 \\
        \hline
    \end{tabular}
    \label{tab:structure}
\end{table}

\subsection{Model of the band structure and magneto-optical studies}\label{section:magnetooptics}

The bulk band structure of {\aSn} close to the $\Gamma$ point, presented schematically in \autoref{fig:BandStr}, can be described by the {\kp} model based on Pidgeon-Brown \cite{Groves_JPhysChemSol1970, Pidgeon_PhysRev1966} and Bir-Pikus \cite{BirPikus_1974, Laude_PRB1971} Hamiltonians. Details of the model can be found in the Supporting Information (SI). The Pidgeon-Brown Hamiltonian depicts the eight bands that are close to the Fermi energy $E_{F}$ and to the $\Gamma$-point of a bulk {\aSn} while taking the remote bands effects up to the second order in $k$. The resulting modelled band structure consists of four bands that are twofold spin degenerate: the split-off (P) $\Gamma_7^+$ band, the heavy hole (HH) $\Gamma_8^+$ band, the p-type inverted iLH $\Gamma_8^+$ band, and the s-type (S) $\Gamma_7^-$ band, whose energies at $\Gamma$, before taking the strain into account, are $-\Delta$, 0, 0, and $E_{g}$, respectively. The interaction between the s-type band and the other three p-type bands is defined by the interband momentum matrix element $P = \sqrt{3/2} \hbar \nu$, where $\nu$ is the electron velocity. In particular, it is responsible for the conversion of the LH band (\ref{fig:BandStr}a) to the p-type inverted iLH band (\ref{fig:BandStr}b). Remote band effects are described by the parameters $\gamma_1$ and $\gamma$ that account for a parabolic correction to the effective masses of the iLH and HH bands. A non-zero magnetic field introduces an extra dimensionless parameter arising from remote far band effects, $\kappa$, which acts as a small correction to the g-factor. To account for the strain induced in the samples, the Bir-Pikus Hamiltonian was added. The compressive biaxial strain has three effects on the band structure: (i) renormalization of the s-p gap $E_{g}$ depending on the hydrostatic deformation potential; (ii) a \textit{k}-independent mixing of the iLH and P bands that is very small in the case of {\aSn} because the energy distance between iLH and P is large, as will be determined later; and most importantly (iii) lifting of the degeneracy of iLH and HH at $\Gamma$ that is directly related to the strain magnitude using the shear deformation potential \textit{b}. The iLH band is then shifted in energy by $\Omega = - 2b(\epsilon_{\parallel} - \epsilon_{\perp})$. Note that this model is more complete than previous {\kp} calculations for strained {\aSn} that consider the iLH, HH \cite{Roman_PRB1972, HuangLiu_PRB2017, Zhang_PRB2018} and S \cite{deCoster_PRB2018} bands using the Luttinger Hamiltonian \cite{Luttinger_PhysRev1956}. All the parameters mentioned above, describing the complete band structure of the strained {\aSn}, can be determined by magneto-optics. Although such a study was performed for bulk {\aSn} \cite{Groves_JPhysChemSol1970}, measurements on strained MBE-grown layers did not provide the complete set of band structure parameters \cite{Hoffman_PRB1989, Wojtowicz_SemSciTech1990, Yuen_JourCrystGrowth1991}. For instance, the strain-induced separation between the iLH and HH bands remains unknown. We have selected 4 layers with the highest carrier mobility, having thicknesses of \qty{50}{\nano\meter}, \qty{100}{\nano\meter}, \qty{150}{\nano\meter} and \qty{200}{\nano\meter} (samples A to D, respectively), for magneto-optical characterization. We present here the results obtained for sample D with a magnetic field normal to the sample plane ($B^{\text{normal}}$). Samples A, B, and C exhibit similar results (see SI for more details.).

\begin{figure}[htbp]
    \includegraphics[width = \textwidth]{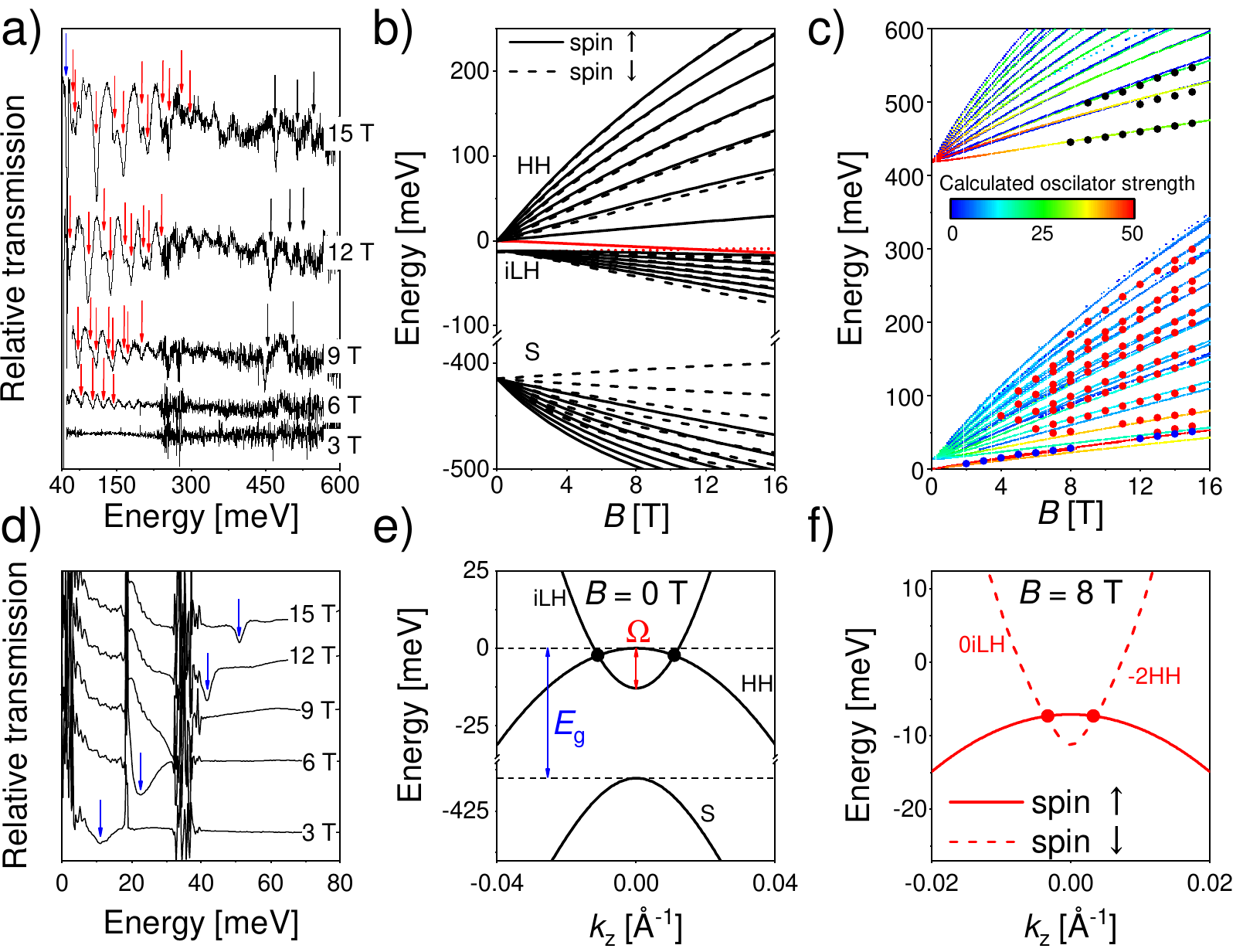}
    \caption{a) Magneto-optical spectra at $T$ = \qty{4}{\kelvin} and different magnetic fields for sample D. Red and black arrows denote the two interband series, and the blue arrow indicates cyclotron resonance (shown in more details in d). b) Calculated Landau levels using the Pidgeon-Brown-Bir-Pikus Hamiltonian and the parameters listed in Table 2. Solid and dashed lines refer to the spins. The two Landau levels ($N$ = -2 from the HH band, and $N$ = 0 from the iLH band) are highlighted in red. c) Fan chart gathering the absorption data presented in a) (dots) and the modelled transitions between Landau levels of b) (solid lines). The calculated oscillator strengths of the transitions are represented by the color scale in units of $10^{12} \text{ cm}^2 \text{s}^{-2}$. d) Far-infrared part of the transmission spectra, with cyclotron resonance indicated by blue arrows. e) Deduced band structure of a strained bulk {\aSn} at zero fields. The P band is not shown for scaling reasons. Dirac nodes are represented by black circles. f) Calculated $k_z$-dispersion of the two Landau levels highlighted in a), at $B = \qty{8}{\tesla}$. Weyl nodes are represented by red circles. Solid and dashed lines refer to the two spin components.}
    \label{fig:magnetooptics}
\end{figure}

\autoref{fig:magnetooptics}a shows the impressively sharp and numerous magneto-optical oscillations that were measured at different magnetic fields. Their field dependence indicates that they originate from transitions between Landau levels. One intraband cyclotron resonance is observed at low energy and displays an extremely sharp absorption line (\autoref{fig:magnetooptics}d). The full width at half-maximum of \qty{2}{\meV} is evidence of high crystalline quality and high carrier mobility. In addition to the cyclotron resonance, two distinct series of interband transitions are unravelled in the magneto-optical spectra (see absorption lines indicated by red and black arrows in \autoref{fig:magnetooptics}a). The presence of two series indicates that three bands are directly involved in the observed magneto-optical oscillations. Furthermore, the spectra clearly show that absorptions come in pairs, which is the signature of the Zeeman splitting.

Under $B^{\text{normal}}$, the modelled bands disperse as Landau levels that are calculated and shown in \autoref{fig:magnetooptics}b. They are used to calculate the magneto-optical transitions that fit the absorption observed in \autoref{fig:magnetooptics}a. This fit is plotted in \autoref{fig:magnetooptics}c and shows perfect agreement with the experimental data. The two interband series are attributed to HH – iLH transitions at low energy and to S – iLH transitions above \qty{400}{\meV}. The band parameters used for the fit are listed in \autoref{tab:magnetooptics} and the determined band structure, calculated accordingly, is plotted in \autoref{fig:magnetooptics}e.

The band structure of {\aSn} is of the same type as that of HgTe. Indeed, the gap $E_g$ is found to be large and negative, meaning that the S band lies under the iLH and HH bands, i.e., the band structure is inverted. Therefore, properties similar to HgTe are expected, in particular, our results suggest that a quantum spin Hall phase occurs in {\aSn} quantum wells (see SI for additional details) \cite{Kufner_PRB2016}. The gap of {\aSn} is found to be relatively large as indicated by the nearly linearly-dispersive magneto-optical transitions, and more directly by the zero-field extrapolation of the second interband series (red arrows in \autoref{fig:magnetooptics}a). The effective mass of the HH band is given by $\gamma_{1}^{-1} = 0.195 m_0$, and that of the iLH band writes:

\begin{align*}
    \frac{m_0}{m_{eff}} = -\gamma_{1} + \frac{2 E_{P}}{3 (\Omega - E_{g})},
\end{align*}

where $m_0$ is the electron rest mass and $E_P = 2 m_0 P^2 = \qty{25.4}{\eV}$ in our case. Therefore, the magneto-optical measurements result in $m_{eff} = 0.027 \pm 0.001 m_0$, which is in reasonable agreement with the transport experiments performed in this work and with the literature \cite{Groves_JPhysChemSol1970, Yuen_JourCrystGrowth1991}. Note that the parameter $\Omega$, which originates from strain, accounts for a 4\% correction to the effective mass. It is determined as $\Omega = \qty{-13}{\meV}$ from the extrapolation at the zero field and the shape of the first interband series (black arrows in \autoref{fig:magnetooptics}a). This parameter is directly related to the strain following $\Omega = - 2 b (\epsilon_{\parallel} - \epsilon_{\perp})$, therefore, using the strain values measured by XRD in this work, we can deduce the shear deformation potential as $b = \qty{-2.5}{\eV}$, in perfect agreement with Ref. \cite{Roman_PRB1972}. This value is very high compared to other semiconductors such as Si, HgTe or III-Vs, which display $\qty{-2}{\eV} \leq b \leq \qty{-1.3}{\eV}$ \cite{Pollak_1990}.

Magneto-optical experiments have demonstrated that strain is a particularly efficient tool to tune the band structure in {\aSn}, and a value as small as $-0.14\%$ is sufficient to induce a DSM phase. Note that $\Omega < 0$ in our measurements, which means that the iLH band goes under the HH band near the $\Gamma$-point (see \autoref{fig:magnetooptics}e). Therefore, it is responsible for the emergence of two Dirac nodes located around

\begin{align*}
  \boldsymbol{k} = (0;0;\pm k_{D}) = \left(0;0;\pm \frac{1}{P}\sqrt{\frac{3 \Omega E_g}{2}} \right).
\end{align*}

The distance between the two Dirac nodes is then determined as \qty{0.018}{\text{\AA}^{-1}} in the investigated strained {\aSn} epilayers with $\epsilon_{\parallel} = -0.14\%$. These nodes are two-fold degenerate at $B = 0$ and are thus Dirac in nature. When the magnetic field is applied, the two spin components are split and Dirac nodes are turned into Weyl nodes. They are formed by the remaining $k_z \parallel B$ dispersion of the $N = -2$ Landau level originating from the HH band, and the $N = 0$ Landau level from the iLH band. These Landau levels are highlighted in \autoref{fig:magnetooptics}b and their $k_z$-dispersion has been calculated (see \autoref{fig:magnetooptics}f). Note that while the magnetic field $B^{\text{normal}}$ changes the nature of the nodes from Dirac to Weyl, it also brings them closer and eventually leads to their annihilation.

\begin{table}
    \caption{The band parameters of the strained {\aSn} determined by magneto-optics at \qty{4}{\kelvin}.}
    \centering
    \begin{tabular}{c c}
        \hline
        Parameter & Value \\
        \hline
        $E_{g}$ & $\qty{-415 \pm 5}{\meV}$\\
        $\nu$ & $\qty{1.22 \pm 0.01 e6}{\metre\per\second}$ \\
        $P$ & $\qty{9.805 \pm 0.075}{\eV \times \text{\AA}}$ \\
        $\Delta$ & $\qty{800 \pm 50}{\meV}$ \\
        $\gamma_1$ & $5.15 \pm 0.25$ \\
        $\gamma$ & $0$ \\
        $\kappa$ & $-3.4 \pm 0.5$ \\
        $\Omega$ & $\qty{-13 \pm 2}{\meV}$ \\
        $b$ & $\qty{-2.5 \pm 0.4}{\eV}$ \\
        $2 k_{D}$ & $\qty{0.018 \pm 0.002}{\per\text{\AA}}$\\
        \hline
    \end{tabular}
    \label{tab:magnetooptics}
\end{table}

Our magneto-optical experiments demonstrate the presence of a Dirac semimetal phase in compressively strained {\aSn} by accurately measuring $\Omega = \qty{-13}{\meV}$ for $\epsilon_{\parallel} = -0.14\%$. Additionally, the band structure has been accurately calculated by experimentally determining the set of band structure parameters.

\subsection{ARPES}\label{section:arpes}

\begin{figure}[htbp]
    \centering
    \includegraphics[width = \textwidth]{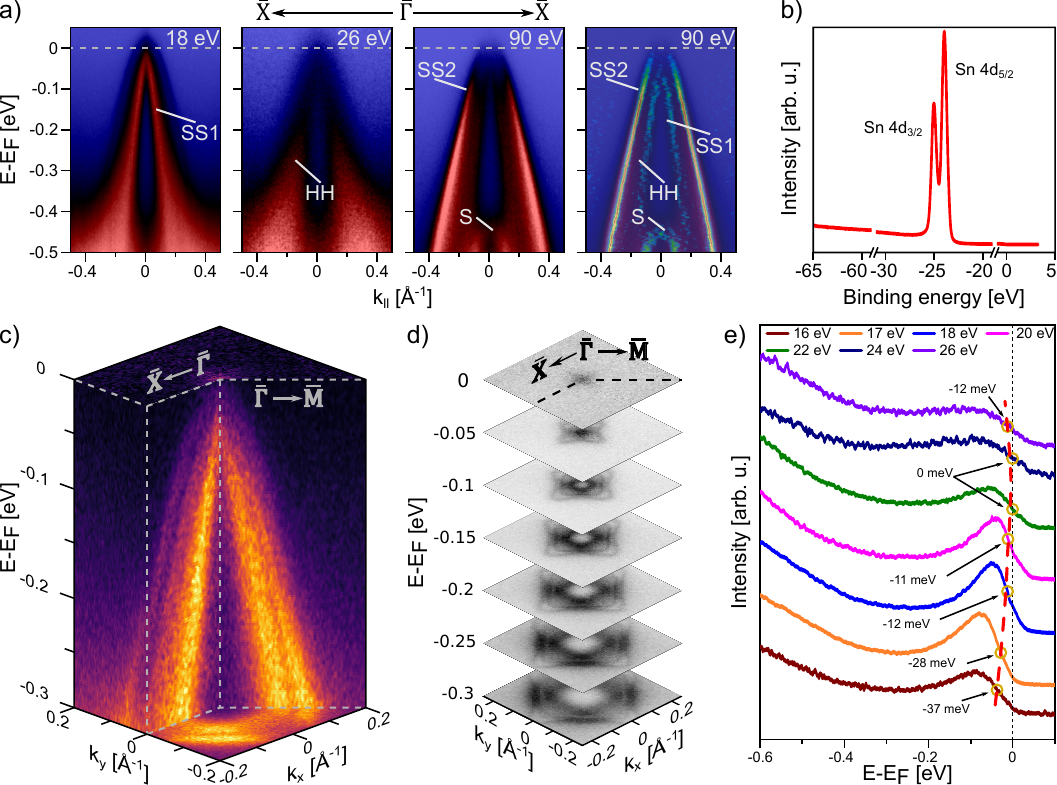}
    \caption{ARPES studies of the 150 nm thick {\aSn} epilayer grown on (001) CdTe/GaAs substrate. a) ARPES images at various photon energies; Dirac-like surface states (SS1), Dyakonov-Khaetski states (SS2), bulk states (HH and S) are marked respectively (the last image is 2D curvature at $E_{ph}$ = \qty{90}{\eV}. b) Core-level spectrum of the film, c) 3D ARPES map, and d) corresponding constant energy contours ($E_{ph}$ = \qty{20}{\eV}). e) Constant energy contour of c) obtained at $E-E_{F}$ = \qty{0.15}{\eV}, demonstrating HH and SS1 states as well as a pair of DK SS2 states.}
    \label{fig:ARPES}
\end{figure}

The band structure, described in \autoref{section:magnetooptics}, was directly investigated in ARPES experiments. Despite multiple reports for {\aSn} layers grown on InSb \cite{Barfuss_PRL2013, Ohtsubo_PRL2013, RojasSanchez_PRL2016, Xu_PRL2017, Rogalev_PRB2017, Scholz_PRB2018, Barbedienne_PRB2018, Rogalev_PRB2019, Madarevic_APLMat2020, Chen_PRB2022}, ARPES studies for samples fabricated on CdTe are still lacking. Here, we present an analysis of photoemission spectra obtained for the dedicated series of (001) {\aSn}/CdTe/GaAs heterostructures with varied thicknesses. The samples were transferred to the synchrotron facility without breaking ultra-high vacuum (UHV) (see \autoref{section:experimental}), before performing any characterization other than RHEED. The observed spectra, primarily acquired at room temperature, turned out to be thickness-independent and their properties are summarized in \autoref{fig:ARPES}. The remaining data can be found in the SI. The observed features are consistent with the previous results, obtained for layers grown on InSb, and with the model described in \autoref{section:magnetooptics}. Regardless of their dispersion and topological character, all of the states revealed by our ARPES studies originate only from {\aSn}, as can be seen from the core level spectrum in \autoref{fig:ARPES}b.

The band structure close to the $\Gamma$ point of the Brillouin zone consists of the bulk $\Gamma_8^+$ heavy hole (HH) and $\Gamma_7^-$ s-type (S) bands. The energy splitting between the S and HH bands at $k_{\parallel} = 0$, $|E_{g}|=0.40\pm 0.02$~eV, is consistent with our magneto-optical measurements. These bulk bands are accompanied by two types of surface states, labelled SS1 and SS2. As can be seen in \autoref{fig:ARPES}a, they have different spectral weights depending on the photon energy and can be resolved after applying the curvature procedure \cite{Zhang_RevSciInst2011}. Both SS1 and SS2, as well as bulk HH states, are also clearly seen along $\Bar{\Gamma} - \Bar{M}$ direction (\autoref{fig:ARPES}c) and show anisotropy consistent with previous reports of {\aSn}/InSb (001) \cite{Scholz_PRB2018, Chen_PRB2022}. To elucidate the topography of the states in more detail, we present a set of experimental constant energy contours (CECs) (\autoref{fig:ARPES}d). While SS1 is isotropic and shows a circular cross-section, the SS2 and HH states show anisotropic dispersion. We suggest that SS1, with linear Dirac-like dispersion, emerges from the band inversion and therefore has a topological origin, similar to another surface state (SS3) presented in SI and previous experimental works for strained {\aSn} on (001) InSb \cite{Rogalev_PRB2017,Chen_PRB2022}. The origin of SS2 has recently been attributed to the Rashba effect, originating from inversion symmetry breaking at the surface, as demonstrated by the investigation performed on \aSn/InSb (001) \cite{Chen_PRB2022}. Indeed, SS2 states come in pairs as one expects for Rashba states, as clearly seen in \autoref{fig:ARPES}e. However, the absence of the Rashba split states with a reduced {\aSn} thickness to several monolayers, while observing the TI phase, raises questions. Moreover, no reliable evidence of the Rashba effect has been observed in other experiments, such as magneto-transport measurement (see \autoref{section:transport}). Consequently, alternative origins should be considered, such as the presence of Dyakonov-Khaetskii (DK) or Volkov-Pankratov (VP) states \cite{Dyakonov_JETP1981, Pankratov_SolStateComm1981}. The existence of such massive surface states has recently been proposed for strained Luttinger semimetals \cite{Khaetskii_PRB2022, Kharitonov_arXiv2022} and the physical origin of their appearance is the hybridization of Dirac spectrum with HH states.

As determined in \autoref{section:magnetooptics}, the tetragonal distortion present in the studied {\aSn} layers ($\epsilon_t=\epsilon_{\perp}-\epsilon_{\parallel}=0.26 \%$ according to the RSM presented in \autoref{fig:structure}g,h) results in the presence of a DSM phase. Therefore, spectra of the bulk HH Dirac-like states are expected to change with the variation of the photon energy. While our data shows a modest dispersion with varying photon energies (detailed in SI), it's crucial to acknowledge that this could arise from the suppression of the SS2 state intensity due to changes in the matrix element. Moreover, the close proximity of the two Dirac points along $k_z$ precludes their definitive resolution with ARPES. On the other hand, the band structure analogous to the observed structure has been reported in recent studies elucidating the DSM phase in {\aSn} grown on InSb (001) substrates, where akin strains of similar magnitude and sign are present \cite{Xu_PRL2017, Scholz_PRB2018, Chen_PRB2022}. Additionally, certain features observed in our experiment, such as point-like Fermi surface and the existence of SS1, SS2, and SS3 are theoretically predicted for the DSM phase \cite{HuangLiu_PRB2017, Zhang_PRB2018, Carrasco_APL2018, Rogalev_PRB2017}. Thus, we believe that, in agreement with our magneto-optics measurements, considering previous experimental and theoretical works we unequivocally validated the existence of DSM in (001) {\aSn}/CdTe/GaAs heterostructures. The consistency between ARPES and magneto-optics data stems from the expectation that the band structure of {\aSn} near the $\Gamma$ point should remain highly similar at both room temperature and low temperature (4~K). Consequently, the results obtained from ARPES and magneto-optics inherently align with each other. Furthermore, the GaAs substrate, CdTe buffer, and {\aSn} exhibit similar coefficients of thermal expansion, minimizing temperature-induced changes in the band structure. Thus, we believe that, in agreement with our magneto-optics measurements, considering previous experimental and theoretical works we unequivocally validated the existence of DSM in (001) {\aSn}/CdTe/GaAs heterostructures.

However, the notable temperature disparity between room temperature and 4~K introduces complexities when comparing the two methods. Such variations can result in changes in band dispersions and the Fermi level, complicating direct comparisons. In our study, we have employed a simple {\kp} model to interpret our magneto-optical data, which provides adequate descriptions of the band structure only near high-symmetry points of the Brillouin zone. Thus, it may yield significant discrepancies for larger wave vector ($k$) values. Furthermore, the {\kp} model relies on phenomenological band parameters (\autoref{tab:magnetooptics}), which are not known a priori at room temperature and do not account for surface states.

\subsection{Transport properties}\label{section:transport}

\begin{figure}[htbp]
  \captionsetup[subfigure]{labelformat = empty}
\centering
  \subfloat[\label{fig:sample_sketch}]{\includegraphics[width = 0.33\textwidth]{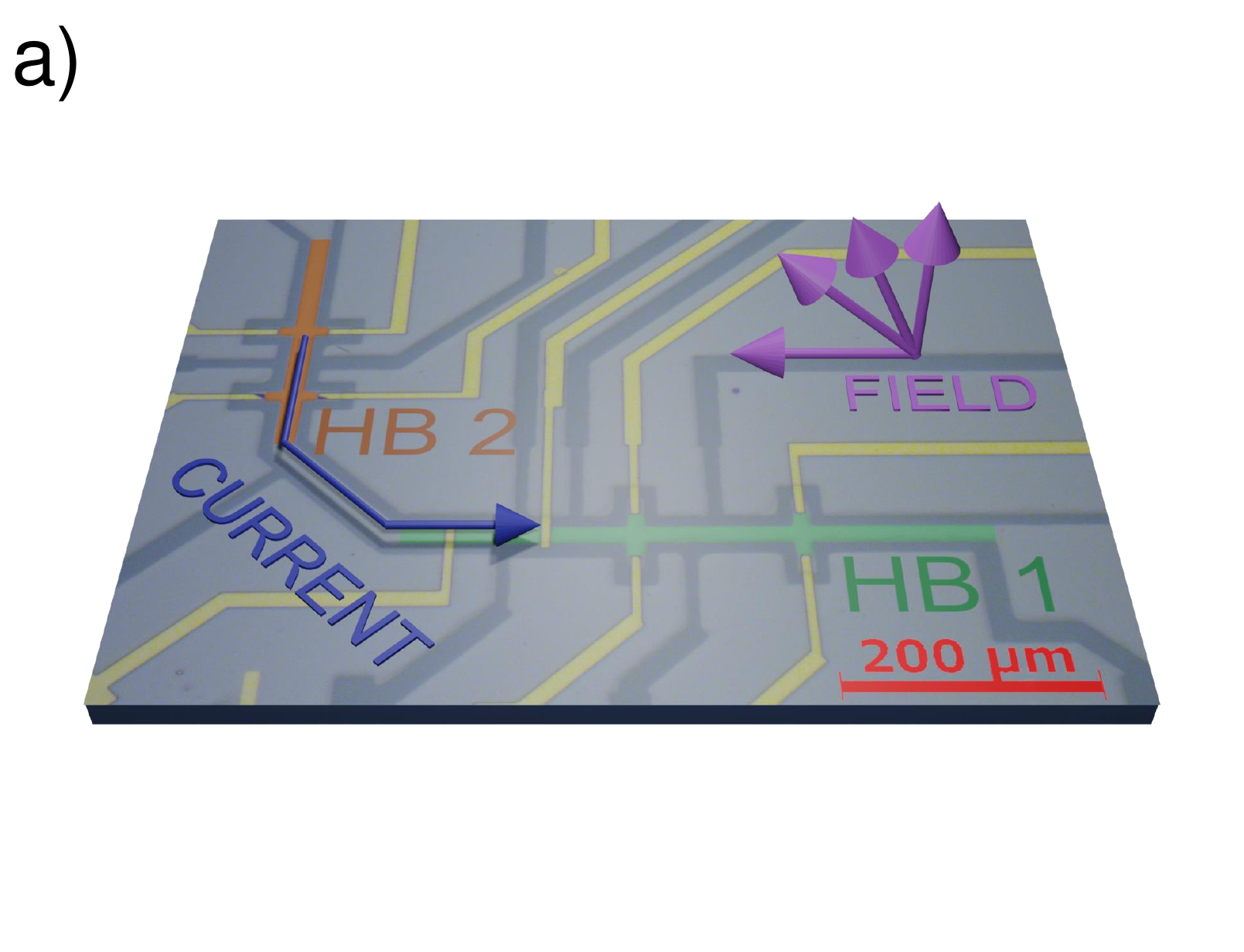}}\hfill
  \subfloat[\label{fig:Hall_200nm}]{\includegraphics[width = 0.33\textwidth]{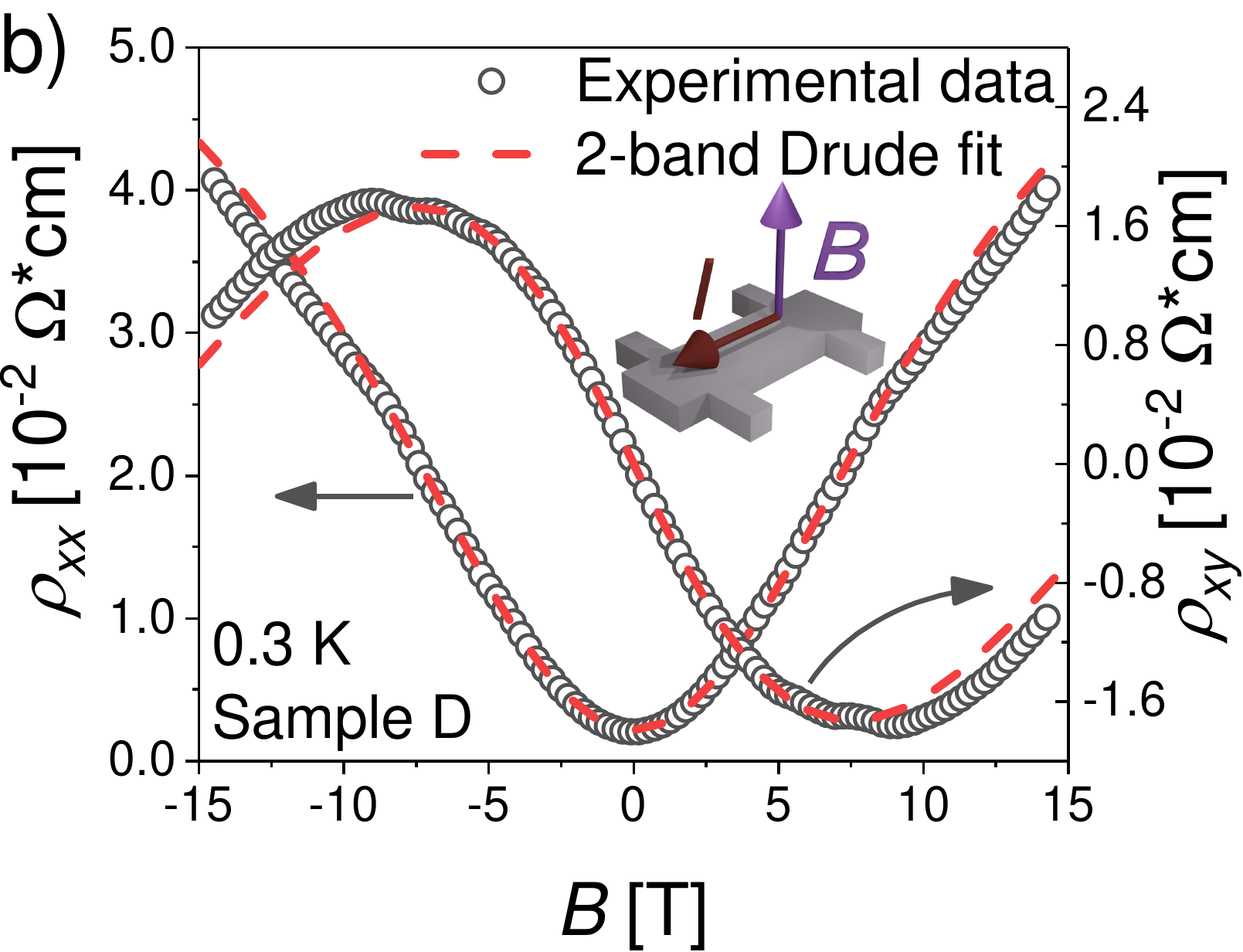}}\hfill
  \subfloat[\label{fig:SheetDensity_vs_Thickness}]{\includegraphics[width = 0.33\textwidth]{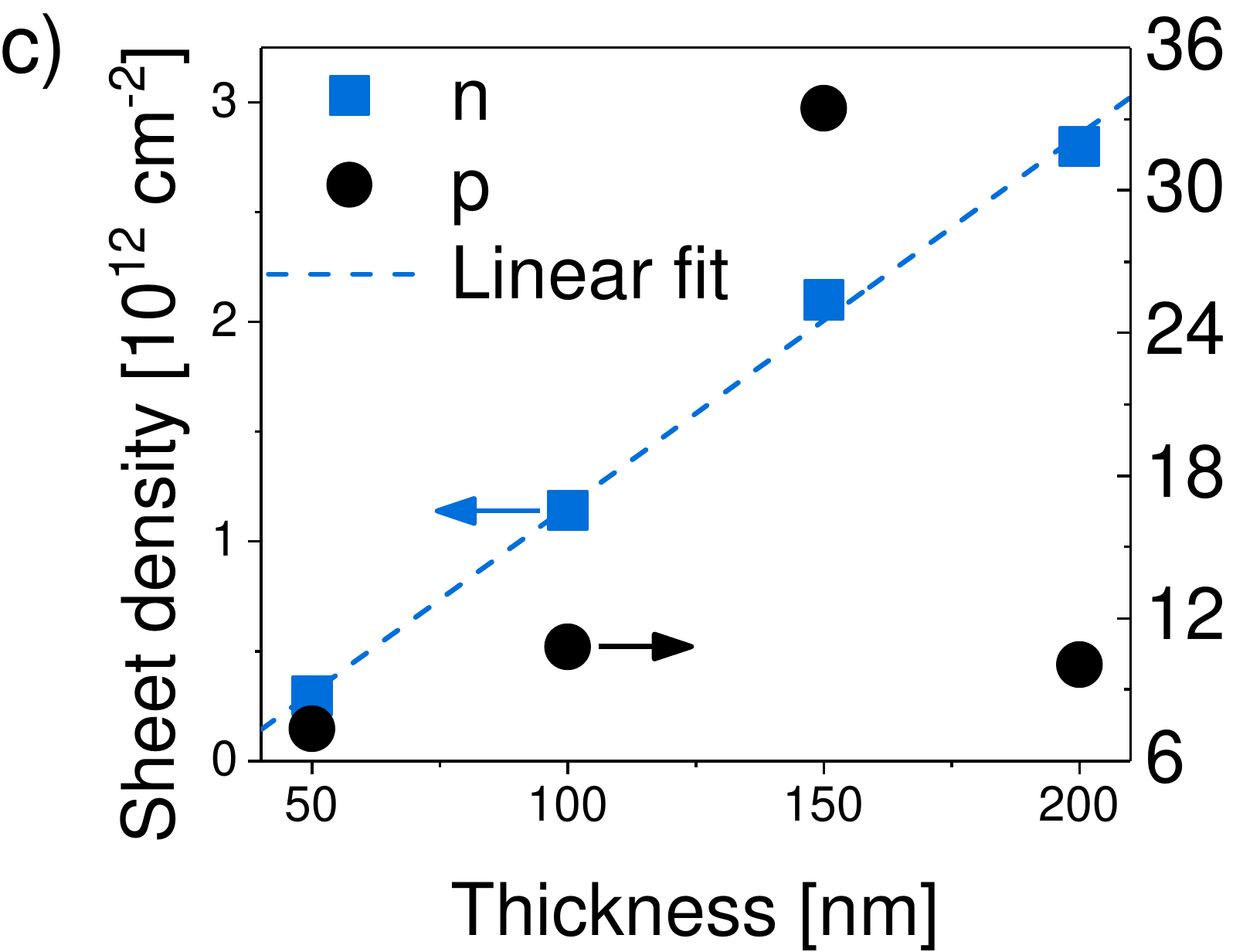}}\hfill
  \caption{\protect\subref{fig:sample_sketch}) False color optical image of a 2-arm microstructure patterned for magneto-transport studies. Light-gray: the surface of {\aSn}, dark-gray: etched mesa, yellow: Ti/Au contacts. Blue and violet arrows show the possible directions of the magnetic field and the direction of the current flow. Perpendicular arms, shown in light green and orange, are denoted as HB1 and HB2. In $B^{\text{in-plane}}$, the current in HB1 is parallel to the magnetic field ($B^{\text{in-plane}}_{\parallel}$), and perpendicular to the field in HB2 ($B^{\text{in-plane}}_{\perp}$). Both arms are equivalent in $B^{\text{normal}}$. In HB1, voltage probes spanning across the conduction channel (transverse contacts) are fabricated in addition to standard probes at the edges of the structure (edge contacts). \protect\subref{fig:Hall_200nm}) Longitudinal and Hall resistivity (open symbols) of sample D as a function of the magnetic field, measured at $T = \qty{0.3}{\kelvin}$ in HB1. Red, dashed lines represent the result of the fitting of the 2-band Drude model in the range $B = \pm \qty{9}{\tesla}$ (curves were extended to cover the entire range of magnetic fields). \protect\subref{fig:SheetDensity_vs_Thickness}) Thickness dependence of the electron and hole sheet density. The dashed line represents linear scaling for electrons.}
  \label{fig:Results_Bperp}
\end{figure}

To get further insight into the electronic structure of {\aSn}, we performed magneto-transport studies of samples A - D, also examined by magneto-optics. Previous reports show that the high-temperature electronic transport of {\aSn} is influenced by the contribution of thermally excited carriers from the L-band \cite{Vail_PSSB2020, Lavine_JPhysChemSo1971, Tu_APL1989, Hoffman_PRB1989}. This is also true for the studied layers, as discussed in SI. Therefore, field-dependent transport studies presented here were restricted to $T<\qty{100}{\kelvin}$, where only carriers from the vicinity of the $\Gamma$-point contribute to the electronic transport due to the complete freeze-out of electrons from the L-band. Measurements were performed on samples patterned in standard Hall bars or microstructures with two perpendicular arms, presented in \autoref{fig:sample_sketch}, in two distinct configurations: in addition to $B^{\text{normal}}$, examined in magneto-optics, magneto-transport was studied in the orientation with magnetic field laying in-plane ($B^{\text{in-plane}}$). More details on sample preparation and experimental setup can be found in the \autoref{section:experimental}.

In $B^{\text{normal}}$, the longitudinal resistivity, $\rho_{xx}$, of all samples is positive and does not show any signs of saturation up to \qty{14.5}{\tesla}, as presented in \autoref{fig:Hall_200nm} for sample D. Low-field longitudinal and Hall ($\rho_{xy}$) resistivity can be simultaneously described by the 2-band Drude model (see \autoref{fig:Hall_200nm} and SI). Fitting to the experimental data provides carrier densities and mobilities of electrons and holes for all layers. Mobility reaches $\mu_{n} = \qty[mode = text]{1.9e4}{\centi\meter^{2}\volt^{-1}\second^{-1}}$ for \textit{n}-type carriers in sample D, confirming its high quality already observed in magneto-optical spectra. Electron densities are of the order of $n \sim \qty[mode = text]{1e17}{\centi\meter^{-3}}$ for all layers. The observed multicarrier behavior is in line with previous reports on {\aSn} grown on CdTe \cite{Vail_PSSB2020, Lavine_JPhysChemSo1971, Tu_APL1989, Hoffman_PRB1989}. However, so far no clear explanation of the source of \textit{p}-type carriers has been given. We note that the electron sheet density (\autoref{fig:SheetDensity_vs_Thickness}) scales linearly with the thickness of the layer, whereas no scaling is observed for holes. Thus, we have to conclude, that electrons are related to the bulk properties of iLH band in the studied {\aSn} epilayers. Moreover, magneto-optical transmission spectra described above, probe bulk properties and are consistent with the \textit{n}-type bulk carriers. The observed \textit{p}-type carriers may originate from the surface band bending due to the oxide formation, and from the formation of the heavily \textit{p}-doped SnTe monolayer at the {\aSn}/CdTe interface, and also from the (trivial) interface states observed in ARPES. Also, they are not expected to alter our conclusions derived from magneto-optical measurements. Due to their low mobility, they do not contribute to Landau quantization and optical transitions associated with Landau levels. Importantly, the CdTe/GaAs substrate does not contribute to the transport properties of the studied structures, contrary to the layers fabricated on InSb \cite{Barbedienne_PRB2018, LeDucAnh_AdvMat2021, Y-Ding_PhysRevAppl2022, Madarevic_PSS2022, Ding_APL2022}. Magneto-transport features in $B^{\text{normal}}$ beyond the Drude model, e.g. SdH oscillations, will be described later in the text.

\begin{figure}[htbp]
  \captionsetup[subfigure]{labelformat = empty}
  \centering
  \subfloat[\label{fig:NLMR_50nm}]{\includegraphics[width=0.49\textwidth]{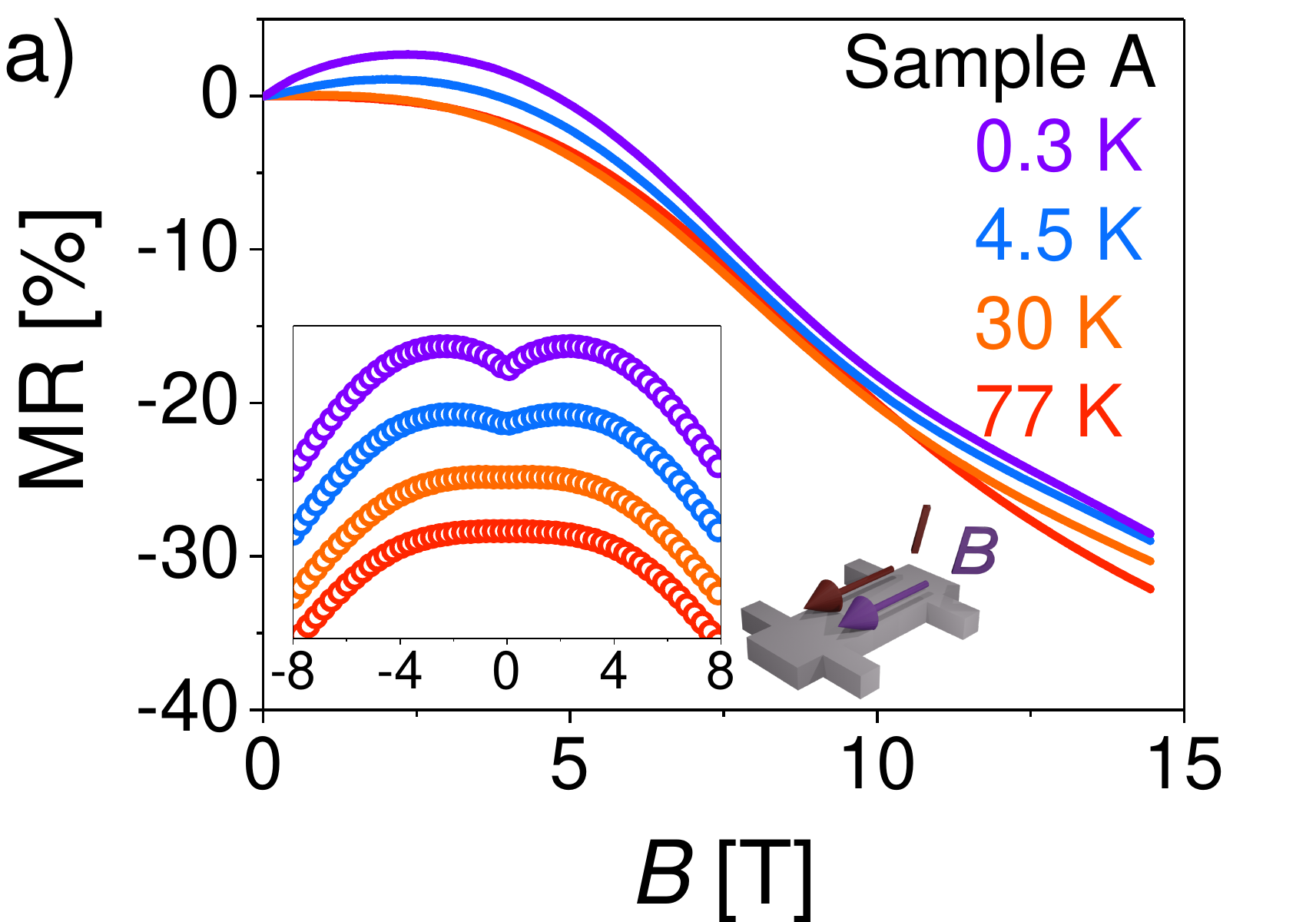}}\hfill
  \subfloat[\label{fig:NLMR_150nm}]{\includegraphics[width=0.49\textwidth]{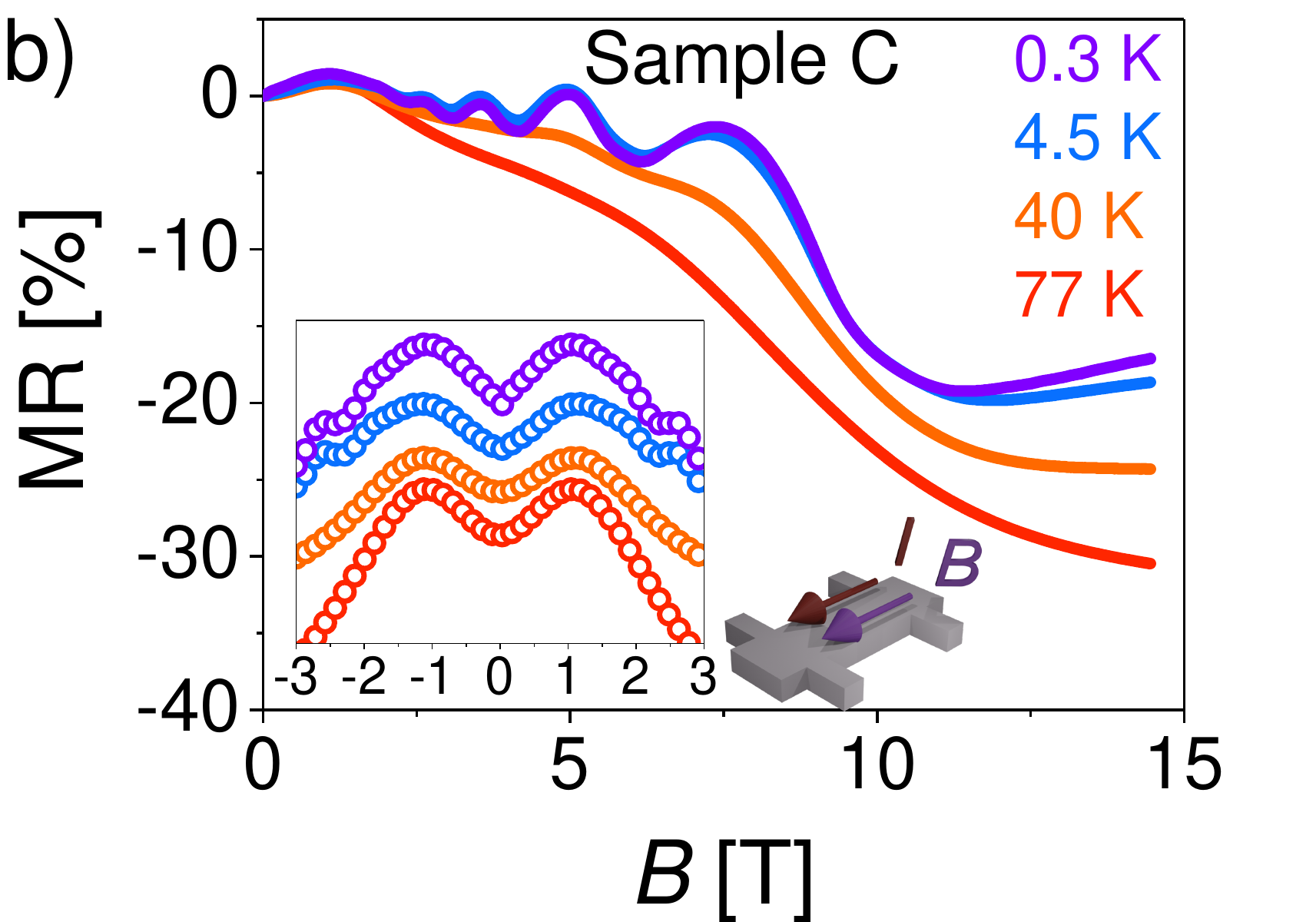}}
  
  \subfloat[\label{fig:NLMR_HB1HB2}]{\includegraphics[width=0.49\textwidth]{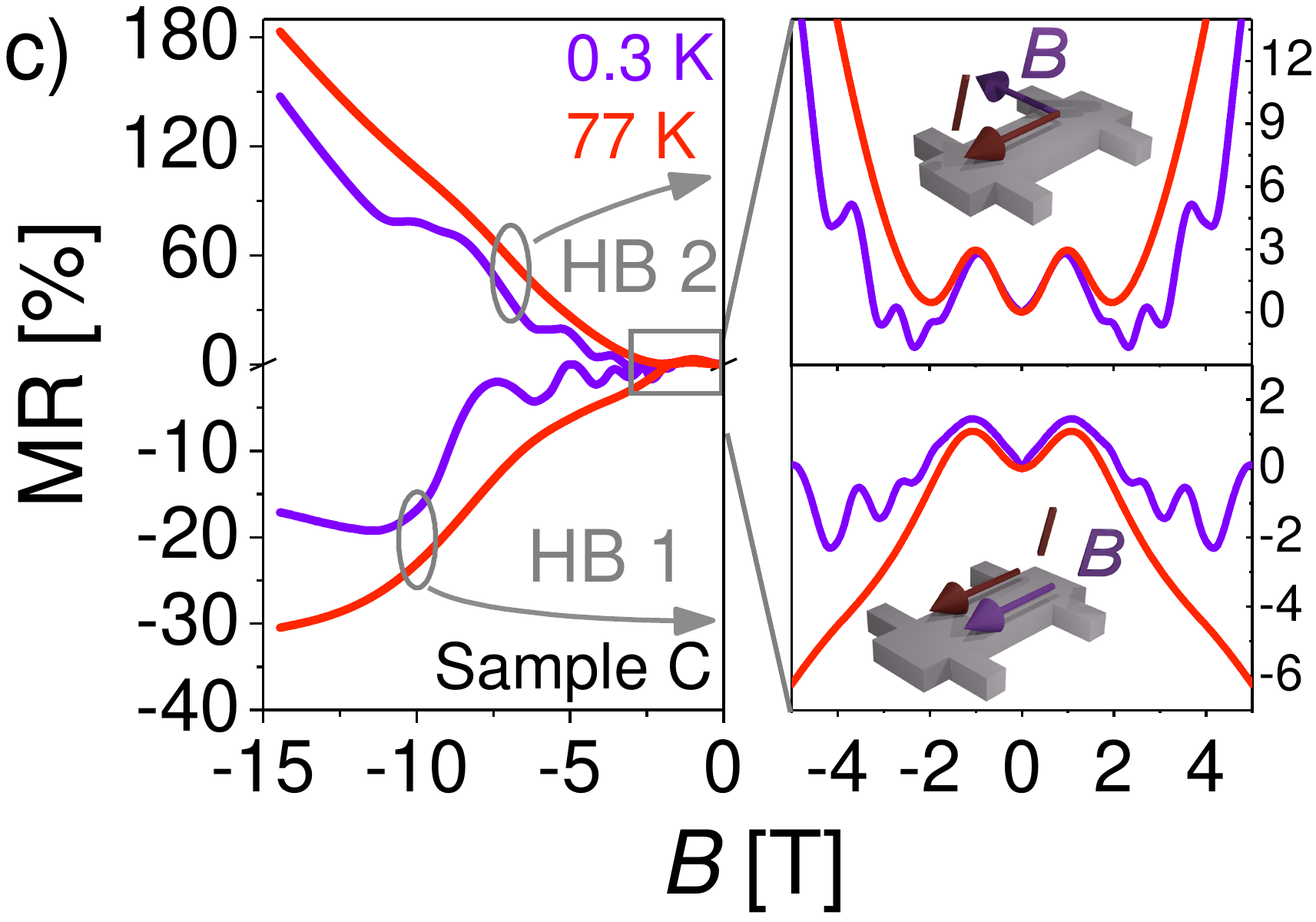}}\hfill
  \subfloat[\label{fig:NLMR_trans_cont_edge_cont}]{\includegraphics[width=0.49\textwidth]{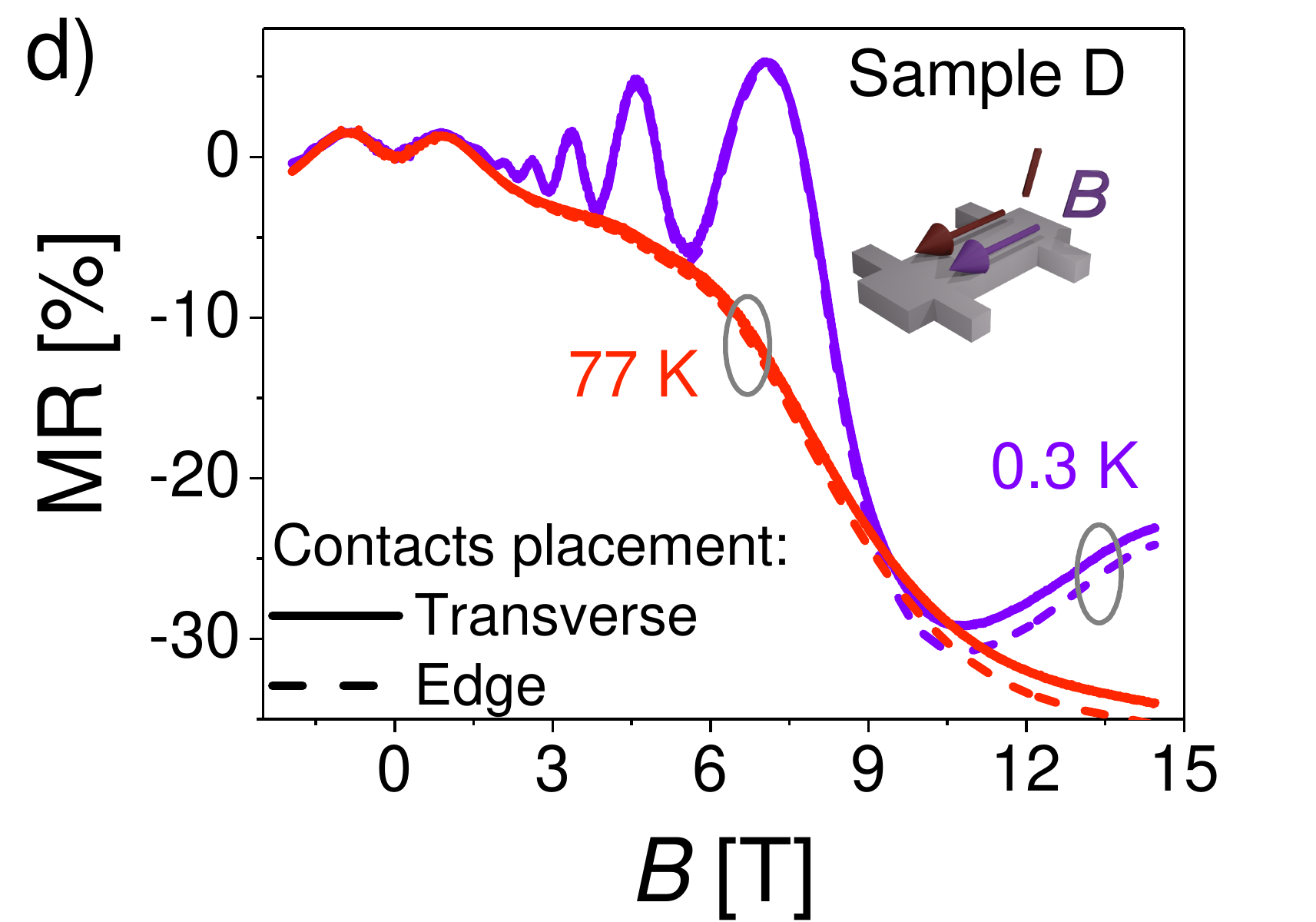}}  
  \caption{Longitudinal magnetoresistance in {\aSn} measured in $B^{\text{in-plane}}$. The main panels of \protect\subref{fig:NLMR_50nm}) and \protect\subref{fig:NLMR_150nm}) show the experimental data for samples A and C, respectively. Insets show the low-field part of NLMR (open symbols; data was shifted for clarity). MR is defined as $\text{MR} = \frac{R(B) - R(B = 0)}{R(B = 0)} * 100\%$ \protect\subref{fig:NLMR_HB1HB2}) Comparison of longitudinal MR for two different field-current orientations for sample C at \qty{0.3}{\kelvin} (violet lines) and \qty{77}{\kelvin} (red lines). \protect\subref{fig:NLMR_trans_cont_edge_cont}) Comparison of NLMR measured with transverse and edge contacts at \qty{0.3}{\kelvin} and \qty{77}{\kelvin} in sample D.}
  \label{fig:NLMR}
\end{figure}

After rotating the sample from $B^{\text{normal}}$ to $B^{\text{in-plane}}$, the character of magnetoresistance (MR) changes. Perpendicular parts of two-arm structures are no longer equivalent, with current flowing either parallel ($B^{\text{in-plane}}_{\parallel}$) or perpendicular ($B^{\text{in-plane}}_{\perp}$) to the magnetic field (see \autoref{fig:sample_sketch}). The dependence of MR on the relative orientation between current and magnetic field is summarized in \autoref{fig:NLMR}a-c for samples A and C (the qualitative behavior of samples B, C and D is the same), which represent the low- and high-mobility cases. In $B^{\text{in-plane}}_{\parallel}$, MR first increases, and after reaching a maximum at $B =  \qtyrange[range-phrase=-]{1}{2.5}{\tesla}$, depending on the thickness, it starts to decrease with an increasing magnetic field, a behavior known as negative longitudinal magnetoresistance (NLMR). This trend persists to 14.5 T without turning back to positive MR. Strong SdH oscillations are superimposed on the low-temperature NLMR of sample C (see \autoref{fig:NLMR_150nm}). Above \qty{40}{\kelvin}, they are strongly damped, and the negative character of the MR is unambiguously seen. In contrast, the overall MR is positive in $B^{\text{in-plane}}_{\perp}$ (\autoref{fig:NLMR_HB1HB2}). The low-field MR, shown in the insets of \autoref{fig:NLMR_50nm},\subref{fig:NLMR_150nm} for the $B^{\text{in-plane}}_{\parallel}$, is discussed in SI.

The NLMR was observed in some of the {\aSn}/InSb samples studied previously \cite{Barbedienne_PRB2018, LeDucAnh_AdvMat2021}, while other reports explicitly claim the absence of this feature \cite{Y-Ding_PhysRevAppl2022, Ding_APL2022}. Its detailed analysis in the grey tin is still lacking. Our study clearly shows that it is an effect related to the material properties of {\aSn} and not a sample-specific feature, as it appears in all examined layers, regardless of their thickness, shape, and dimensions of the patterned microstructure or carrier density and mobility. In other compounds, such as the Weyl semimetal TaAs \cite{Huang_PRX2015, Zhang_NatCom2016} and the Dirac semimetal Na$_{3}$Bi \cite{Xiong_Science2015}, NLMR was explained as a signature of a chiral anomaly. This phenomenon can be effectively described as the creation and annihilation of particles with opposite chirality and was predicted for a crystal with linearly dispersing chiral bands \cite{Nielsen_PhysLett1983, Son_PRB2013, Burkov_PRL2014}, such as in Weyl semimetals. Thus, it is tempting to consider the NLMR as evidence of the presence of a WSM phase. Such interpretation would agree with the results of the ARPES and magneto-optics data, as well as with previous theoretical predictions \cite{HuangLiu_PRB2017}. However, some concerns have been raised about other mechanisms that can result in similar behaviour \cite{dosReis_NewJourPhys2016, Li_FrontPhys2017}. The list includes weak localization (WL) \cite{Kawabata_JPSJap1980, Kawabata_SolStCom1980}, current jetting effect \cite{Pippard2009, Hu_PRL2005}, and magnetism-related phenomena \cite{Ohno_PRL1992, Parkin_Review1995, Ramirez_JourPhysCondMat1997}. The experimental features of NLMR in the investigated structures allow us to rule out all of these effects. Weak localization would be present in any direction of the magnetic field, which is not the case for the NLMR in the current study. As shown in \autoref{fig:NLMR_trans_cont_edge_cont}, various pairs of the voltage probes yield the same result in the $B^{\text{in-plane}}_{\parallel}$, which excludes the current-jetting effect \cite{dosReis_NewJourPhys2016, Liang_PRX2018}. The presence of magnetic impurities was not only highly improbable due to the high purity of the MBE growth process but also excluded by core-level spectroscopy. Therefore, we attribute the NLMR in the grey tin to the chiral anomaly of the Weyl fermions. The description of our results in frames of semiclassical theory \cite{Son_PRB2013, Cano_PRB2017}, together with an analysis of alternative sources of NLMR, is presented in SI.

Finally, we move to the analysis of SdH oscillations, which are present in both $B^{\text{normal}}$ and $B^{\text{in-plane}}$. The analysis of these oscillations proved useful in determining the band topology of multiple materials, including {\aSn} in both the TI \cite{Barbedienne_PRB2018} and DSM \cite{LeDucAnh_AdvMat2021, Vail_PSSB2020} phases. To analyze SdH oscillations, we have converted resistivity $\rho_{xx}$ to the conductivity $\sigma_{xx}$ using conductivity tensor, since fundamentally the quantization occurs in conductivity rather than in resistivity \cite{Ando_JPhysSocJap2013}. The oscillatory part of the $\sigma_{xx}$ can be described by the Lifshitz-Kosevitch equation (LK) \cite{Schoenberg1984, Ando_JPhysSocJap2013, Murakawa_Science2013}, which in the 3D case takes the following form:
\begin{equation}
    \label{eq:L-K_formula}
    \Delta\sigma_{xx} = A_{0} + A R_{T} R_{D} \sqrt{\frac{B}{2F}}
    \cos\left[ 2\pi \left( \frac{F}{B }- \frac{1}{2} + \beta \pm \delta \right) \right].
\end{equation}
$F$ is the frequency of oscillations, $2\pi*\beta$ is the Berry phase of the carriers and $\delta = \frac{1}{8}$ is a phase shift related to the dimensionality of the system and the cross-section of the Fermi surface ($\delta = 0$ for the 2D case). It is expected that $\beta = \frac{1}{2} \left(0\right)$ in topological (normal) materials \cite{Miktik_PRL1999, Miktik_PRB2012}. $A$ and $A_{0}$ are constant; $R_{T}$ and $R_{D}$ are the damping coefficients:
\begin{equation}
    \label{eq:temp_damping}
    R_{T} = \frac{2 \pi^{2} k_{B} T/ \hbar \omega_{c}}{\sinh \left(2 \pi^{2} k_{B} T/ \hbar \omega_{c} \right) }
\end{equation}

\begin{equation}
    \label{eq:dingle}
    R_{D} = \exp \left(-2\pi^{2} k_{B} T_{D}/\hbar \omega_{c} \right).
\end{equation}
Above, $\omega_{c} = \frac{eB}{m_{c}}$ is a cyclotron frequency with $m_{c}$ being the cyclotron mass, and $T_{D} = \frac{\hbar}{2\pi k_{B} \tau}$ is a Dingle temperature, related to the quantum scattering time $\tau$, and therefore to the quantum mobility $\mu_{SdH} = \frac{e\tau}{m_{c}}$.

\begin{figure}[htbp]
    \captionsetup[subfigure]{labelformat = empty}
    \centering
    \subfloat[\label{fig:DSIGxx_200nm_Bperp}]{\includegraphics[width=0.3\textwidth]{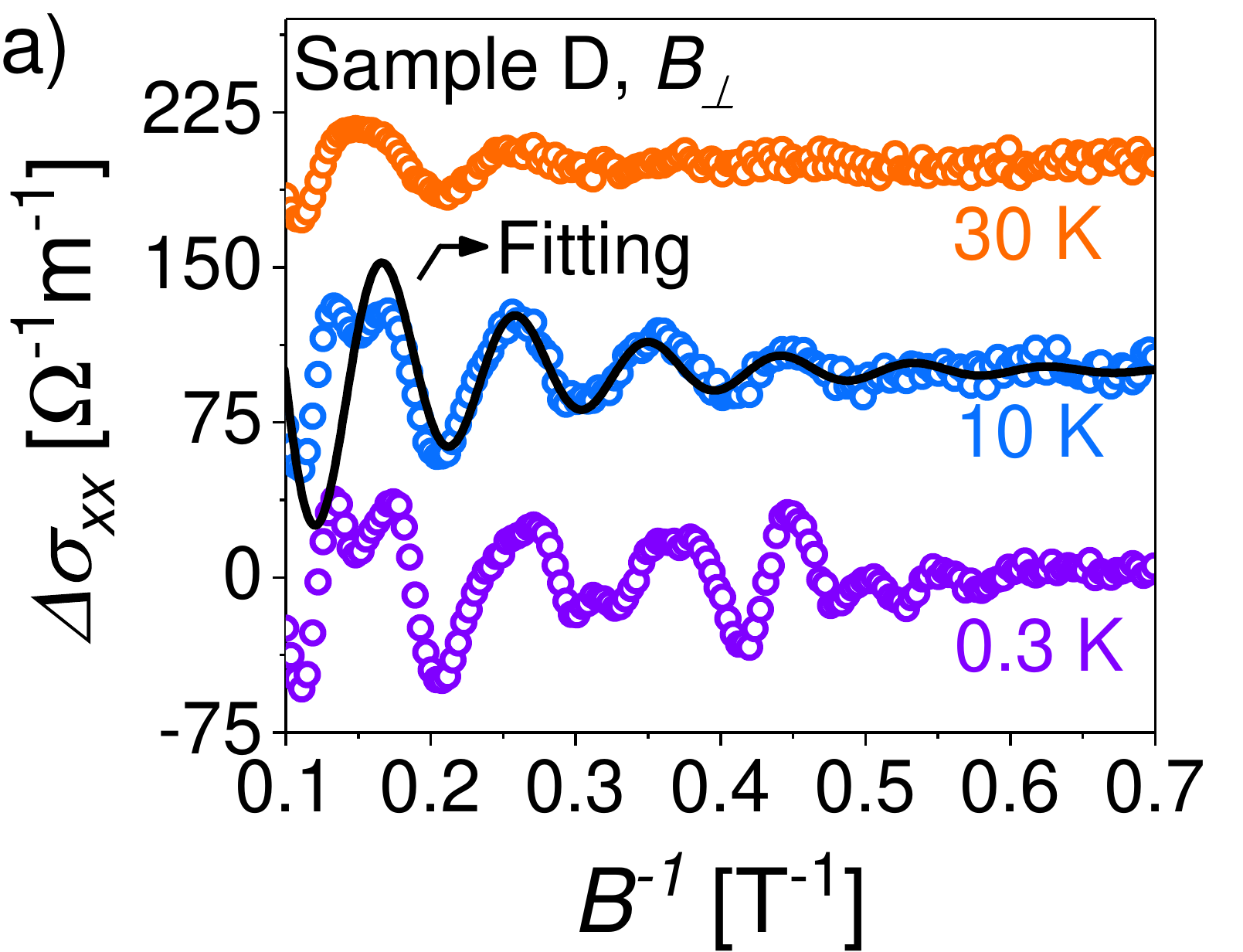}\hspace{0.049\textwidth}}
    \subfloat[\label{fig:FFT_200nm_Bperp}]{\includegraphics[width=0.3\textwidth]{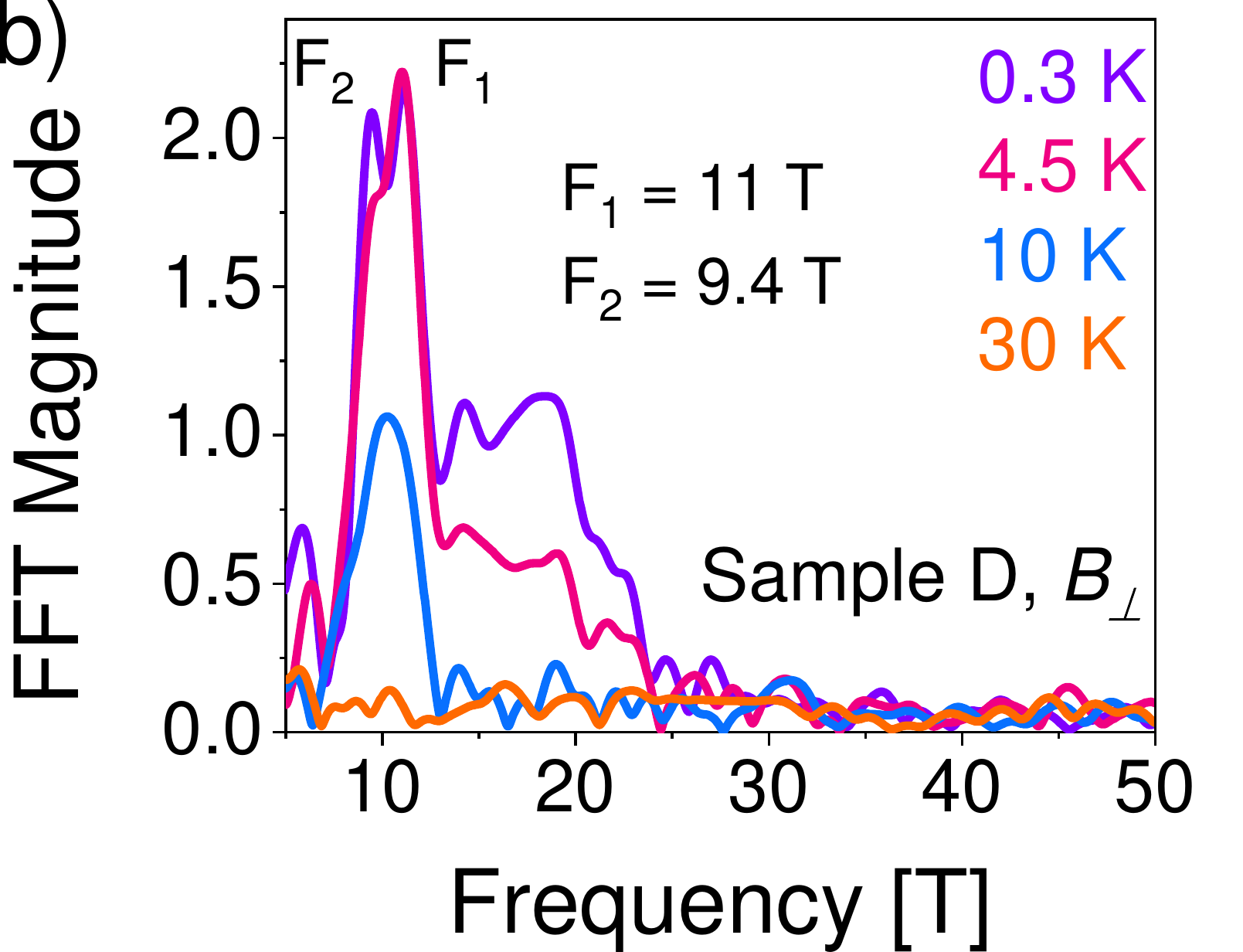}\hspace{0.049\textwidth}}
    \subfloat[\label{fig:LLplot_200nm_BperpBparal}]{\includegraphics[width=0.3\textwidth]{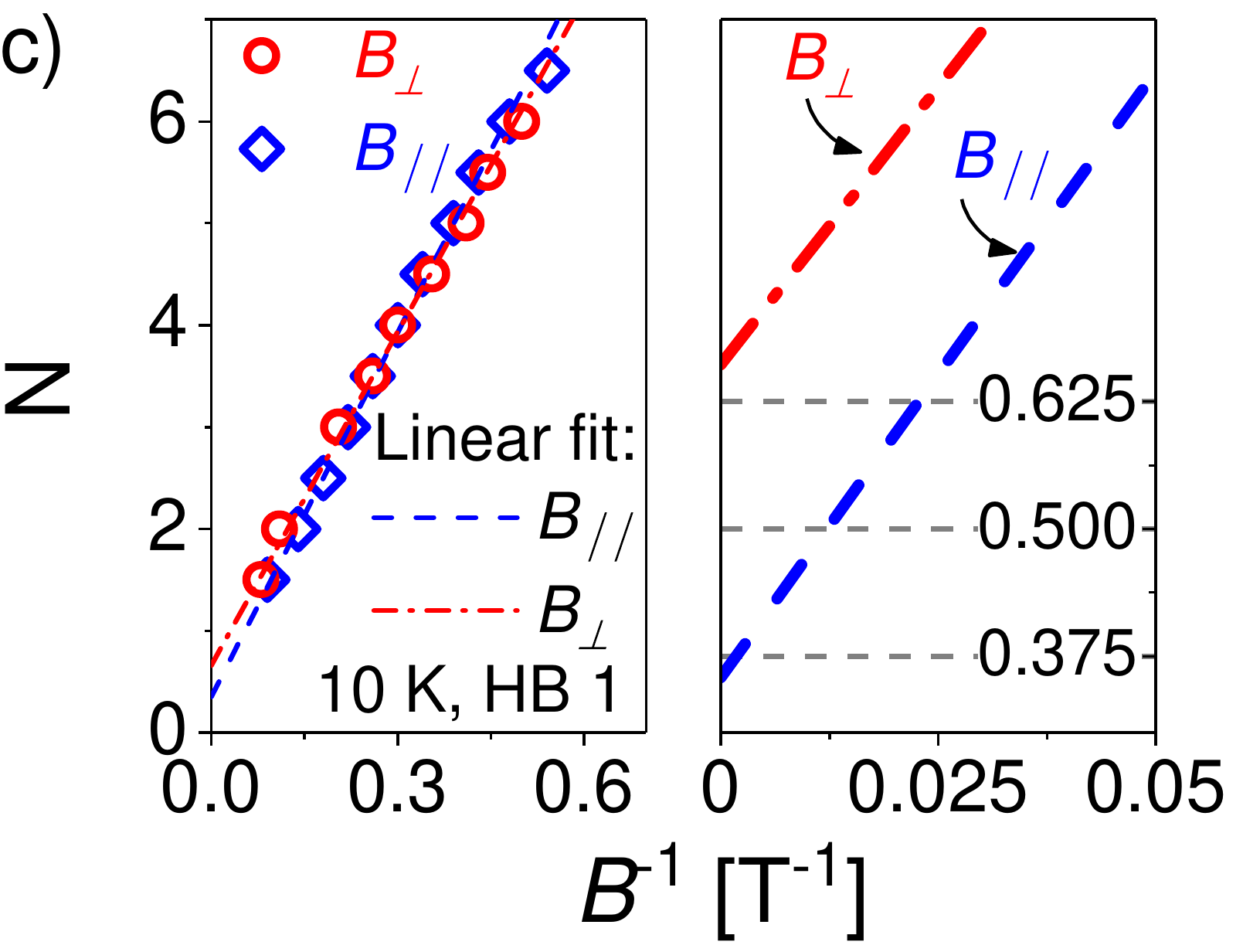}}
    
    \subfloat[\label{fig:DSIGxx_200nm_Bparal}]{\includegraphics[width=0.3\textwidth]{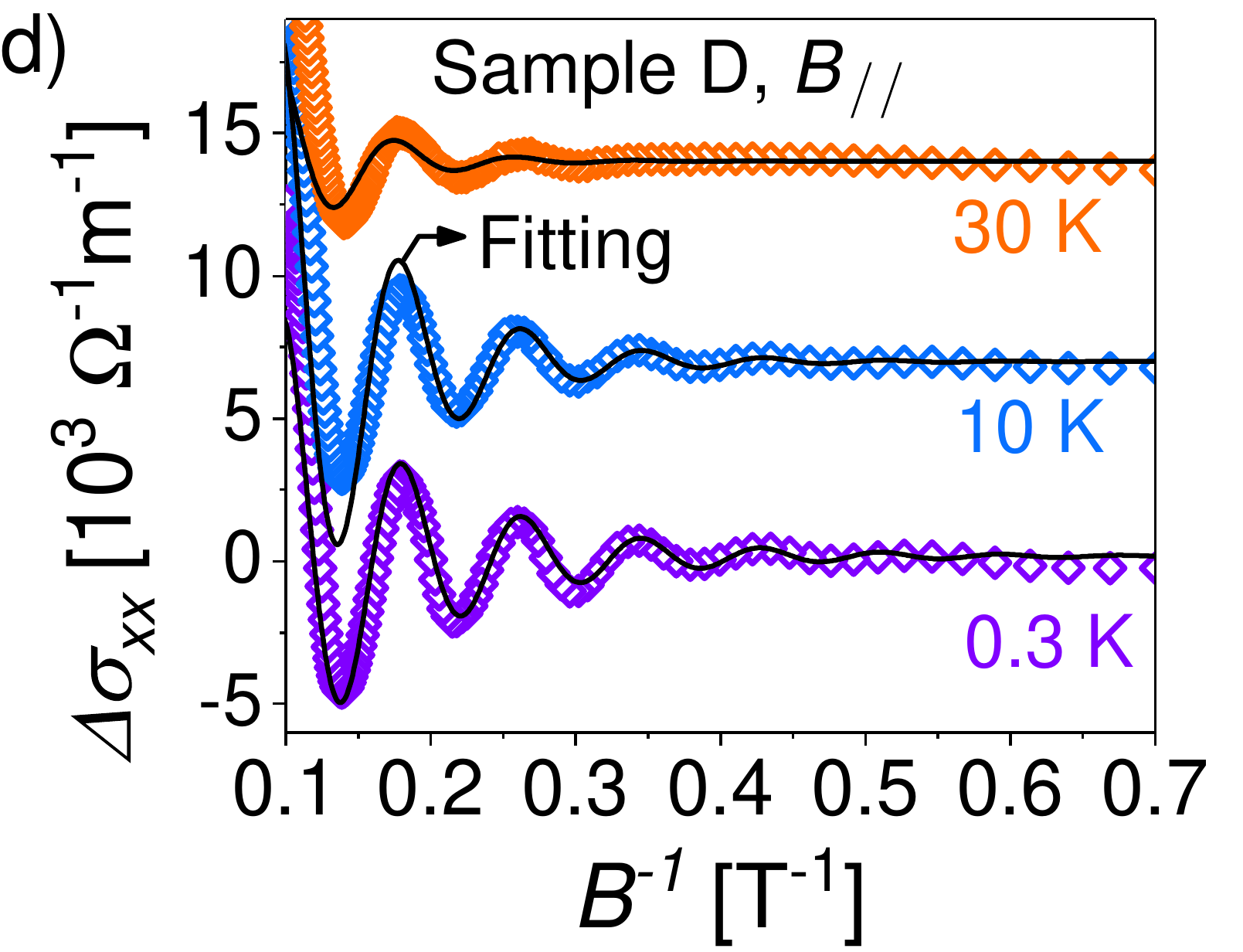}\hspace{0.049\textwidth}}
    \subfloat[\label{fig:FFT_200nm_Bparal}]{\includegraphics[width=0.3\textwidth]{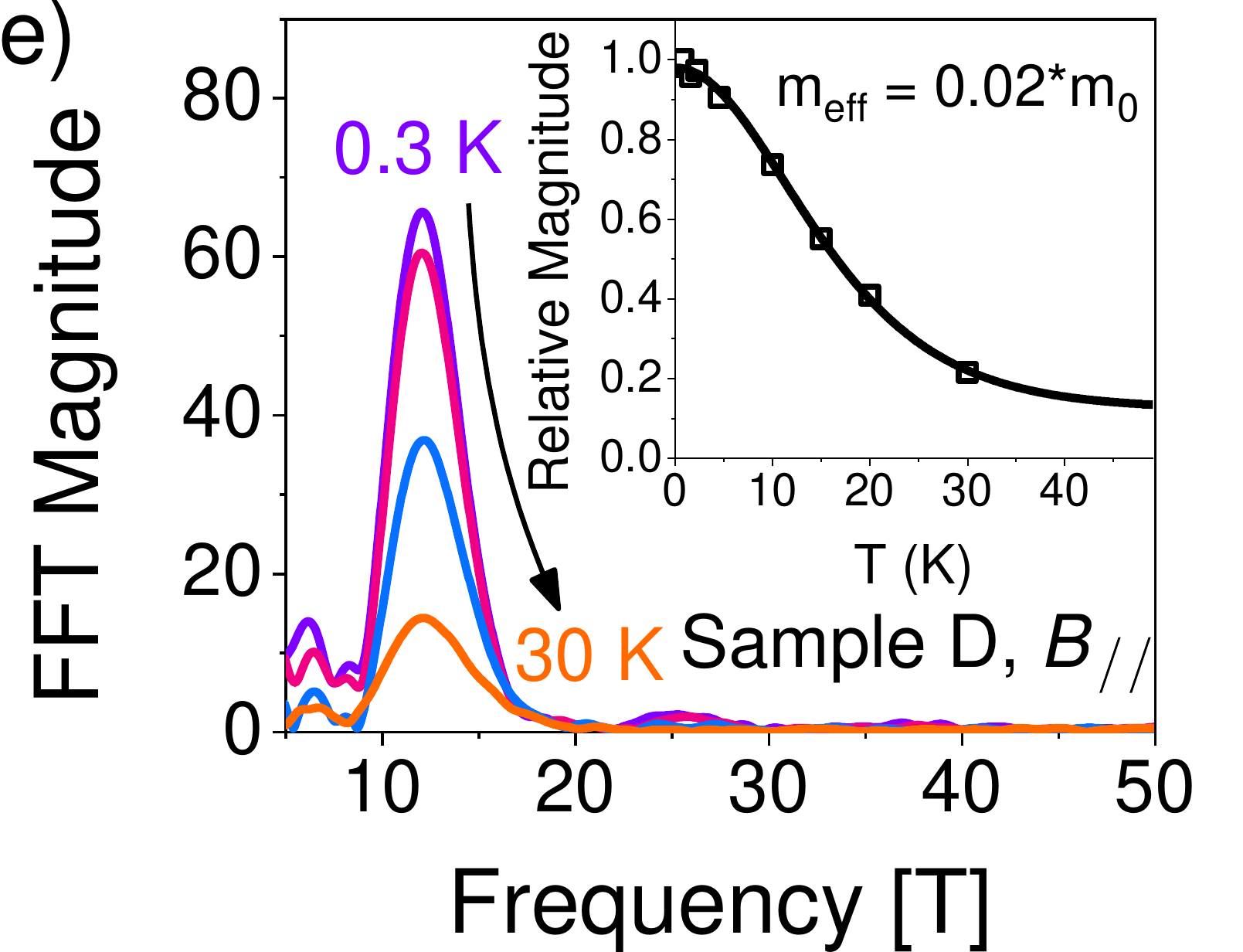}\hspace{0.049\textwidth}}
    \subfloat[\label{fig:LLplot_vs_thickn}]{\includegraphics[width=0.3\textwidth]{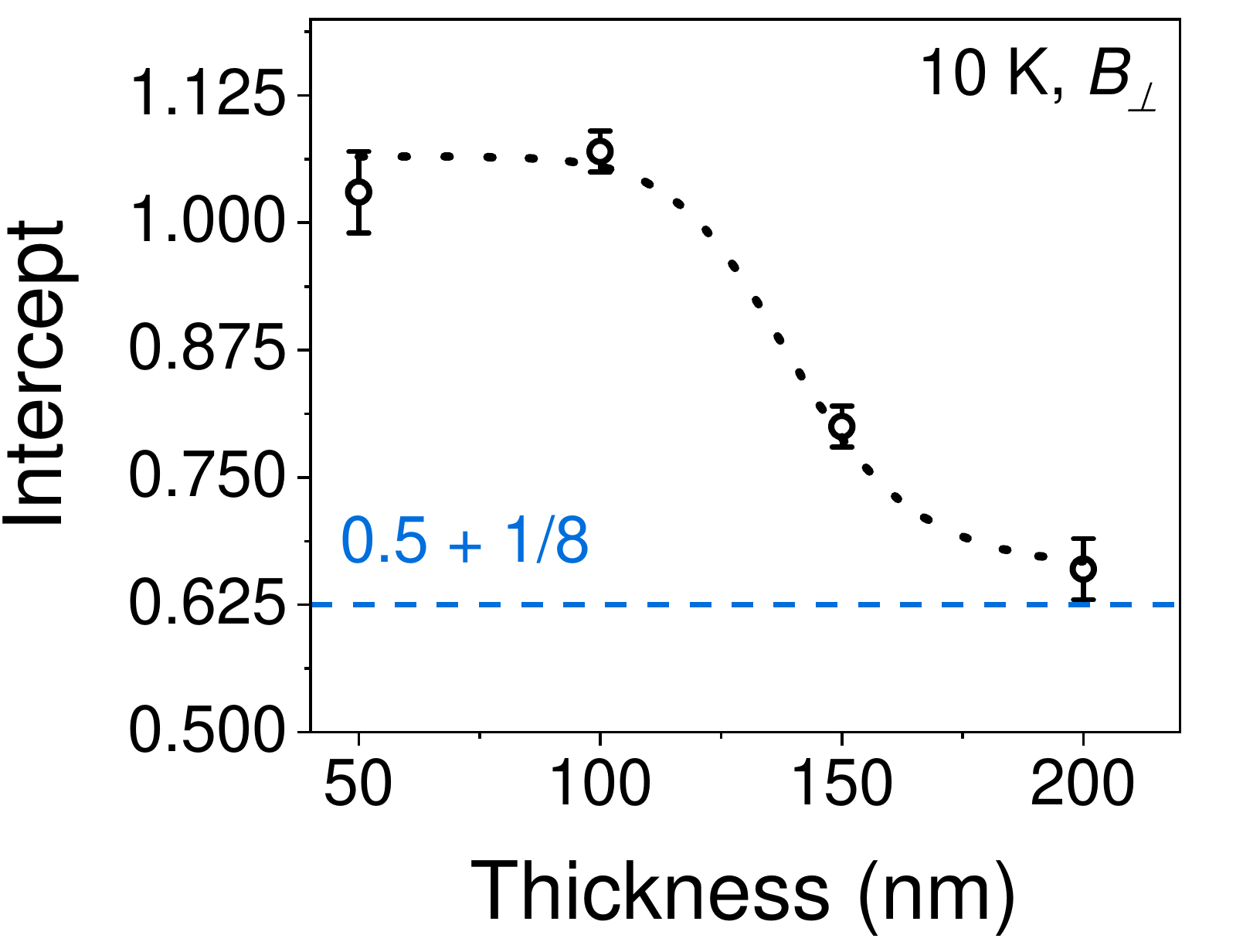}}
    \caption{Shubnikov-de Haas oscillations in {\aSn} thin films. \protect\subref{fig:DSIGxx_200nm_Bperp}), \protect\subref{fig:DSIGxx_200nm_Bparal}) Open symbols - oscillatory part of $\Delta\sigma_{xx}$ in sample D at various temperatures, obtained after removal of the smooth background for $B^{\text{normal}}$ and $B^{\text{in-plane}}_{\parallel}$, respectively. Solid black lines show the result of the fitting of \autoref{eq:L-K_formula} to the experimental data. The fitting parameters were $A$ and $A_0$ for $B_{\parallel}$, and additionally $T_{D}$ for $B^{\text{normal}}$, with the remaining parameters determined as described in the main text. \protect\subref{fig:FFT_200nm_Bperp}), \protect\subref{fig:FFT_200nm_Bparal}) Temperature-dependent FFT spectra of the SdH oscillations of \protect\subref{fig:DSIGxx_200nm_Bperp}) and \protect\subref{fig:DSIGxx_200nm_Bparal}), respectively. Inset of \protect\subref{fig:FFT_200nm_Bparal}) shows the temperature-dependent magnitude of the FFT spectrum for $B^{\text{in-plane}}_{\parallel}$ with the fitting based on \autoref{eq:temp_damping}. The effective mass determined this way is $m_{c} = 0.02 m_{0}$. \protect\subref{fig:LLplot_200nm_BperpBparal}) Landau level fan diagram constructed for both $B^{\text{normal}}$ and $B^{\text{in-plane}}_{\parallel}$ in order to determine the phase of oscillations. The integers (half-integers) were assigned to the minima (maxima) of $\Delta\sigma_{xx}$. \protect\subref{fig:LLplot_vs_thickn}) Intercept, $d$, of the linear fit to the LL index for {\aSn} samples with various thicknesses. The dotted line is a guide for the eye. The dashed blue line marks the value of $d = \frac{1}{2}+\frac{1}{8}$.}
    \label{fig:SdH}
\end{figure}

The oscillatory component of $\sigma_{xx}$ of sample D is shown as open symbols in \autoref{fig:DSIGxx_200nm_Bperp} and \autoref{fig:DSIGxx_200nm_Bparal} for the $B^{\text{normal}}$ and $B^{\text{in-plane}}_{\parallel}$, respectively. Low-temperature oscillations in $B^{\text{normal}}$ exhibit multi-frequency character, confirmed by the fast Fourier transformation (FFT) spectra displayed in \autoref{fig:FFT_200nm_Bperp}. Also, the Zeeman spin splitting is visible for the maximum at $B^{-1} \approx \qty{0.15}{\tesla^{-1}}$. Contrary to $B^{\text{normal}}$, SdH oscillations in $B^{\text{in-plane}}$ show a single-frequency character in the full temperature range, with the FFT peak centered at \qty{12}{\tesla} (\autoref{fig:FFT_200nm_Bparal}). We conclude that the plethora of massive surface states is responsible for the multi-frequency character of the SdH oscillation in $B^{\text{normal}}$ at low temperatures. By fitting \autoref{eq:temp_damping} to the temperature-dependent FFT magnitude we determined the electron cyclotron mass $m_{c} = 0.02 m_{0}$, as presented in the inset of \autoref{fig:FFT_200nm_Bparal}. A similar value was obtained for $B^{\text{normal}}$. We have also calculated the Dingle temperatures for $B^{\text{in-plane}}_{\parallel}$ using \autoref{eq:dingle} \cite{Ando_JPhysSocJap2013}, and obtained values in the range of $T_{D} =$ \qtyrange[range-phrase=--, range-units=single]{30}{40}{\kelvin}, which corresponds to the quantum mobility of \qtyrange[range-phrase=--, range-units=single]{4000}{2500}{\cm^2\volt^{-1}\sec^{-1}}.

The Berry phase of the carriers is usually obtained from analysis of the Landau-level fan diagram. The minima (maxima) of $\Delta\sigma_{xx}$ are labelled with integers (half-integers) and plotted versus their position in $B^{-1}$. According to \autoref{eq:L-K_formula}, the intercept $d$ of the linear fit to the data yields the Berry phase. In the case of sample D, this procedure gives $d = \frac{1}{2} \pm \frac{1}{8}$ depending on the direction of the magnetic field, as shown in \autoref{fig:LLplot_200nm_BperpBparal}. The different sign of the $\delta$ phase shift indicates an anisotropic Fermi surface, with a maximum cross-section at the $B^{\text{in-plane}}_{\parallel}$ and a minimum at the $B^{\text{normal}}$ for electrons \cite{Sun_FrontPhys2019}. It agrees with the angle dependence of the frequency of the SdH oscillations, and the frequency increases when the field orientation is changed from the perpendicular to the parallel.

Since we determined all the parameters of the band structure that affect the LK equation, we can fit it to the experimental data with few adjustable parameters, namely the amplitude $A$ and the offset $A_0$, and the Dingle temperature $T_D$ for $B^{\text{normal}}$. A single cosine term was used for the fitting. The results of this procedure are shown by solid lines in ~\autoref{fig:DSIGxx_200nm_Bperp}, \ref{fig:DSIGxx_200nm_Bparal} for $B^{\text{normal}}$ and $B^{\text{in-plane}}_{\parallel}$, respectively. The good agreement between the fitting curves and the experimental data confirms the correct determination of the parameters. Most importantly, the oscillations in sample D show a nontrivial Berry phase, as expected for topological semimetals. However, we note that a certain thickness dependence of the phase was observed in the studied {\aSn} samples, as shown in \autoref{fig:LLplot_vs_thickn}. Such a dependence can be caused by the hybridization between bulk Dirac electrons and massive surface states. Although the exact source of this dependence requires additional investigation, it is consistent with the previous reports, showing that in thinner layers of grey tin, the Berry phase obtained from the linear fit to the LL fan diagram deviates from the ideal value of $\pi$ \cite{Barbedienne_PRB2018, LeDucAnh_AdvMat2021}.

\section{Conclusions}\label{section:conclusions}

We presented the first comprehensive study of the electronic structure of compressively-strained {\aSn} on insulating substrates by a variety of experimental techniques. The magneto-optical results provide a set of band-structure parameters, which allow us to fully describe grown epilayers as a 3D Dirac semimetal. We found that the band structure is inverted, as expected, and determined the value by which the valence band is shifted to due strain, $\Omega = -\qty{13}{\meV}$. The band structure is also visualized in ARPES experiments, which reveals the presence of several bulk and surface states, distinct in nature. The topological nature of the investigated samples is further proved by magneto-transport studies. After ruling out alternative possibilities, we explain the presence of negative longitudinal magnetoresistance by the chiral anomaly of the Weyl fermions. Shubnikov-de Haas oscillations reveal both the 3D nature of the Fermi surface and its non-trivial character, through observation of the $\pi$ Berry phase shift. We argue that the observation of these peculiar magneto-transport effects was only possible due to the insulating character of the substrates used for the growth.

We also wish to stress the practical aspect of our work. High-quality CdTe substrates of large size are currently not available. On the other hand, the growth of CdTe buffers on GaAs is well-established. It gives the possibility of wafer-scale epitaxy of {\aSn} if hybrid substrates are used. The high crystalline quality and homogeneous strain of the obtained layers clearly indicate that this is a promising route that may evoke progress in applied studies. It also opens new possibilities, as other buffers can be used, such as Cd$_{1-x}$Zn${_x}$Te or magnetically-doped CdTe, giving access to a wider range of biaxial compressive strains and to magnetism-proximatized phenomena. We also note that the fabrication of CdTe/Sn/CdTe quantum wells is a path for exploration of the QSHI state in {\aSn}. We, therefore, conclude that grey tin is a promising material for the exploration of various phenomena related to nontrivial topology and relativistic-like band structure, with the potential for large-scale fabrication for industrial purposes.

\section{Material and methods} \label{section:experimental}

\textbf{MBE growth} Sample synthesis was carried out in two steps in two interconnected MBE Veeco GENxplor growth chambers, with a base pressure below $10^{-10}$~mbar, using effusion cells with elemental Cd, Zn and Te for buffer, and Sn for {\aSn}, as sources. Semi-insulating epi-ready (001) GaAs from AXT, Inc. were used as substrates. First, in chamber one, several nanometer layers of ZnTe ($a_{0} = \qty{6.106}{\text{\AA}}$) were deposited on the GaAs substrate to alleviate the lattice constant difference between GaAs ($a_{0} = \qty{5.654}{\text{\AA}}$) and CdTe ($a_{0} = \qty{6.482}{\text{\AA}}$) and to ensure (001) growth \cite{Shtrikman_JourElMat1988}, as well as to improve misfit stress relaxation of the buffer layer \cite{Nishino_JourApplPhys1996}. Then a $\qty{4}{\micro\meter}$ thick CdTe buffer was grown  at \qtyrange{295}{325}{\celsius} to ensure a fully relaxed layer \cite{Tatsuoka_JourApplPhys1990, khiar2014well}. The material flux ratio was controlled by a beam flux monitor placed in front of the substrate position. The growth was performed under Cd-rich conditions. After the deposition of the buffer layer, the structure was transferred to the second growth chamber, without breaking the vacuum, where {\aSn} deposition took place. Epitaxial growth was carried out at a temperature below \qty{10}{\celsius} with a material flux ratio and deposition rate controlled by a quartz crystal microbalance (QCM) placed in front of the substrate position. In both cases, the structural quality of the films was monitored in-situ in real-time by reflection high-energy electron diffraction (RHEED).

\textbf{Structural characterization} The surface of the grown films was examined by atomic force microscopy (AFM) in tapping mode with a Bruker MultiMode 8-HR microscope. XRD patterns were recorded with the PANalytical X'Pert Pro MRD diffractometer with a 1.6 kW X-ray tube using CuK$\alpha$1 radiation ($\lambda = \qty{1.5406}{\text{\AA}}$), a symmetric 2 $\times$ Ge (220) monochromator and 2D Pixel detector. Samples for TEM observations in the form of cross-sectional lamellas were made using the FIB method. The lamellas were cut perpendicular to the sample surface in the [100] and [110]. Platinum as a protective material was deposited onto the sample surface. For TEM examinations, the Titan Cubed 80-300 transmission electron microscope operated with an acceleration voltage of 300 kV was used.

\textbf{ARPES} The electronic band structure of the films was examined by angle-resolved photoemission spectroscopy at URANOS (former UARPES) beamline at the SOLARIS synchrotron (Krak{\'o}w, Poland) with photon energies ranging from 16 to 90~eV using horizontally polarized light at temperatures ranging from liquid nitrogen to RT. To record high-quality spectra, Scienta Omicron DA30L photoelectron spectrometer with an energy resolution of 1.8~eV and angular resolution of \qty{0.1}{\degree} was used. Core-level (CL) spectra were measured with a photon energy of 90 eV to determine the surface elemental composition. The experiments were carried out at pressures below $2 \times 10^{-10}$~mbar. The investigated {\aSn} epilayers were transported to the ARPES system in a battery-operated ion getter pumped Ferrovac VSN40S ultra-high vacuum (UHV) suitcase \cite{Turowski_ApplSurfSci2023} sustaining a base pressure below $2 \times 10^{-10}$~mbar, without breaking UHV. During measurements the sample was grounded through a manipulator, to which it was attached with an In-Ga eutectic alloy.

\textbf{Magneto-optics} Magneto-optical absorption experiments are performed in an Oxford Instruments 1.5~K/15~T cryostat at 4.5~K. Spectra are acquired using a Bruker Fourier transform spectrometer. All measurements are made in Faraday geometry. A globar and a Hg lamp are employed as the mid-infrared and far-infrared sources to measure the absorptions. Measurements are performed at fixed magnetic fields between 0 and 15~T. A He-cooled bolometer is used to detect the transmitted signal. The relative transmission at a fixed magnetic field, T(B)/T(B = 0), is extracted and analyzed.

\textbf{Hall bars preparation} The samples were patterned in the Hall bars by etching in diluted HCl, with the use of a PMMA mask defined by electron-beam lithography. The dimensions of the Hall bars are $\qty{20}{\micro\meter} \times \qty{400}{\micro\meter}$. Ti/Au contacts were deposited in the ultra-high vacuum e-beam evaporator. The samples were kept at room temperature at all stages of microfabrication. Selected layers were patterned into microstructures with two perpendicular arms. One of them was equipped with voltage probes that span across the channel (transverse contacts) in addition to standard probes at the edges of the Hall bar (edge contacts).

\textbf{Magneto-transport} Magneto-transport measurements were performed in a He-3 refrigerator (HelioxVL, Oxford Instruments, temperature range \qtyrange{0.3}{100}{\kelvin}), equipped with a piezoelectric motor that allows \textit{in-situ} rotation of the sample in the magnetic field. Tree distinct orientations of magnetic field were applied: $B^{\text{normal}}$, with magnetic field normal to the sample plane (i.e. standard Hall configuration),  $B^{\text{in-plane}}_{\parallel}$, with an in-plane magnetic field parallel to the current flow and $B^{\text{in-plane}}_{\perp}$, with in-plane magnetic field perpendicular to the current. The standard low-frequency lock-in technique, with an excitation current not exceeding $\qty{500}{\nano\ampere}$, was used to collect the signal.

\medskip
\textbf{Acknowledgments} \par
The authors thank T. Dietl for valuable discussions. The research was supported by the Foundation for Polish Science through the IRA Program co-financed by EU within SG OP and EFSE, by ANR-19-CE30-022–01 (ENS) and the Austrian Science Funds FWF (Project I-4493). V.V.V. acknowledges long-term program of support of the Ukrainian research teams at the Polish Academy of Sciences carried out in collaboration with the U.S. National Academy of Sciences with the financial support of external partners. This publication was partially developed under the provision of the Polish Ministry of Science and Higher Education project “Support for research and development with the use of research infrastructure of the National Synchrotron Radiation Centre SOLARIS” under contract nr 1/SOL/2021/2. We acknowledge SOLARIS Centre for the access to the Beamline URANOS (former UARPES) where the measurements were performed.

\medskip
\textbf{Author contributions} \par
J.P.: Investigation - magneto-transport \& processing, Analysis, Writing - original draft;
G.K.: Investigation - magneto-optics, Analysis, Writing - original draft;
A.K.: Investigation - magneto-transport, Analysis, Writing – review \& editing, Conceptualization, Supervision;
B.T.: Investigation - MBE growth \& ARPES, Analysis, Writing - original draft;
J.B.O.: Investigation - magneto-optics, Analysis;
R.R.: Investigation - MBE growth \& ARPES;
T.W.: Investigation - Processing;
P.D.: Investigation - TEM;
M.A.: Investigation - AFM;
W.Z.: Investigation - MBE growth;
B.K.: Investigation - TEM;
Z.M.: Investigation - ARPES;
M.R.: Investigation - ARPES;
N.O.: Investigation - ARPES;
L.-A.de V.: Investigation - magneto-optics, Analysis;
Y.G.: Investigation - magneto-optics, Analysis;
T.Wojtowicz: Supervision, Funding acquisition, Writing – review \& editing;
V.V.V.: Investigation - MBE growth, ARPES, XRD, Analysis, Writing – original draft, review \& editing, Conceptualization, Supervision.

\medskip

\printbibliography

\end{document}


\maketitle
\clearpage
\tableofcontents
\clearpage


\section{Extended structural data}
\label{section:structure}

\begin{figure}[htbp]
    \includegraphics[width=\columnwidth]{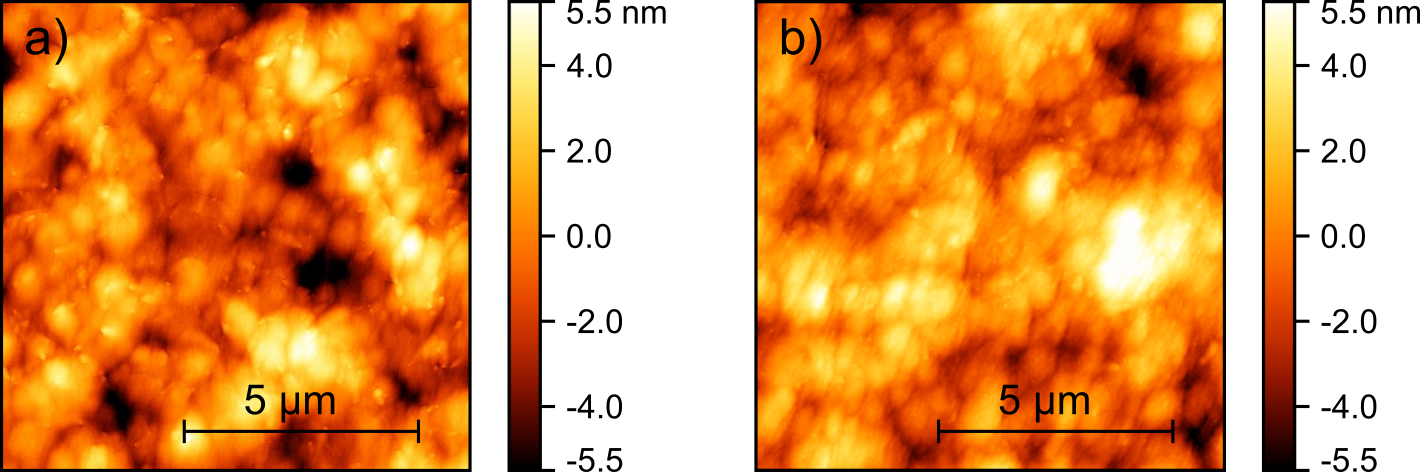}
    \caption{Atomic Force Microscopy micrographs of (a) \qty{5}{\micro\meter} CdTe buffer and (b) \qty{50}{\nano\meter} {\aSn} film demonstrating similar surface structure and roughness.}
    \label{fig:AFM}
\end{figure}

\begin{figure}[htbp]
    \includegraphics[width=\columnwidth]{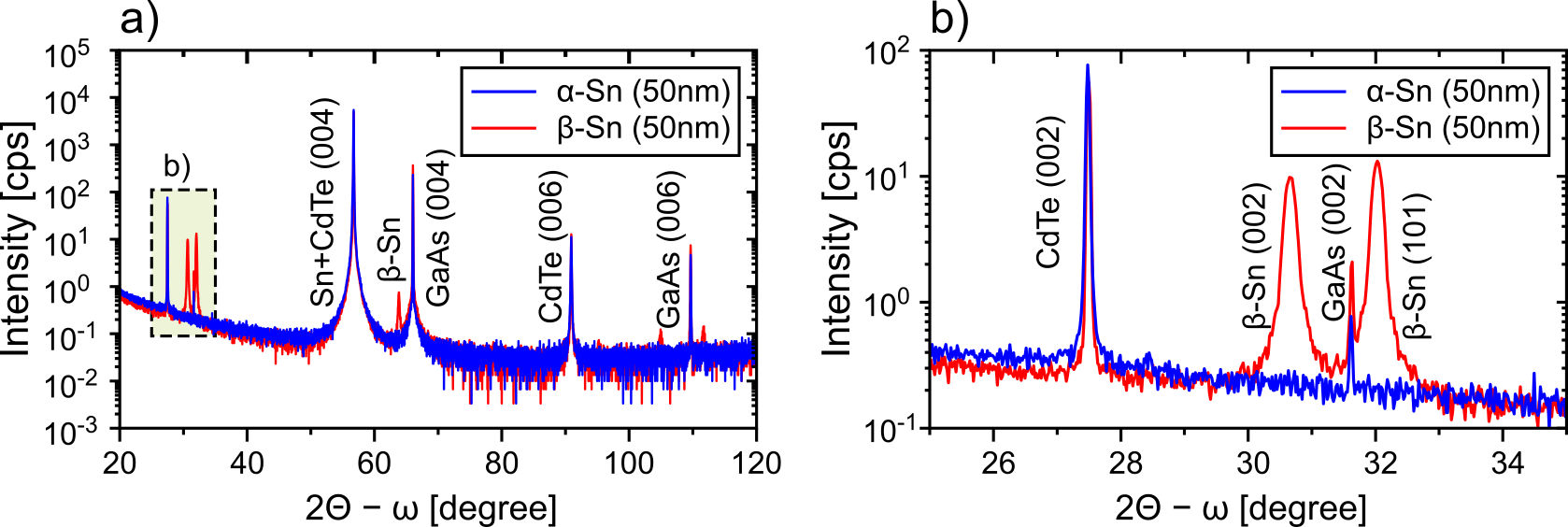}
    \caption{a) XRD $2\Theta-\omega$ scans of two Sn (50 nm) epilayers on (001) CdTe/GaAs (blue line - pure {\aSn} phase, red line - the epilayer containing $\beta$-Sn inclusions). b) Magnified high intensity peaks of $\beta$-Sn phase.}
    \label{fig:XRD}
\end{figure}

\begin{figure}[htbp]
    \includegraphics[width=\textwidth]{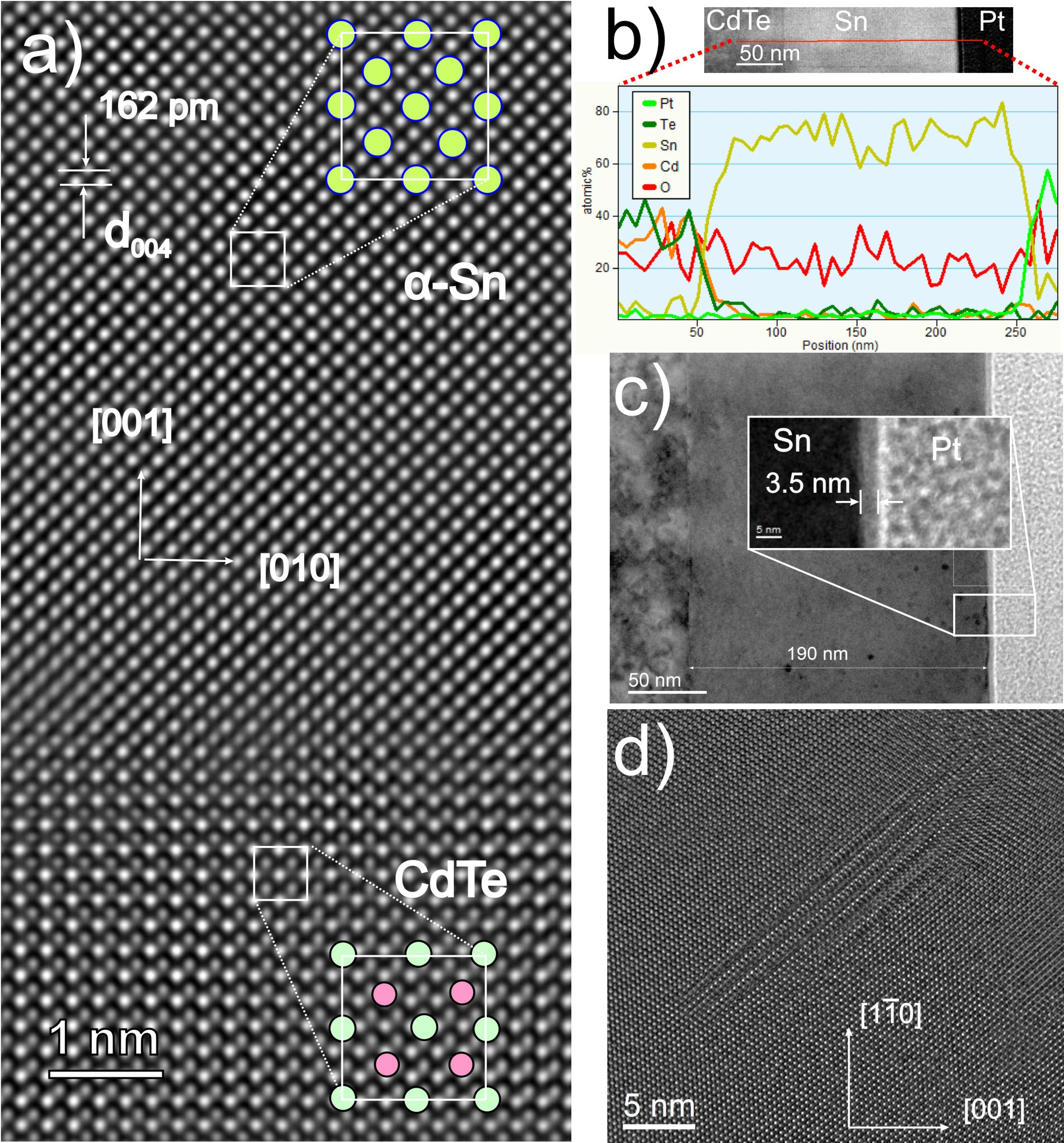}
    \caption{
    a) Fourier filtered high-resolution transmission electron microscope image with the [100] zone axis of the {\aSn} on (001) CdTe (sample D from the main text). The arrows indicate {\it b} and {\it c} crystallographic directions. The insets present a projection of atomic columns in the unit cells of {\aSn} and CdTe. b) The top panel presents STEM/HAADF cross-sectional image of (001) {\aSn} (200~nm)/CdTe heterostructure (the top Pt layer is a result of FIB preparation).  The bottom panel shows the EDX elemental concentrations line profile along the red line indicated in the top panel. c) STEM/HAADF image of the same sample. The inset demonstrates the presence of about 3.5~nm thick Sn oxide on the top of {\aSn} layer, d) HR-TEM image with the [110] zone axis of the same sample demonstrating the presence of stacking faults in the {\aSn} layer.}
    \label{fig:TEM}
\end{figure}

The reproducible growth of excellent-grade CdTe on (001) GaAs has been achieved as reported in many previous works (see e.g. \cite{Wichrowska_ActPhysPolA2014}). The high structural quality of \qty{4}{\micro\meter} thick CdTe buffer obtained in the present work is confirmed by several techniques including reflection high-energy electron diffraction (RHEED) and X-ray diffraction (XRD). The value of full width at half maximum (FWHM) obtained by XRD for (004) $\omega$-rocking curve typically does not exceed 300 arcsec which is in agreement with previous works \cite{Wichrowska_ActPhysPolA2014}. Further improvement of the structural quality of grey tin epilayers is expected through enhancement of the surface quality of the CdTe buffer layer since the surface of {\aSn} epilayers replicates the roughness and defects of CdTe (\autoref{fig:AFM}). Previous reports have indicated that grey Sn quality, as well as transport, is very sensitive to the surface quality of CdTe and nucleation of $\beta$-Sn is possible on CdTe surface imperfections \cite{Asom_APL1989, Tu_APL1989, Reno_APL1989} even at temperatures lower than $\beta \rightarrow \alpha$ transition point (\qty{13.2}{\celsius}) (see \autoref{fig:XRD}). It is worth noting that $\beta$-Sn reflections are highly visible by XRD even for a relatively thin film of Sn and thus expected to be detectable even when a relatively small amount of $\beta$-Sn is present. In samples A, B, C, and D further investigated by magneto-optics and magneto-transport in the current work no traces of $\beta$-Sn are detected neither by XRD nor by TEM. The diamond structure of Sn films and lattice constant of $6.48\pm 0.02$ \AA deduced from HR-STEM images (\autoref{fig:TEM}a) are in good agreement with the parameters obtained from XRD. Sharp CdTe/{\aSn} and $\alpha$-Sn/Pt interfaces are revealed by TEM/EDX measurements (\autoref{fig:TEM}b).  Furthermore, a \qty{3.5}{\nano\meter} thick layer on the top of {\aSn} with an amorphous structure is identified as Sn oxide using TEM (\autoref{fig:TEM}c). Ultra-thin oxide layer is of advantage for subsequent transport measurements since it protects the film interior from further unwanted oxidation and other sorts of contamination coming from the sample surface. Despite the absence of structural defects e.g. dislocations and inclusions of other phases, traces of stacking faults are discovered in the obtained films. Taking into account a tiny amount of these defects we expect no significant contribution from them to the transport. 

\FloatBarrier
\newpage

\section{{\kp} band structure under strain}
The Pidgeon-Brown \cite{Groves_JPhysChemSol1970, Pidgeon_PhysRev1966} Hamiltonian writes:

\begin{align*}
    \renewcommand{\arraystretch}{2.5}
    H_{k}=
    \begin{pmatrix}
    E_{g} & 0 & -\frac{P k_{+}}{\sqrt{2}} & \sqrt{\frac{2}{3}} P k_{z} & \frac{P k_{-}}{\sqrt{6}} & 0 & \frac{P k_{z}}{\sqrt{3}} & \frac{P k_{-}}{\sqrt{3}} \\
    0 & E_{g} & 0 & -\frac{P k_{+}}{\sqrt{6}} & \sqrt{\frac{2}{3}} P k_{z} & \frac{P k_{-}}{\sqrt{2}} & \frac{P k_{+}}{\sqrt{3}} & -\frac{P k_{z}}{\sqrt{3}} \\
    -\frac{P k_{-}}{\sqrt{2}} & 0 & c_{-} & 2 \sqrt{3} \gamma k_{z} k_{-} & \sqrt{3} \gamma k_{-}^{2} & 0 & 0 & 0 \\
    \sqrt{ \frac{2}{3} } P k_{z} & -\frac{P k_{-}}{\sqrt{6}} & 2 \sqrt{3} \gamma k_{z} k_{-} & c_{+} & 0 & \sqrt{3} \gamma k_{-}^{2} & 0 & 0 \\
    \frac{P k_{+}}{\sqrt{6}} & \sqrt{ \frac{2}{3} } P k_{z} & \sqrt{3} \gamma k_{+}^{2} & 0 & c_{+} & -2 \sqrt{3} \gamma k_{z} k_{-} & 0 & 0 \\
    0 & \frac{P k_{+}}{\sqrt{2}} & 0 & \sqrt{3} \gamma k_{+}^{2} & -2 \sqrt{3} \gamma k_{z} k_{+} & c_{-} & 0 & 0 \\
    \frac{P k_{z}}{\sqrt{3}} & \frac{P k_{-}}{\sqrt{3}} & 0 & 0 & 0 & 0 & -\Delta & 0 \\
    \frac{P k_{+}}{\sqrt{3}} & -\frac{P k_{z}}{\sqrt{3}} & 0 & 0 &0 & 0 & 0 & -\Delta
    \end{pmatrix}
\end{align*}

with $k_{\pm}=k_{x} \pm i k_{y}$ and $k_{z}$ along the [001] axis. The matrix is written in the basis:

\begin{align*}
    &\ket{+} \equiv \ket{S \uparrow}, \ket{-} \equiv \ket{S \downarrow}, \\
    &\ket{3/2} \equiv \ket{\frac{1}{\sqrt{2}} \left( X + iY \right) \uparrow},\\
    &\ket{1/2} \equiv \ket{\frac{i}{\sqrt{6}} \left[ \left( X + iY \right) \downarrow - 2Z \uparrow \right]},\\
    &\ket{-1/2} \equiv \ket{\frac{1}{\sqrt{6}} \left[ \left( X - iY \right) \uparrow + 2Z \downarrow \right]},\\
    &\ket{-3/2} \equiv \ket{\frac{i}{\sqrt{2}} \left( X - iY \right) \downarrow},\\
    &\ket{7/2} \equiv \ket{\frac{i}{\sqrt{3}} \left[- \left( X - iY \right) \uparrow + Z \downarrow \right]},\\
    &\ket{-7/2} \equiv \ket{\frac{1}{\sqrt{3}} \left[ \left( X + iY \right) \downarrow + Z \uparrow \right]}
\end{align*}

This Hamiltonian describes four bands that are twofold spin degenerated each, the s-type (S) band, and the three P-type bands: the heavy hole (HH) band, the inversed light hole (iLH) band and the split-off (P) band. $P$ is the interband momentum matrix element accounting for the interaction between the $S$ band and the three p-type bands. Here, $c_{\pm} = -\left( \gamma_{1} \mp \gamma\right) \hbar^{2}(k_x^2+k_y^2) / 2m_{0}-\left(\gamma_{1} \mp 2 \gamma\right) \hbar^{2} k_{z}^{2} / 2 m_{0}$. It introduces dimensionless parameters $\left(\gamma_{i}\right)$ accounting for the remote band effects and in the spherical approximation, $\gamma_{2}=\gamma_{3}=\gamma$.

To account for the compressive biaxial strain along the [001] axis, the Bir-Pikus \cite{BirPikus_1974, Laude_PRB1971} Hamiltonian, which is written in the same basis, is added, so that the comprehensive matrix writes:

\begin{align*}
    \renewcommand{\arraystretch}{2.5}
    H_{k}=
    \begin{pmatrix}
    \tilde{E}_{g} & 0 & -\frac{P k_{+}}{\sqrt{2}} & \sqrt{\frac{2}{3}} P k_{z} & \frac{P k_{-}}{\sqrt{6}} & 0 & \frac{P k_{z}}{\sqrt{3}} & \frac{P k_{-}}{\sqrt{3}} \\
    0 & \tilde{E}_{g} & 0 & -\frac{P k_{+}}{\sqrt{6}} & \sqrt{\frac{2}{3}} P k_{z} & \frac{P k_{-}}{\sqrt{2}} & \frac{P k_{+}}{\sqrt{3}} & -\frac{P k_{z}}{\sqrt{3}} \\
    -\frac{P k_{-}}{\sqrt{2}} & 0 & c_{-} & 2 \sqrt{3} \gamma k_{z} k_{-} & \sqrt{3} \gamma k_{-}^{2} & 0 & 0 & 0 \\
    \sqrt{ \frac{2}{3} } P k_{z} & -\frac{P k_{-}}{\sqrt{6}} & 2 \sqrt{3} \gamma k_{z} k_{-} & c_{+}+\Omega & 0 & \sqrt{3} \gamma k_{-}^{2} & \frac{\Omega}{\sqrt{2}} & 0 \\
    \frac{P k_{+}}{\sqrt{6}} & \sqrt{ \frac{2}{3} } P k_{z} & \sqrt{3} \gamma k_{+}^{2} & 0 & c_{+}+\Omega & -2 \sqrt{3} \gamma k_{z} k_{-} & 0 & -\frac{\Omega}{\sqrt{2}} \\
    0 & \frac{P k_{+}}{\sqrt{2}} & 0 & \sqrt{3} \gamma k_{+}^{2} & -2 \sqrt{3} \gamma k_{z} k_{+} & c_{-} & 0 & 0 \\
    \frac{P k_{z}}{\sqrt{3}} & \frac{P k_{-}}{\sqrt{3}} & 0 & \frac{\Omega}{\sqrt{2}} & 0 & 0 & \frac{\Omega}{2}-\Delta & 0 \\
    \frac{P k_{+}}{\sqrt{3}} & -\frac{P k_{z}}{\sqrt{3}} & 0 & 0 & -\frac{\Omega}{\sqrt{2}} & 0 & 0 & \frac{\Omega}{2}-\Delta
    \end{pmatrix}
\end{align*}

where the origin of the energy has been taken to the $\mathrm{HH}$ band. There is a renormalization of the S-band gap, $ \tilde{E}_{g} = E_{g} + a \varepsilon - b\left(\varepsilon_{\parallel}-\varepsilon_{\perp}\right)$ with $a$ the hydrostatic deformation potential; and $\Omega=-2b\left(\varepsilon_{\parallel}-\varepsilon_{\perp}\right)$ which denotes the iLH - HH band splitting at $k=0$ that is responsible for the emergence of the Dirac cones. $b$ is the shear deformation potential and $\varepsilon_{\parallel}, \varepsilon_{\perp}$ are the in-plane and out-of-plane strain values $\left(\varepsilon_{x x}=\varepsilon_{y y}=\varepsilon_{\parallel}\right)$.

\section{Landau levels and magneto-optical oscillator strength}

Under an applied magnetic field along [001] $\parallel$ z, we perform the following Peierls substitution in the Hamiltonian (A1): $k_{+}=\sqrt{2 \xi} a^{\dagger}$ and $k_{-}=\sqrt{2 \xi} a$, with $\xi=e B / \hbar \equiv m \omega_{o} / \hbar$ the inverse square magnetic length and $a$ and $a^{\dagger}$ the usual ladder operators. The final Hamiltonian is then projected on a harmonic oscillator functions basis and decouples in two blocks I and II (Hamiltonians A2 and A3)at $k_{z}=0$. The two blocks correspond to the Kramers pair and write:

\begin{landscape}
\begin{align*}
    \renewcommand{\arraystretch}{2.5}
    H_{B}^{I} =
    \begin{pmatrix}
    \tilde{E}_{g} & -P \sqrt{\hbar \omega_{o}(n+1)} & P \sqrt{\frac{\hbar \omega_{o} n}{3}} & P \sqrt{\frac{2 \hbar \omega_{o} n}{3}} \\
    -P \sqrt{\hbar \omega_{o}(n+1)} & -\hbar \omega_{o}\left[\left(\gamma_{1}+\gamma\right)(2 n-3)+\frac{3}{2} \kappa\right] & 2 \gamma \hbar \omega_{o} \sqrt{3 n(n-1)} & 0 \\
    P \sqrt{\frac{\hbar \omega_{o} n}{3}} & 2 \gamma \hbar \omega_{o} \sqrt{3 n(n-1)} & \Omega-\hbar \omega_{o}\left[\left(\gamma_{1}-\gamma\right)(2 n+1)-\frac{1}{2} \kappa\right] & -\frac{1}{\sqrt{2}}\left[\Omega+\hbar \omega_{o}(\kappa+1)\right] \\
    P \sqrt{\frac{2 \hbar \omega_{o} n}{3}} & 0 & -\frac{1}{\sqrt{2}}\left[\Omega+\hbar \omega_{o}(\kappa+1)\right] & \frac{\Omega}{2}-\Delta+\hbar \omega_{o}\left(\kappa+\frac{1}{2}\right)
    \end{pmatrix}
    \text { \#(A2) } \\
\end{align*}

\begin{align*}
    \renewcommand{\arraystretch}{2.5}
    H_{B}^{I I} = \left( 
    \begin{array}{cccc}
    \tilde{E}_{g} & -P \sqrt{\frac{\hbar \omega_{o}(n+1)}{3}} & P \sqrt{\hbar \omega_{o}(n+2)} & P \sqrt{\frac{2 \hbar \omega_{o} n}{3}} \\
    -P \sqrt{\frac{\hbar \omega_{o}(n+1)}{3}} & \Omega-\hbar \omega_{o}\left[\left(\gamma_{1}-\gamma\right)(2 n+1)+\frac{1}{2} \kappa\right] & 2 \gamma \hbar \omega_{o} \sqrt{3(n+1)(n+2)} & \frac{1}{\sqrt{2}}\left[\Omega+\hbar \omega_{o}(\kappa+1)\right. \\
    P \sqrt{\frac{\hbar \omega_{o} n}{3}} & 2 \gamma \hbar \omega_{o} \sqrt{3(n+1)(n+2)} & -\hbar \omega_{o}\left[\left(\gamma_{1}+\gamma\right)(2 n+5)-\frac{3}{2} \kappa\right] & 0 \\
    P \sqrt{\frac{2 \hbar \omega_{o} n}{3}} & \frac{1}{\sqrt{2}}\left[\Omega+\hbar \omega_{o}(\kappa+1)\right. & 0 & \frac{\Omega}{2}-\Delta-\hbar \omega_{o}\left(\kappa+\frac{1}{2}\right)
    \end{array} 
    \right)
    \text { \#(A3) } 
\end{align*}

\end{landscape}

The first block (A2) is written in the basis $\ket{+;(n-1)}, \ket{3 / 2; (n-2)}, \ket{-1 / 2; n}, \ket{-7 / 2; n}$ while the second block (A3) is written in the basis $\ket{-; (n+1)}, \ket{1 / 2 ; n}, \ket{-3 / 2 ;(n+2)}, \ket{7 / 2 ; n}$. The diagonalization of these matrices gives the Landau levels. For $n<2$, one would have to restrain the basis, leading to pathological Landau levels in the sense that they acquire an unnatural dispersion versus magnetic field. A striking evidence for this is the $k_{z}$-dispersion of Landau levels $n=-2$ and $n=0$ originated from HH and iLH bands respectively. The $k_{z}$-dispersion of the Landau levels can be calculated by adding the $k_{z}$-terms into the Hamiltonian that is no longer decoupled in two blocks.

In order to calculate the magneto-optical oscillator strengths of the transitions, we write the initial and final Landau levels involved in magneto-optical transitions as follows:

$$
\begin{gathered}
    \left|i^{I}\right\rangle=\left(\begin{array}{c}
    \alpha_{+, n}|n-1\rangle \\
    \alpha_{3 / 2, n}|n-2\rangle \\
    \alpha_{-1 / 2, n}|n\rangle \\
    \alpha_{-7 / 2, n}|n\rangle
    \end{array}\right) \quad \text { and } \quad\left|f^{I}\right\rangle=\left(\begin{array}{c}
    \beta_{+, m}|m-1\rangle \\
    \beta_{3 / 2, m}|m-2\rangle \\
    \beta_{-1 / 2, m}|m\rangle \\
    \beta_{-7 / 2, m}|m\rangle
    \end{array}\right) \\
    \left|i^{I I}\right\rangle=\left(\begin{array}{c}
    \alpha_{-, n}|n+1\rangle \\
    \alpha_{1 / 2, n}|n\rangle \\
    \alpha_{-3 / 2, n}|n+2\rangle \\
    \alpha_{7 / 2, n}|n\rangle
    \end{array}\right) \quad \text { and } \quad\left|f^{I I}\right\rangle=\left(\begin{array}{c}
    \beta_{+, m}|m+1\rangle \\
    \beta_{1 / 2, m}|m\rangle \\
    \beta_{-3 / 2, m}|m+2\rangle \\
    \beta_{7 / 2, m}|m\rangle
    \end{array}\right)
\end{gathered}
$$

The oscillator strengths of the transitions are proportional to $\left|M_{\pm, n \rightarrow m}\right|^{2} \equiv\left|\left\langle f\left|d_{\pm}\right| i\right\rangle\right|^{2}$, where $d_{\pm}$is the dipole operators of the $\sigma^{+}$and $\sigma^{-}$light polarization used in our case (Faraday geometry). These operators write:

$$
d_{\pm}=\boldsymbol{\epsilon}_{\pm} \cdot \boldsymbol{v}=\frac{1}{\hbar} \sum_{j=x, y, z} \epsilon_{\pm, j} \frac{\partial H_{k}}{\partial k_{j}}
$$

With $\boldsymbol{\epsilon}_{\pm}=(1 / \sqrt{2} ; \pm i / \sqrt{2} ; 0)$ for the $\sigma^{+}$and $\sigma^{-}$polarization. The calculations give four series of allowed transitions. They occur between levels of the same block, thus, conserving spin and with indices $n$ differing by $\pm 1$ for the polarizations $\sigma^{\pm}$. The oscillator strengths of these four series are:

$$
\begin{aligned}
    &M_{+, n \rightarrow n+1}^{I}=\frac{2}{\hbar}\left[P\left(\sqrt{\frac{1}{12}} \beta_{+, n+1}^{*} \alpha_{-1 / 2, n}+\frac{1}{\sqrt{6}} \beta_{+, n+1}^{*} \alpha_{-7 / 2, n}-\beta_{3 / 2, n+1}^{*} \alpha_{+, n}\right)+2 \sqrt{6 n} \gamma \hbar \omega_{o} \beta_{3 / 2, n+1}^{*} \alpha_{-1 / 2, n}-\right. \\
    &\left.\sqrt{(n-1)} \hbar \omega_{o}\left(\gamma_{1}+\gamma\right) \beta_{3 / 2, n+1}^{*} \alpha_{3 / 2, n}-\sqrt{(n+1)} \hbar \omega_{o}\left(\gamma_{1}-\gamma\right) \beta_{-1 / 2, n+1}^{*} \alpha_{-1 / 2, n}\right] \\
    &M_{-, n \rightarrow n-1}^{I}=\frac{2}{\hbar}\left[P\left(\sqrt{\frac{1}{12}} \beta_{-1 / 2, n-1}^{*} \alpha_{+, n}+\frac{1}{\sqrt{6}} \beta_{-7 / 2, n-1}^{*} \alpha_{+, n}-\beta_{+, n-1}^{*} \alpha_{3 / 2, n}\right)+2 \sqrt{6(n-1)} \gamma \hbar \omega_{o} \beta_{-\frac{1}{2}, n-1}^{*} \alpha_{3 / 2, n}-\right. \\
    &\left.\sqrt{(n-2)} \hbar \omega_{o}\left(\gamma_{1}+\gamma\right) \beta_{3 / 2, n-1}^{*} \alpha_{3 / 2, n}-\sqrt{n} \hbar \omega_{o}\left(\gamma_{1}-\gamma\right) \beta_{-1 / 2, n-1}^{*} \alpha_{-1 / 2, n}\right] \\
    &M_{+, n \rightarrow n+1}^{I I}=\frac{2}{\hbar}\left[P\left(-\sqrt{\frac{1}{12}} \beta_{1 / 2, n+1}^{*} \alpha_{-, n}+\frac{1}{\sqrt{6}} \beta_{\frac{7}{2}, n+1}^{*} \alpha_{-, n}+\beta_{-, n+1}^{*} \alpha_{-3 / 2, n}\right)+2 \sqrt{6(n+2)} \gamma \hbar \omega_{o} \beta_{1 / 2, n+1}^{*} \alpha_{-3 / 2, n}-\right. \\
    &\left.\sqrt{(n+3)} \hbar \omega_{o}\left(\gamma_{1}+\gamma\right) \beta_{-3 / 2, n+1}^{*} \alpha_{-3 / 2, n}-\sqrt{(n+1)} \hbar \omega_{o}\left(\gamma_{1}-\gamma\right) \beta_{1 / 2, n+1}^{*} \alpha_{1 / 2, n}\right] \\
    &M_{-, n \rightarrow n-1}^{I I}=\frac{2}{\hbar}\left[P\left(-\sqrt{\frac{1}{12}} \beta_{-, n-1}^{*} \alpha_{1 / 2, n}+\frac{1}{\sqrt{6}} \beta_{-, n-1}^{*} \alpha_{7 / 2, n}+\beta_{-3 / 2, n-1}^{*} \alpha_{-, n}\right)+2 \sqrt{6(n+1)} \gamma \hbar \omega_{o} \beta_{-3 / 2, n-1}^{*} \alpha_{1 / 2, n}-\right. \\
    &\left.\sqrt{(n+2)} \hbar \omega_{o}\left(\gamma_{1}+\gamma\right) \beta_{-3 / 2, n-1}^{*} \alpha_{-3 / 2, n}-\sqrt{n} \hbar \omega_{o}\left(\gamma_{1}-\gamma\right) \beta_{1 / 2, n-1}^{*} \alpha_{1 / 2, n}\right]
\end{aligned}
$$

The magneto-optical strengths using the parameters listed in Table I are shown in the color scale of Fig. 1 (c) of the main text.

\section{{\aSn} thin film}

\begin{figure}[htbp]
    \includegraphics[width=\textwidth]{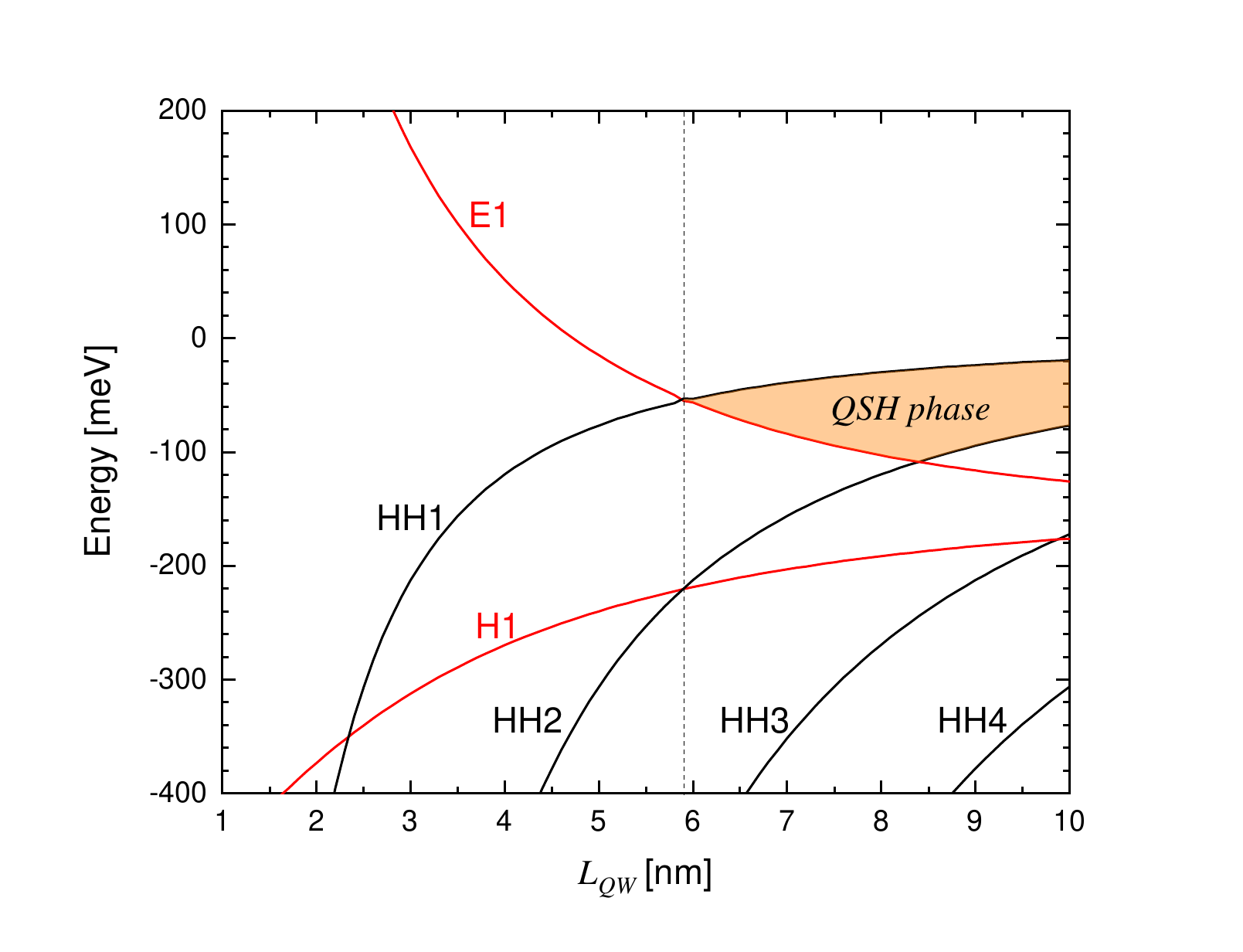}
    \caption{Calculated quantum confined states in {\aSn} with the parameters listed in Table I.}
    \label{fig:QSHI}
\end{figure}

\autoref{fig:QSHI} shows the calculated energy levels for a {\aSn} thin film (or quantum well) as a function of the thickness using the parameters listed in Table I of the main text. A confinement with infinite barriers is considered in the z-direction, thus, the substitution $k_{z} \rightarrow-i \frac{\partial}{\partial z}$ is performed. At $k_{x}=k_{y}=0$, the Hamiltonian (A1) decouples in two identical $4 \times 4$ blocs. The HH band is not interacting with the three other bands in $k_{z}$ so the confinement energies of the states $\mathrm{HH}_{\mathrm{n}}$ are simply :

$$
E_{n}=\frac{\hbar^{2} \pi^{2} n^{2} \gamma_{1}}{2 m_{0} L_{Q W}^{2}}
$$

The problem on the three other bands is treated numerically within the frame of the envelope function theory \cite{Bastard1996}. It results in the states $E_{1}$ and $H_{1}$, among others.

We observe that the states $\mathrm{E}_{1}$ and $\mathrm{HH}_{1}$ are inverted beyond a critical thickness $d_{c}=5.8 \mathrm{~nm}$, thus, a Quantum Spin Hall phase is expected, similarly to the case of HgTe.

\section{Extended magnetooptical data}
\label{section:MO}

\begin{figure}[htbp]
    \centering
    \includegraphics[width=0.7\textwidth]{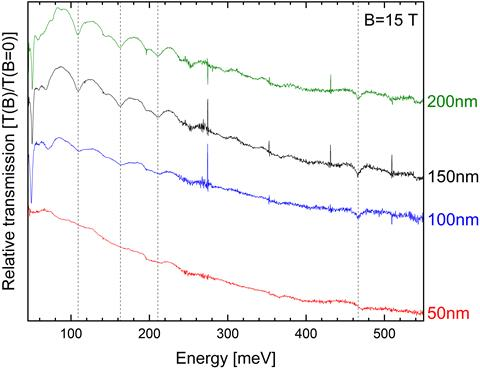}
    \caption{Comparison of the relative transmission for films with 50, 100, 150, and 200 nm. Dash vertical lines indicate absorption lines, which are the same for all thicknesses.}
    \label{fig:SM_MO}
\end{figure}

As it was said in the main text, MO spectra show the same behaviour for all studied thicknesses, see Fig.\ref{fig:SM_MO}. The 200~nm thick film was chosen for the detailed analysis since its superior quality allowed us to indicate more transitions, which in turn allowed us to fit the band parameters with higher accuracy.

\section{Extended ARPES data}
\label{section:ARPES}

\begin{figure}[htbp]
    \includegraphics[width=\columnwidth]{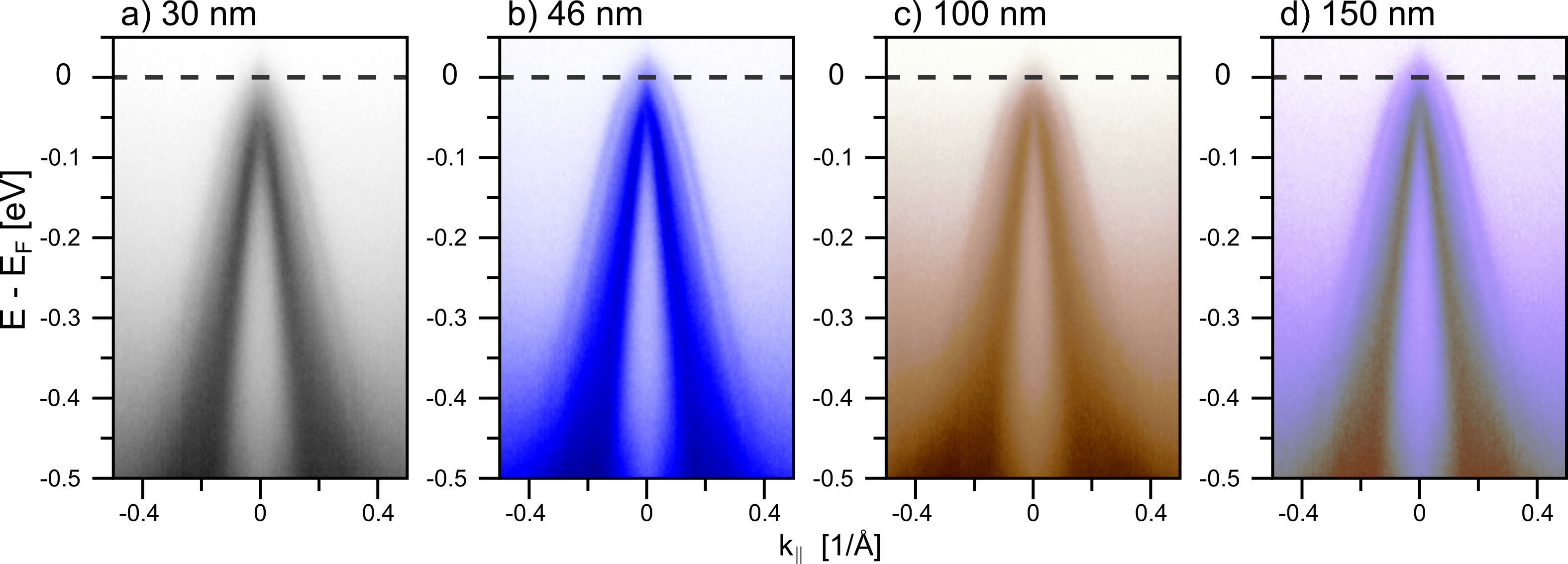}
    \caption{ARPES spectra of the $\alpha$-Sn epilayers of different thickness (30-200 nm) obtained in the vicinity of $\bar\Gamma$-point of surface BZ.}
    \label{fig:ARPES_thickness}
\end{figure}

\begin{figure}[htbp]
    \includegraphics[width=\columnwidth]{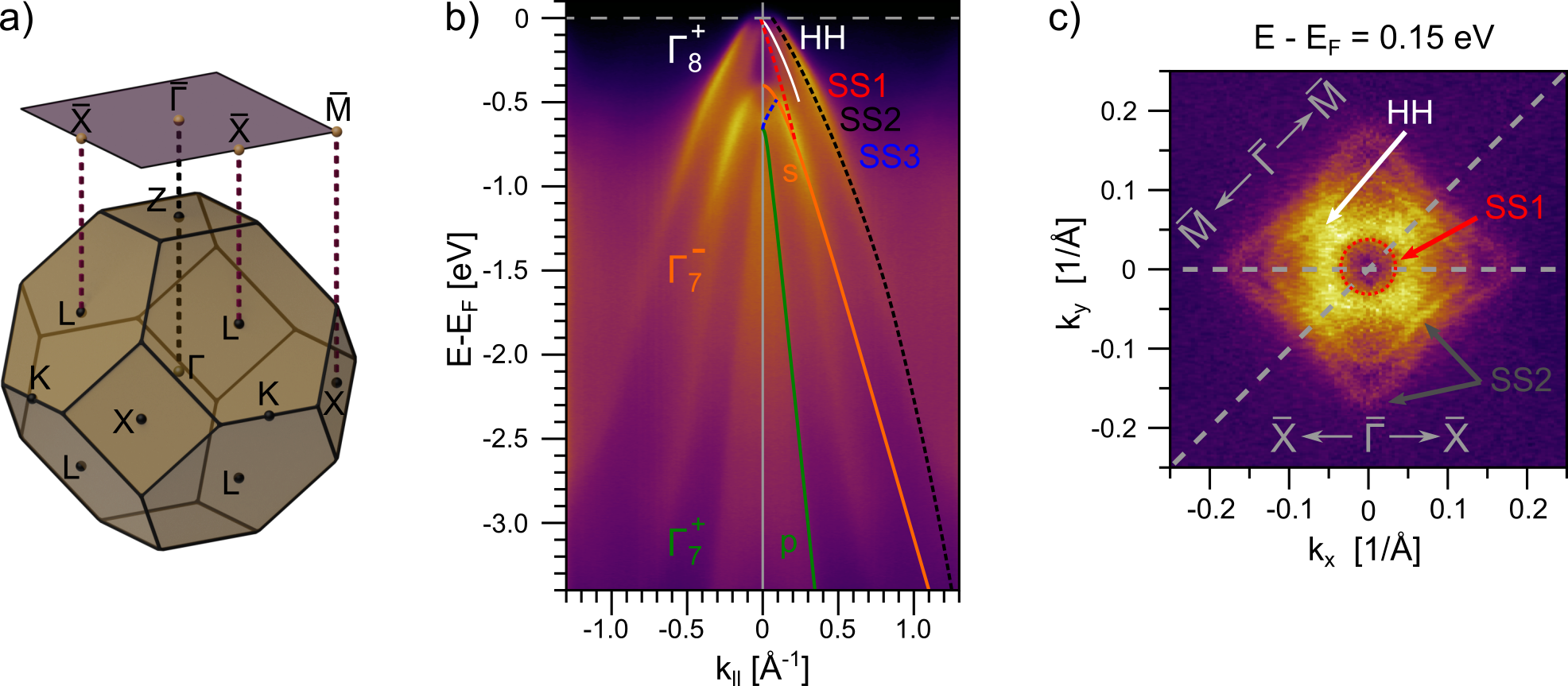}
    \caption{(a) 1st bulk Brillouin zone of $\alpha$-Sn and its projection to (001) surface, (b) ARPES spectrum close to $\Gamma$-point demonstrating  bulk and surface bands (the spectrum obtained at $E_{ph}$ = 90 eV, surface and bulk bands are marked with corresponding colors) (c) ARPES constant energy contour of b) demonstrating cross-section of bulk HH and surface SS1 and SS2 states (obtained at $E-E_{F}$ = 0.15 eV)}
    \label{fig:ARPES_bands}
\end{figure}

\begin{figure}[htbp]
    \includegraphics[width=\columnwidth]{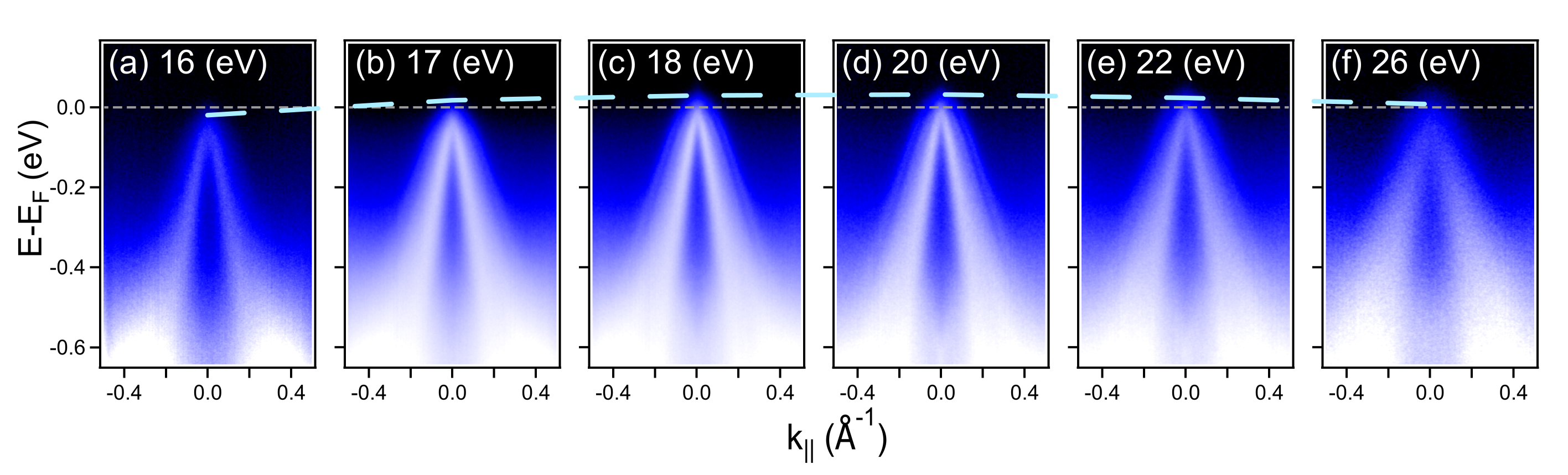}
    \caption{High-resolution ARPES spectra of {\aSn} epilayer grown on (001) CdTe/GaAs at different photon energies. The dashed line showing HH dispersion is a guide for the eyes.}
    \label{fig:Eph}
\end{figure}

In the ARPES study, four samples of (001) {\aSn}/CdTe, having a thickness of \qty{30}{\nano\meter}, \qty{46}{\nano\meter}, \qty{100}{\nano\meter} and \qty{150}{\nano\meter}, show identical band structure (\autoref{fig:ARPES_thickness}) in the vicinity of $\Bar{\Gamma}$ point along $\Bar{\Gamma} - \Bar{X}$ direction of surface Brillouin zone (BZ) (\autoref{fig:ARPES_bands}a). The band structure of one of the samples (\autoref{fig:ARPES_bands}b) including deep bands is presented. The Fermi energy ($E_{F}$) intersects the Dirac point, resulting in the observation of only valence bands (VB). In addition to bands close to the top of the valence band (VB), namely bulk $\Gamma_8^+$ heavy hole (HH) band, as well as surface states SS1 and SS2, deep bands, namely $\Gamma_7^-$ {\it s}-type (S), $\Gamma_7^+$ {\it p}-states (P) and SS3 are observed.  The presence of these states and bands is in agreement with previous experimental works for strained {\aSn} on (001) InSb \cite{Rogalev_PRB2017, Chen_PRB2022}. SS1 exhibits linear Dirac-like dispersion, while SS2 has a more parabolic shape. The bulk HH band is situated in between the surface states. The SS2 is doubled as seen in \autoref{fig:ARPES_bands}c. The energy splitting between the S and HH bands at $k_{\parallel} = 0$, $|E_{g}|=0.40\pm 0.02$ eV is consistent with the value obtained by ellipsometry magneto-optic measurements \cite{Carrasco_APL2018} (see Table 2 of main text). However, the obtained energy separation between the P and HH bands, denoted as $\Delta=0.66\pm 0.15$ eV, differs from magneto-optics and ellipsometry values due to the large broadening of the P band and correspondingly substantial error in determining its position by ARPES.\\  
The surface states play an important role in spintronic applications \cite{Ding_AdvFunMat2021, Ding_AdvMat2021, Binda_PRB2021, RojasSanchez_PRL2016, Zhang_SciAdv2022}. The SS1 and SS3 have a topological origin and arise due to band inversion between $\Gamma_8^+$ and $\Gamma_7^-$ as well as $\Gamma_7^-$ and $\Gamma_7^+$ respectively, as shown in \cite{Rogalev_PRB2017}. The topological nature of SS1 has been proved by spin-resolved ARPES where their helical spin texture had been detected \cite{Rogalev_PRB2017, Ohtsubo_PRL2013}. The SS2 states have been experimentally observed only recently \cite{Chen_PRB2022} and have been interpreted as Rashba-split states. However, theoretically, the existence of additional massive SS (Dyakonov-Khaetski (DK) or Volkov-Pankratov (VP) states) was predicted in the '80s at the interface between normal insulator and Luttinger semimetal i.e {\aSn}, HgTe \cite{Dyakonov_JETP1981, Pankratov_SolStateComm1981}. More recently, there has been a renewed interest in the band structure of strained Luttinger semimetal, and the emergence of surface DK or VP states has been predicted for the strained semimetal phase \cite{Khaetskii_PRB2022, Kharitonov_arXiv2022}. Experimentally, surface states encircling bulk Dirac states, similar to {\aSn} have been observed recently in other DSM materials, e.g. Na$_3$Bi \cite{Xu_Science2015} by ARPES. The existence of VP states has been demonstrated recently in HgTe \cite{Mahler_PRX2019} and PbSnSe \cite{Bermejo-Ortiz_PRB2023} by transport and magneto-optics respectively. Therefore we believe that the origin of SS2 can be attributed to massive DK or VP states.\\ 
The photon energy dependence of the {\aSn} epilayer spectrum in the vicinity of $\Gamma$ point is illustrated in \autoref{fig:Eph}. Although SS1 is seen almost at any energy, SS2 completely vanished at $E_{ph}$ = 16 and 26 eV which can be attributed to matrix-element effects. Such periodic suppression of photoemission intensity has been often observed previously for low-energy ARPES \cite{Louie_PRL1980}. The weak dispersion of HH states shown in the figure by a dashed line is anticipated as evidence of the DSM phase.

\FloatBarrier
\newpage

\section{Extended magneto-transport data}
\subsection{Temperature dependence and MR in $B^{\text{normal}}$}
\begin{figure}[htbp]
  \captionsetup[subfigure]{labelformat = empty}
  \centering
    \subfloat[\label{fig:RHO_vs_T}]{\includegraphics[width=0.49\textwidth]{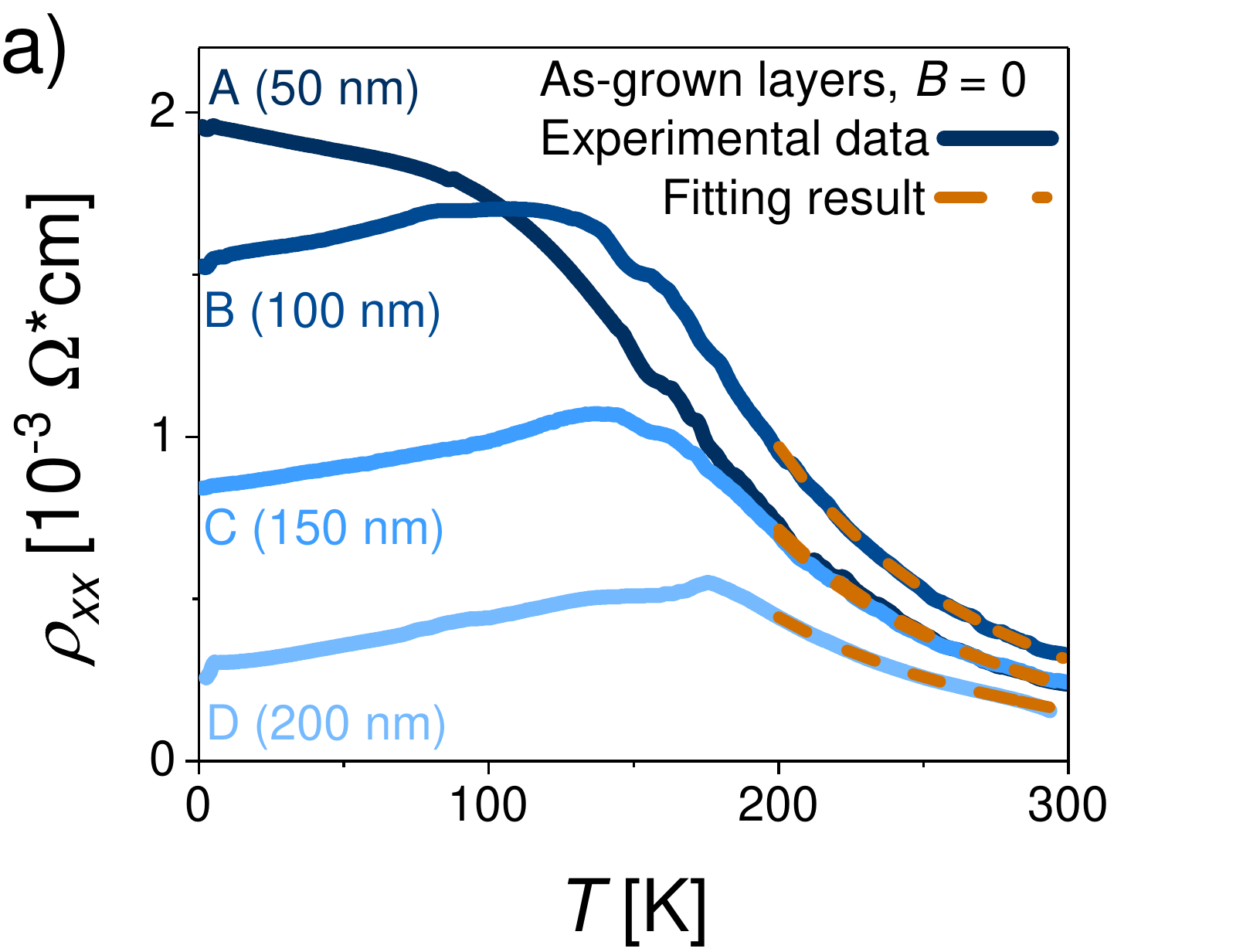}}\hfill
    \subfloat[\label{fig:RHOxx_RHOxy_50nm}]{\includegraphics[width = 0.49\textwidth]{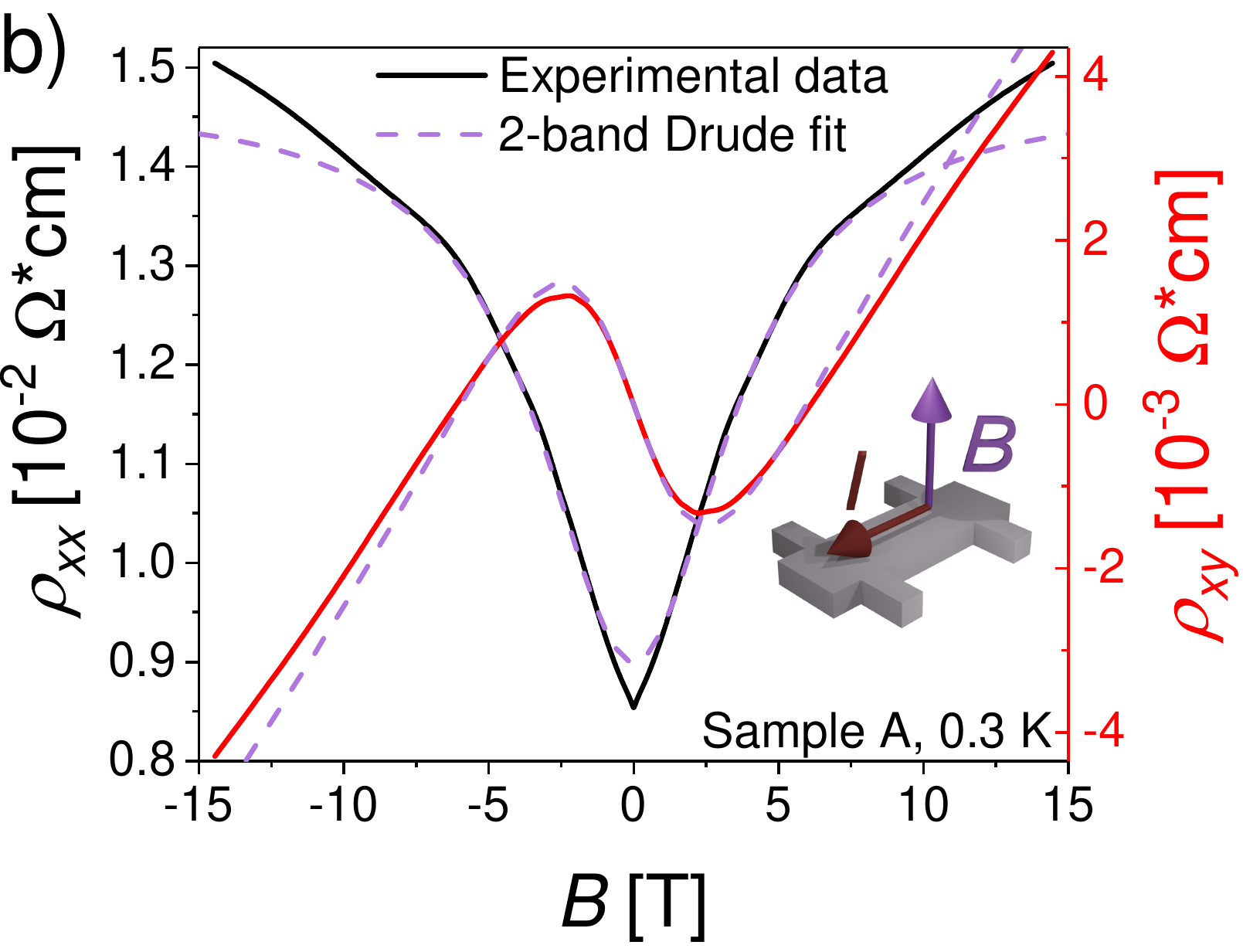}}
    
    \subfloat[\label{fig:RHOxx_RHOxy_100nm}]{\includegraphics[width = 0.49\textwidth]{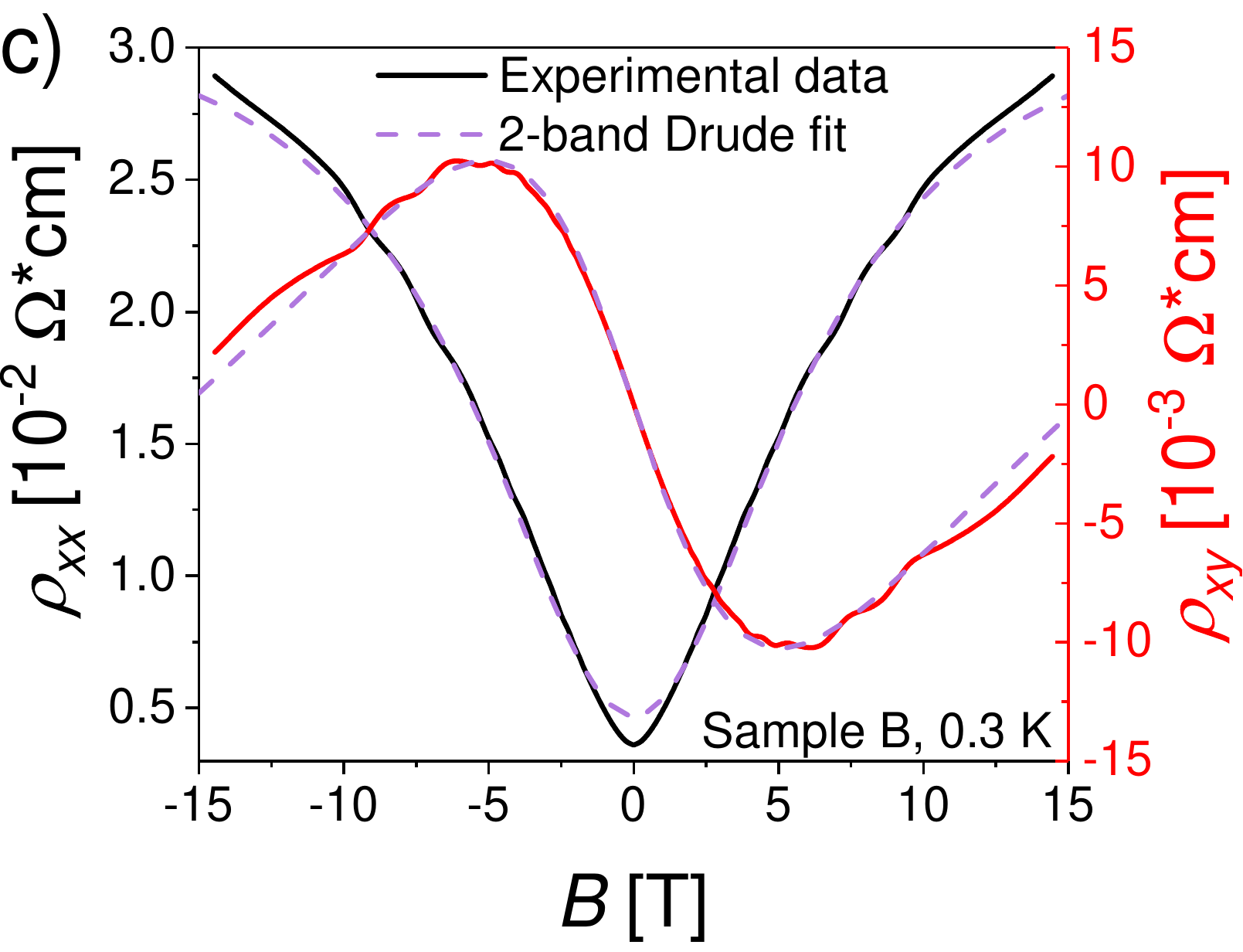}}\hfill
    \subfloat[\label{fig:RHOxx_RHOxy_150nm}]{\includegraphics[width = 0.49\textwidth]{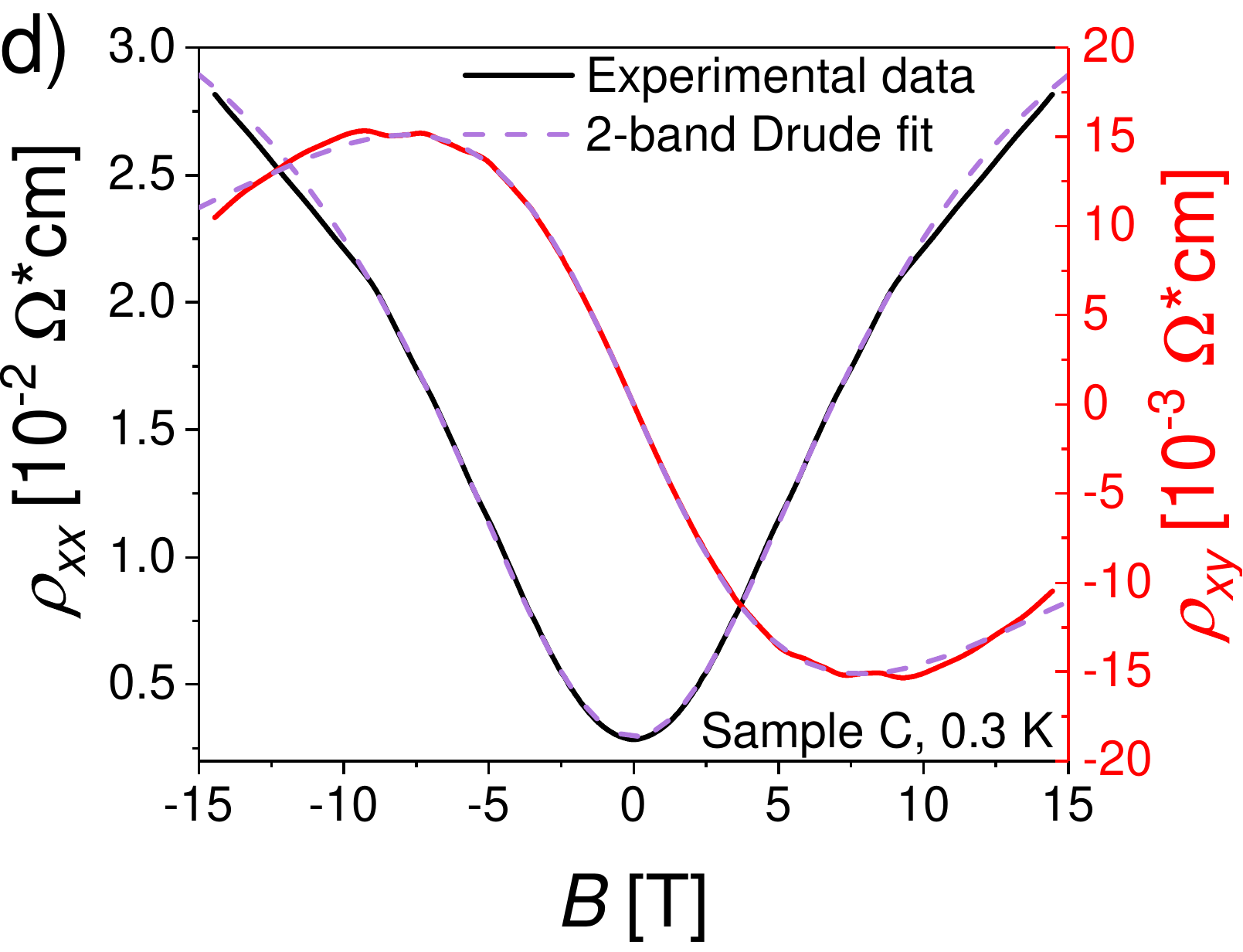}}  
    \caption{\protect\subref{fig:RHO_vs_T}) Temperature-dependent resistivity of {\aSn} layers with various thicknesses (blue lines) and fitting of the thermal activation model (dashed, orange lines) in the range \qtyrange{200}{300}{\kelvin}. \protect\subref{fig:RHOxx_RHOxy_50nm}) - \protect\subref{fig:RHOxx_RHOxy_150nm}) Longitudinal and Hall resistivities (solid lines) of samples A-C, respectively, measured at $T = \qty{0.3}{\kelvin}$. Violet, dashed lines show the results of the fitting of the 2-band Drude model as explained in the text. Fitting was performed in the limited range of magnetic field, curves were expanded to cover the full range.}
    \label{fig:Hall}
\end{figure}

Temperature-dependent resistivity, measured for all layers prior to the etching, is shown in \autoref{fig:RHO_vs_T}. Between \qtyrange{200}{300}{\kelvin}, resistivity shows exponential behavior, typical for the thermally excited carriers in the semiconductor. Fitting of the thermal activation model, $\rho (T) = \rho_{\infty} + \rho_{0} \exp{(E_{act}/2kT)}$ yields the activation energy $E_{act} = \qtyrange{35}{60}{\meV}$. We attribute these values to the distance between the $E_{F}$ and the bottom of the L band. Below \qty{200}{\kelvin} the resistivity continues to increase with decreasing temperature and reaches a maximum for samples B -D. At even lower temperatures, we observe a metallic behavior for those layers, and we, therefore, conclude that thermally excited carriers from the L band do not contribute to the transport of {\aSn} in this temperature range.\\

\FloatBarrier

\begin{figure}[htbp]
\captionsetup[subfigure]{labelformat = empty}
    \centering
    \subfloat[\label{fig:DRHO_50nm}]{\includegraphics[width = 0.49\textwidth]{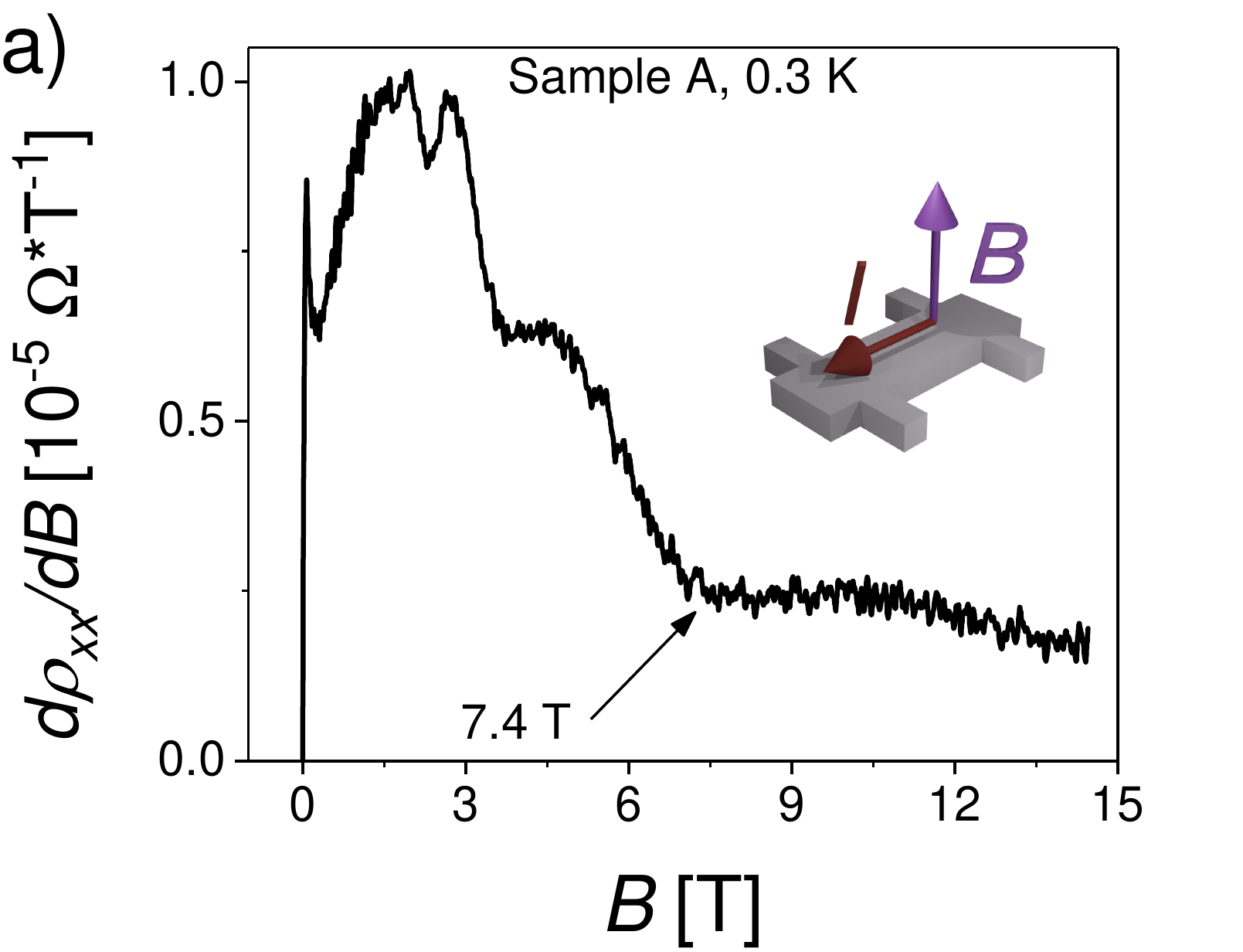}}\hfill
    \subfloat[\label{fig:DRHO_100nm}]{\includegraphics[width = 0.49\textwidth]{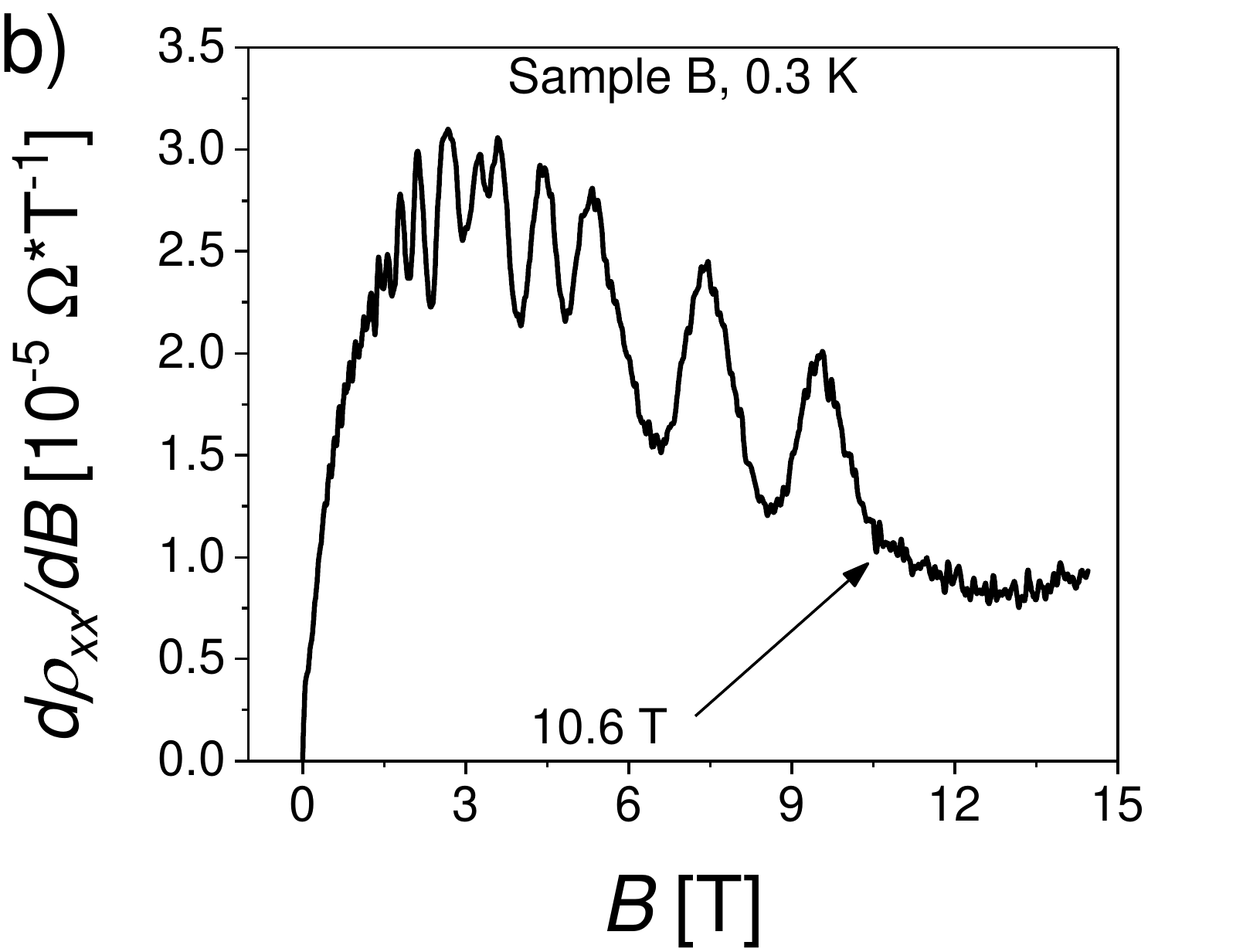}}
    
    \subfloat[\label{fig:DRHO_150nm}]{\includegraphics[width = 0.49\textwidth]{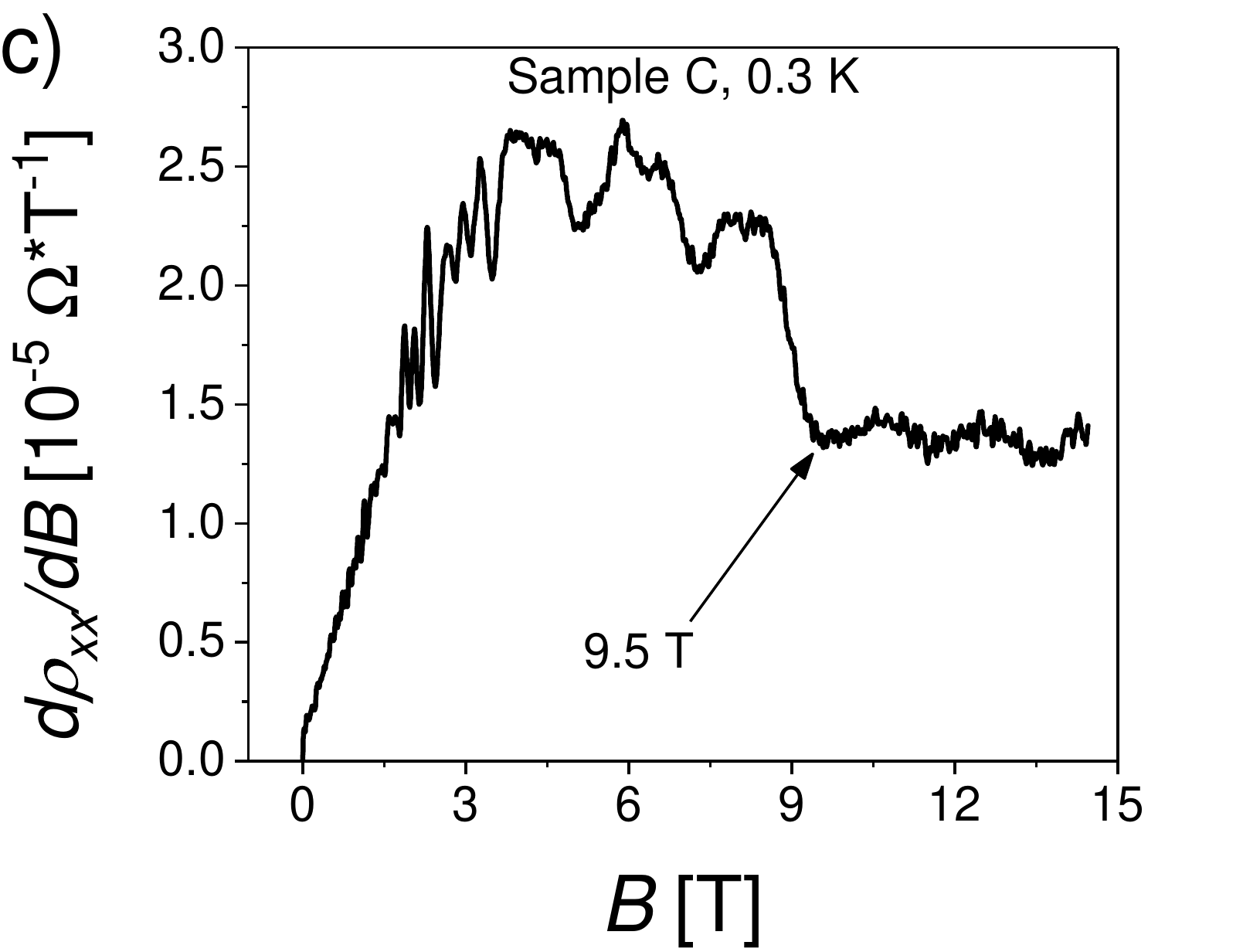}}\hfill
    \subfloat[\label{fig:DRHO_200nm}]{\includegraphics[width = 0.49\textwidth]{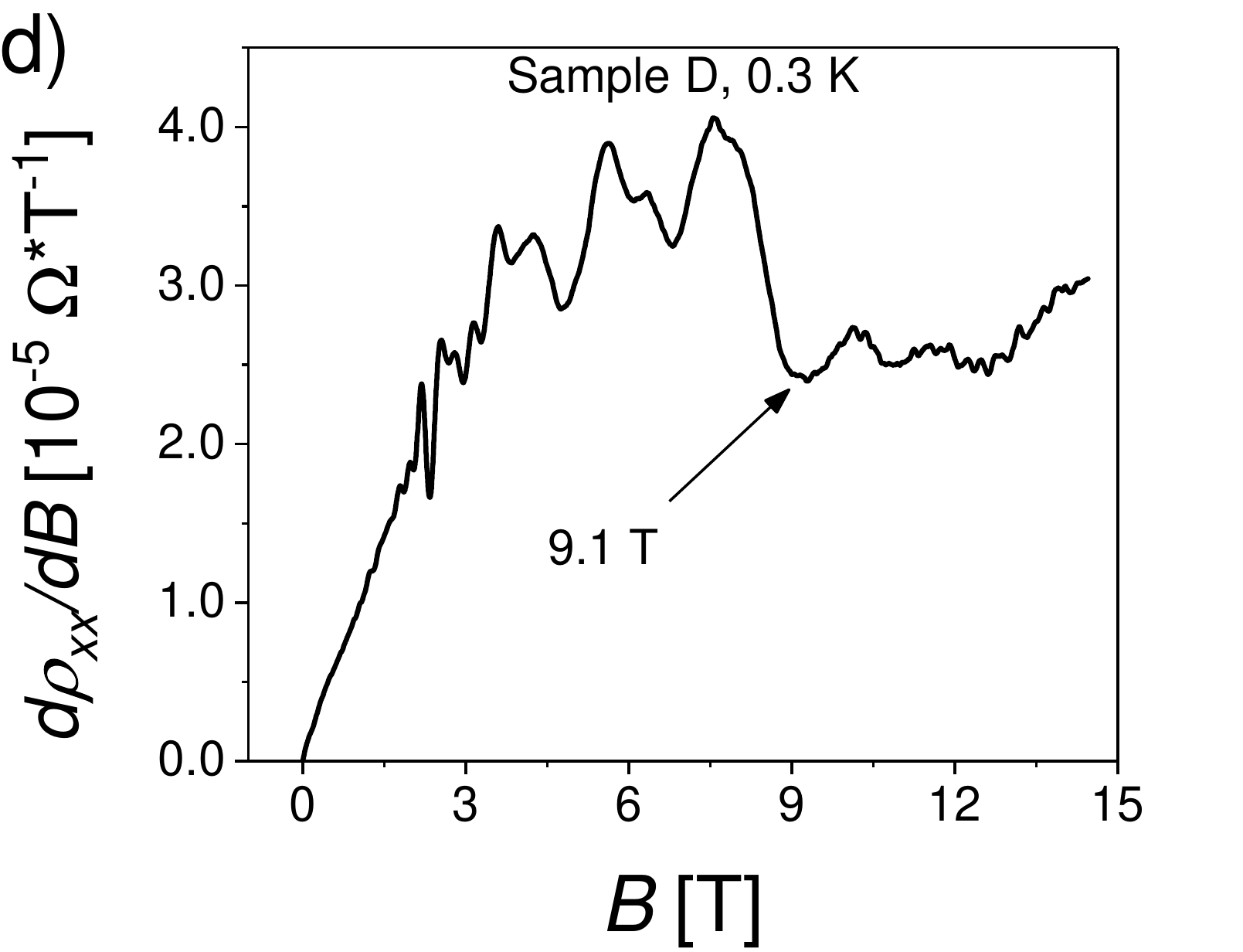}}
    \caption{First derivative of the longitudinal resistivity of {\aSn} layers in $B^{\text{normal}}$ at $T = \qty{0.3}{\kelvin}$. Arrows mark the value of $B$ where oscillations get damped and the derivative becomes almost constant.}
    \label{fig:DRHO}
\end{figure}

In $B^{\text{normal}}$, longitudinal magnetoresistance is positive and Hall magnetoresistance is non-linear, as presented in \autoref{fig:Hall}b - d. In the low-field regime, $\rho_{xx}$ and $\rho_{xy}$ can be simultaneously described by the 2-band Drude model:
\begin{subequations}
    \begin{align}
        \rho_{xx} = \frac{1}{e}\frac{n_{e}\mu_{e}+(n_{h}\mu_{h}\mu_{e}^{2}+n_{e}\mu_{e}\mu_{h}^{2})B^2}{(n_{e}\mu_{e}+n_{h}\mu_{h})^{2}+\mu_{e}^{2}\mu_{h}^{2}B^{2}(n_{e}-n_{h})^{2}}\\
        \rho_{xy} = \frac{B}{e}\frac{n_{e}\mu_{e}^{2}-n_{h}\mu_{h}^{2}+\mu_{e}^{2}\mu_{h}^{2}B^{2}(n_{e}-n_{h})}{(n_{e}\mu_{e}+n_{h}\mu_{h})^{2}+\mu_{e}^{2}\mu_{h}^{2}B^{2}(n_{e}-n_{h})^{2}}.
    \end{align}
    \label{eq:2-band_Drude}
\end{subequations}
Above, $e$ is the elementary charge and $n_{e, p}$, $\mu_{e, p}$ - density and mobility of electrons and holes, respectively. According to \autoref{eq:2-band_Drude}, longitudinal magnetoresistance saturates at the high magnetic field. However, this is not true for the studied samples. It can be noted that at the high-field region, $\rho_{xx}$ becomes nearly linear. This feature is clearly seen after plotting $\frac{d\rho_{xx}}{dB}$, as presented in \autoref{fig:DRHO}. Fitting of \autoref{eq:2-band_Drude} to experimental data was therefore performed in the limited range of magnetic fields, with the maximum value determined by the derivative, as marked by arrows in \autoref{fig:DRHO}. Resulting curves (\autoref{fig:Hall}b - d) were extended to cover the full range of magnetic fields. $\rho_{xx}$ and $\rho_{xy}$ were fitted simultaneously. It is clearly seen that MR deviates from the 2-band Drude model above the magnetic field determined in \autoref{fig:DRHO}. Another deviation from the Drude model, more pronounced for thinner layers, can be seen around $B = 0$ for $\rho_{xx}$. We attribute this feature to the weak antilocalization.\\

\begin{table}
    \caption{Value of carrier densities (hole $p$ and electron $n$) and mobilities ($\mu_p$ and $\mu_n$ for hole and electron mobilities, respectively), obtained for studied samples through the fitting the magneto-transport data at 0.3~K with the 2-band Drude model \autoref{eq:2-band_Drude}.}
    \centering
    \begin{tabular}{c c c c c c}
        \hline
        Sample & Thickness [nm] & $p$, [10$^{17}$, cm$^{-3}$] & $n$, [10$^{17}$, cm$^{-3}$] & $\mu_p$, [cm$^{2}$/Vs] & $\mu_n$, [cm$^{2}$/Vs] \\
        \hline
        A & 50 & 14.7 & 0.59 & 310 & 3960 \\
        B & 100 & 10.8 & 1.14 & 220 & 9500 \\
        C & 150 & 22.3 & 1.4 & 840 & 13440 \\
        D & 200 & 5.0 & 1.4 & 370 & 19050 \\
        \hline
    \end{tabular}
    \label{tab:electrophys}
\end{table}

\FloatBarrier

\subsection{Additional NLMR data}

\begin{figure}[htbp]
    \captionsetup[subfigure]{labelformat = empty}
    \subfloat[\label{fig:NLMR_100nm}]{\includegraphics[width = 0.49\textwidth]{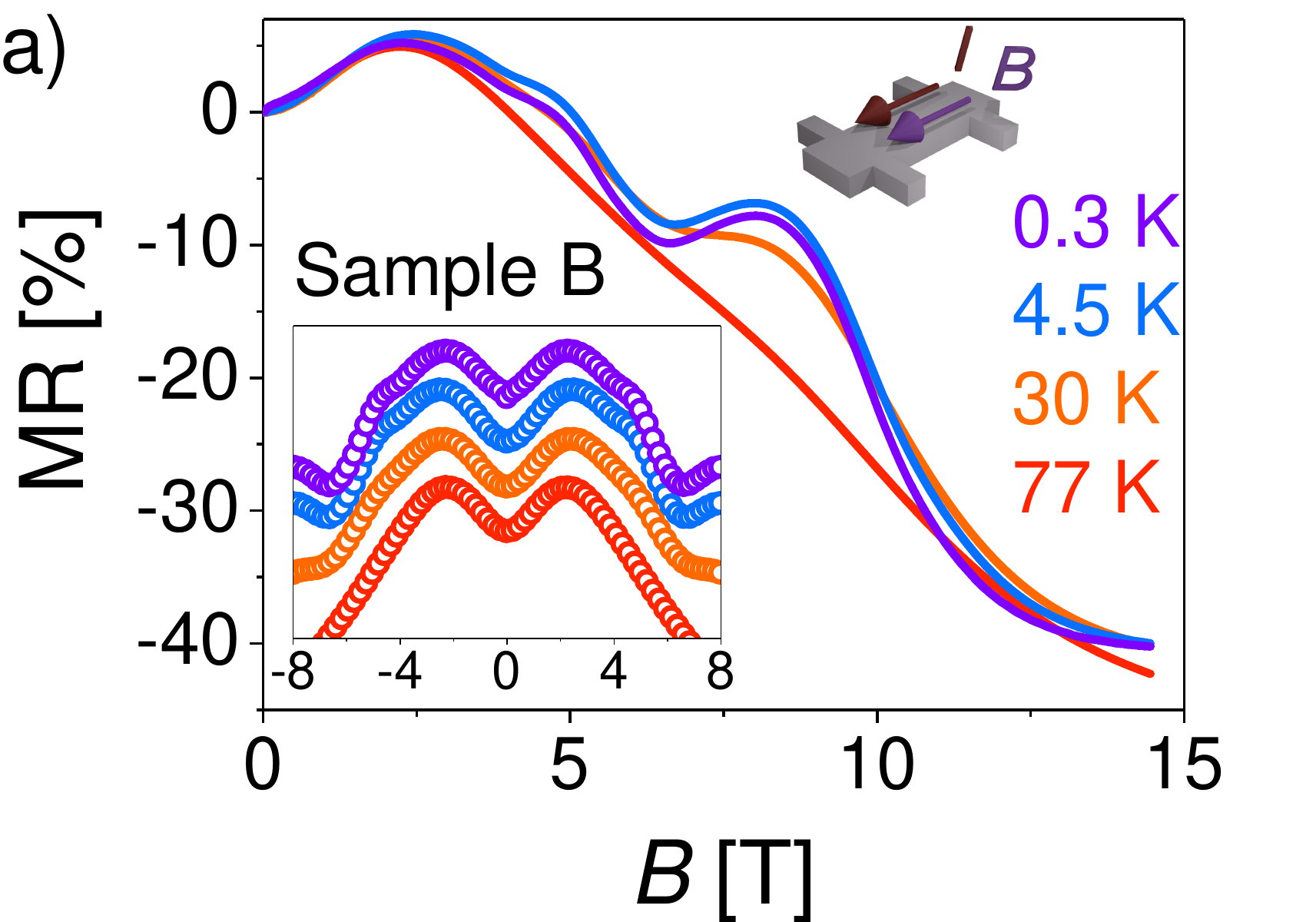}}\hfill
    \subfloat[\label{fig:NLMR_200nm}]{\includegraphics[width = 0.49\textwidth]{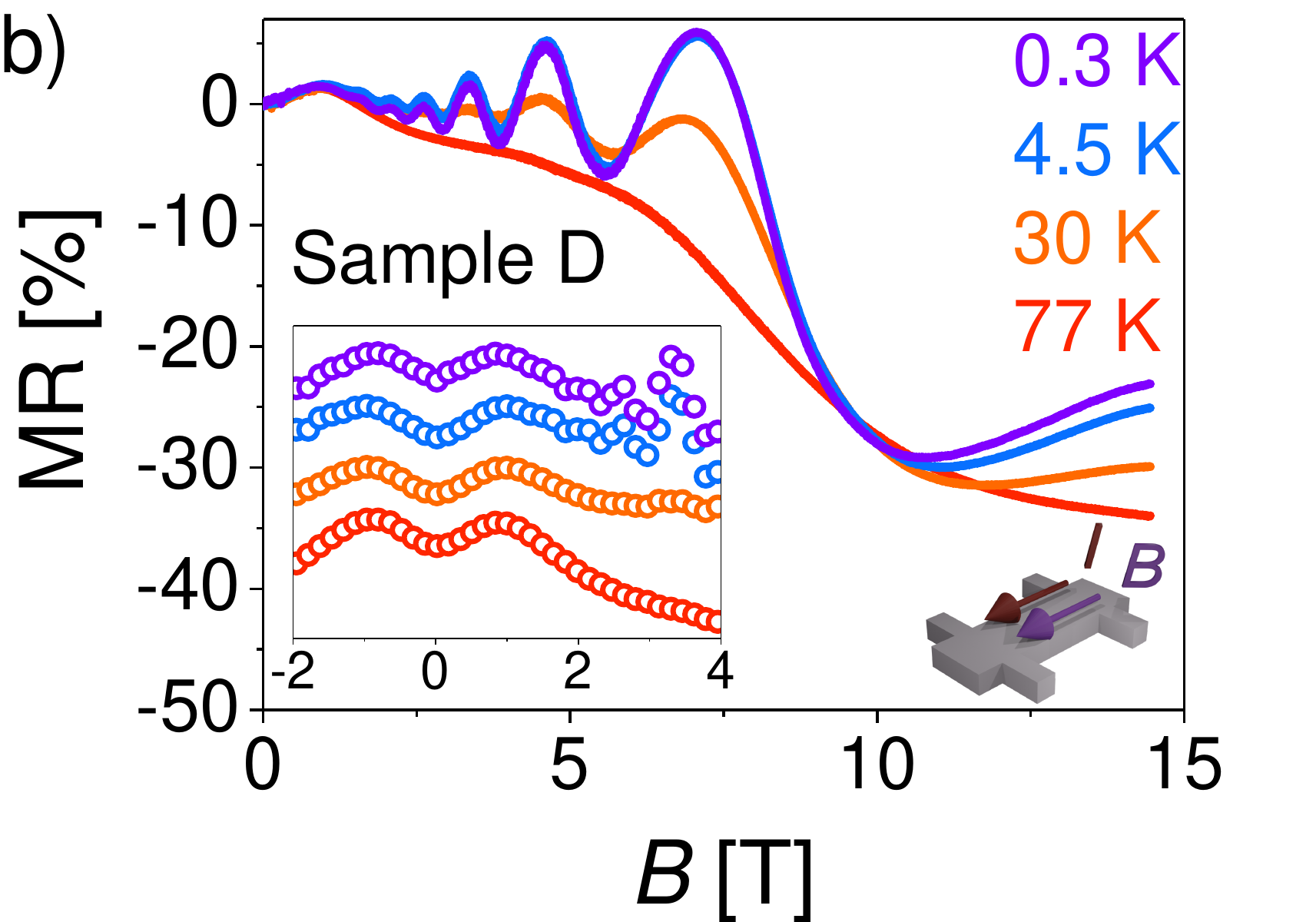}}
    
    \subfloat[\label{fig:SI_Bparal_HB2_100nm}]{\includegraphics[width = 0.49\textwidth]{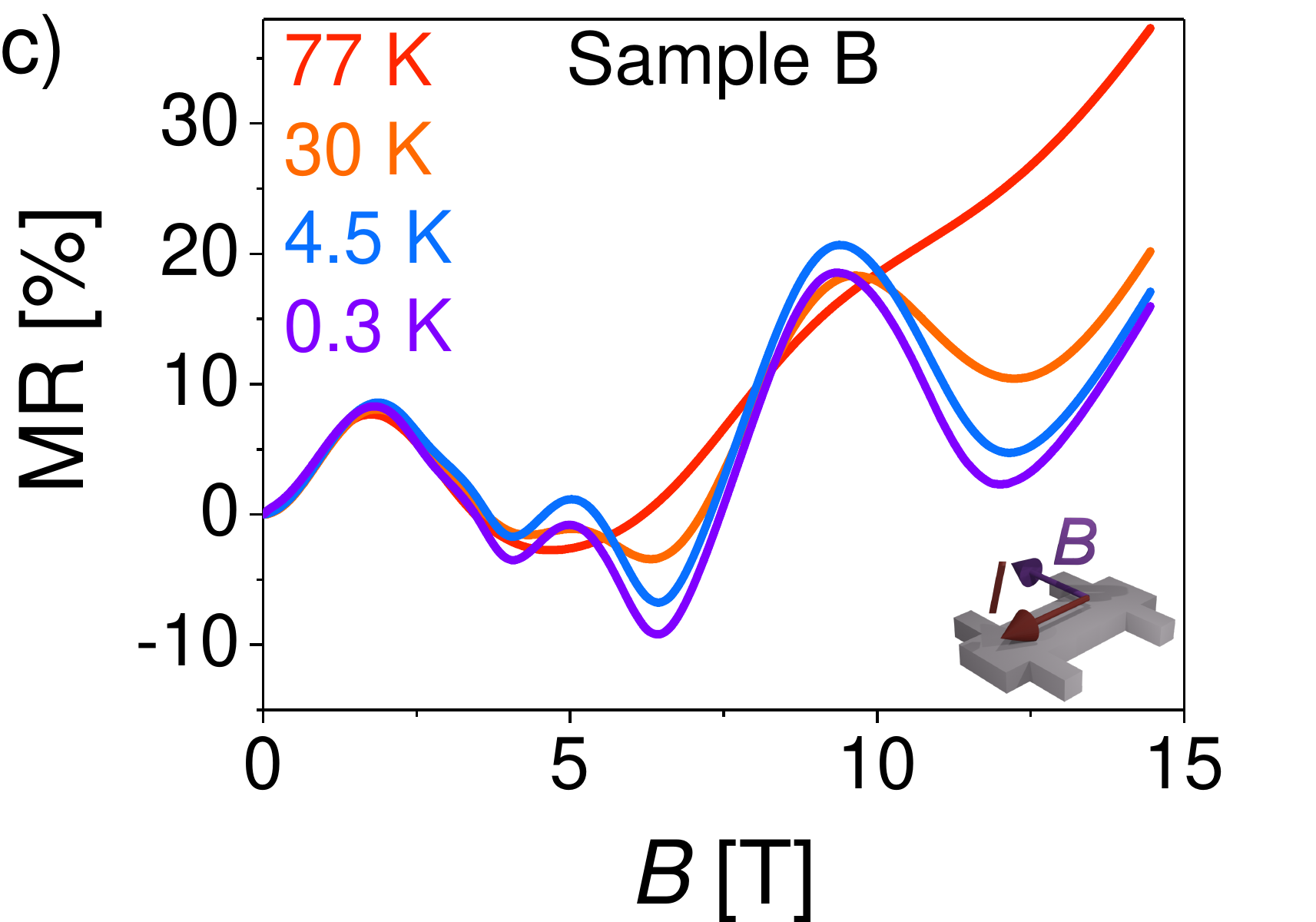}}\hfill
    \subfloat[\label{fig:SI_Bparal_HB2_200nm}]{\includegraphics[width = 0.49\textwidth]{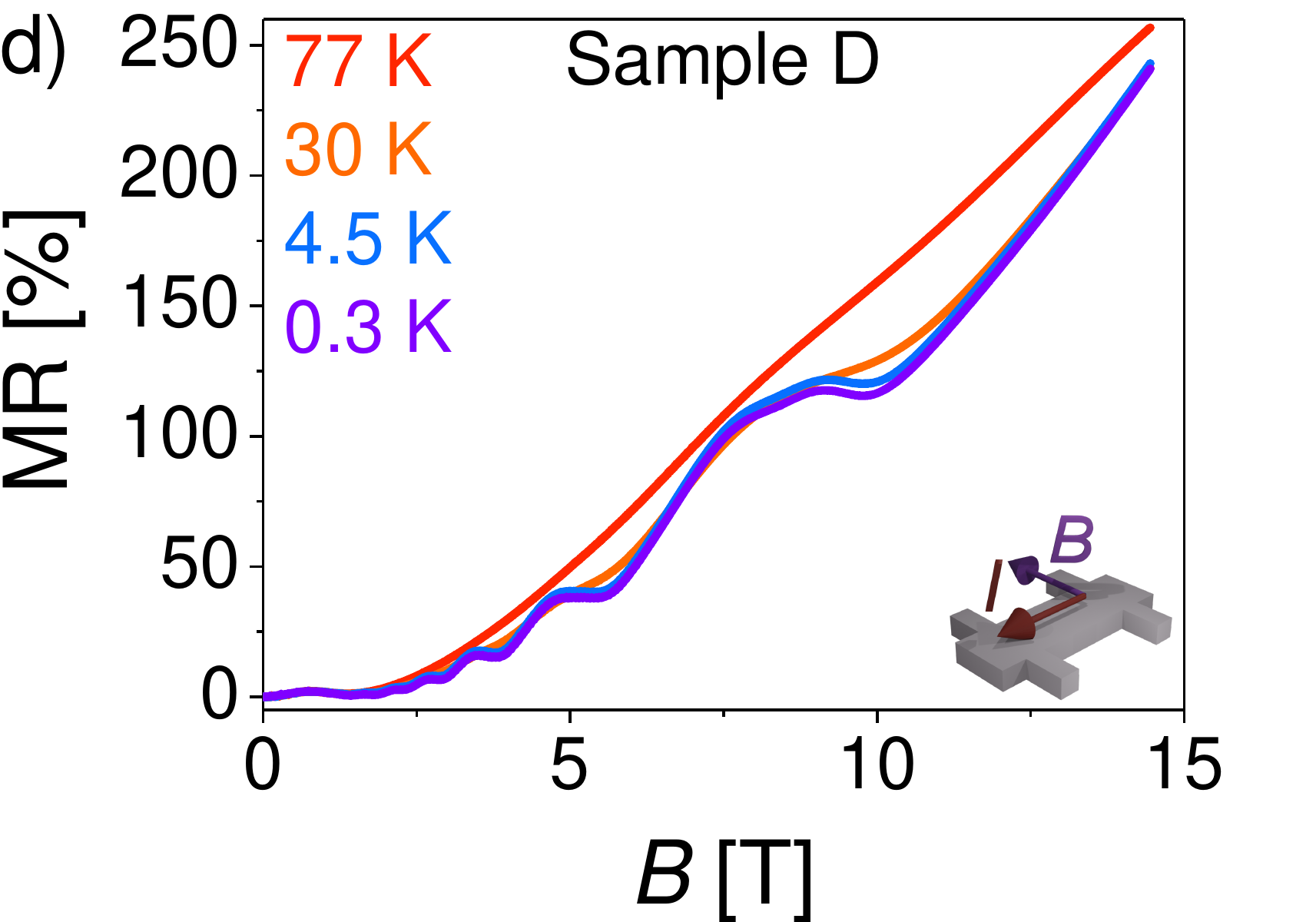}}
    \caption{Magnetoresistance of samples B and D in $B^{\text{in-plane}}$.}
    \label{fig:Bparal}
\end{figure}

\autoref{fig:Bparal} presents MR in $B^{\text{in-plane}}$ of samples B and D not included in the main text. The magnetoresistance of sample D strongly resembles the results obtained for sample C. For $B^{\text{in-plane}}_{\parallel}$, SdH oscillations dominate the low-temperature MR. At higher temperatures, where oscillations are damped, non-saturating negative MR is clearly seen. Positive low-field contribution is also present, without significant temperature dependence. The overall MR is positive for $B^{\text{in-plane}}_{\perp}$. Slightly different observations are made for sample B. Here, the SdH oscillations have lower amplitude than in samples C and D, and positive low-field MR in  $B^{\text{in-plane}}_{\parallel}$ is broader. We also notice a more pronounced negative part of the MR curve in $B^{\text{in-plane}}_{\perp}$.

\FloatBarrier

\subsection{Alternative sources of NLMR}

Here, we briefly discuss the alternative sources of NLMR, which could rule out the chiral anomaly.

The presence of \textbf{magnetic impurities} can lead to negative magnetoresistance because the ordering of magnetic moments in the magnetic field would result in reduced carrier scattering. We did not intentionally introduce any magnetic impurities, nor were they detected by the core-level spectroscopy. The obtained MR data does not support this mechanism, either. First, we do not observe any anomalous signal in the Hall resistance. Second, the magnitude of the NLMR increases at higher temperatures, contrary to the behavior expected for the magnetic ordering. We also note that NLMR is present in the large number of samples fabricated in the time span of more than one year, which makes unintentional doping highly unlikely.

\textbf{Current jetting effect} occurs in a medium with anisotropic conductivity tensor and results in the increased current density along a line parallel to the magnetic field (so-called ``jet'') \cite{Pippard2009, dosReis_NewJourPhys2016, Liang_PRX2018}. It is enhanced if the current is injected through the point contacts. When the magnetic field is increasing, the ``jet'' narrows down, which leads to an artificial negative magnetoresistance measured at the edges of the sample. It has also been shown \cite{dosReis_NewJourPhys2016} that the measured signal can change when the position of the voltage probes is modified. We argue that this effect is unlikely to be the source of NLMR in the investigated samples. First, the current was injected through the contact pads 20 times larger than the width of the studied Hall bars, which ensures its homogeneous distribution. Second, the signal measured with the voltage probes placed at the edges of the Hall bar and spanning across the Hall bar is the same. This is equivalent to the ``squeeze test'' proposed in the ref. \cite{Liang_PRX2018}. We have also measured a number of non-etched samples with various positions of voltage probes and always observed the same NLMR.

\begin{figure}[htbp]
    \captionsetup[subfigure]{labelformat = empty}
    \subfloat[\label{fig:WAL-test_a}]{\includegraphics[width = 0.49\textwidth]{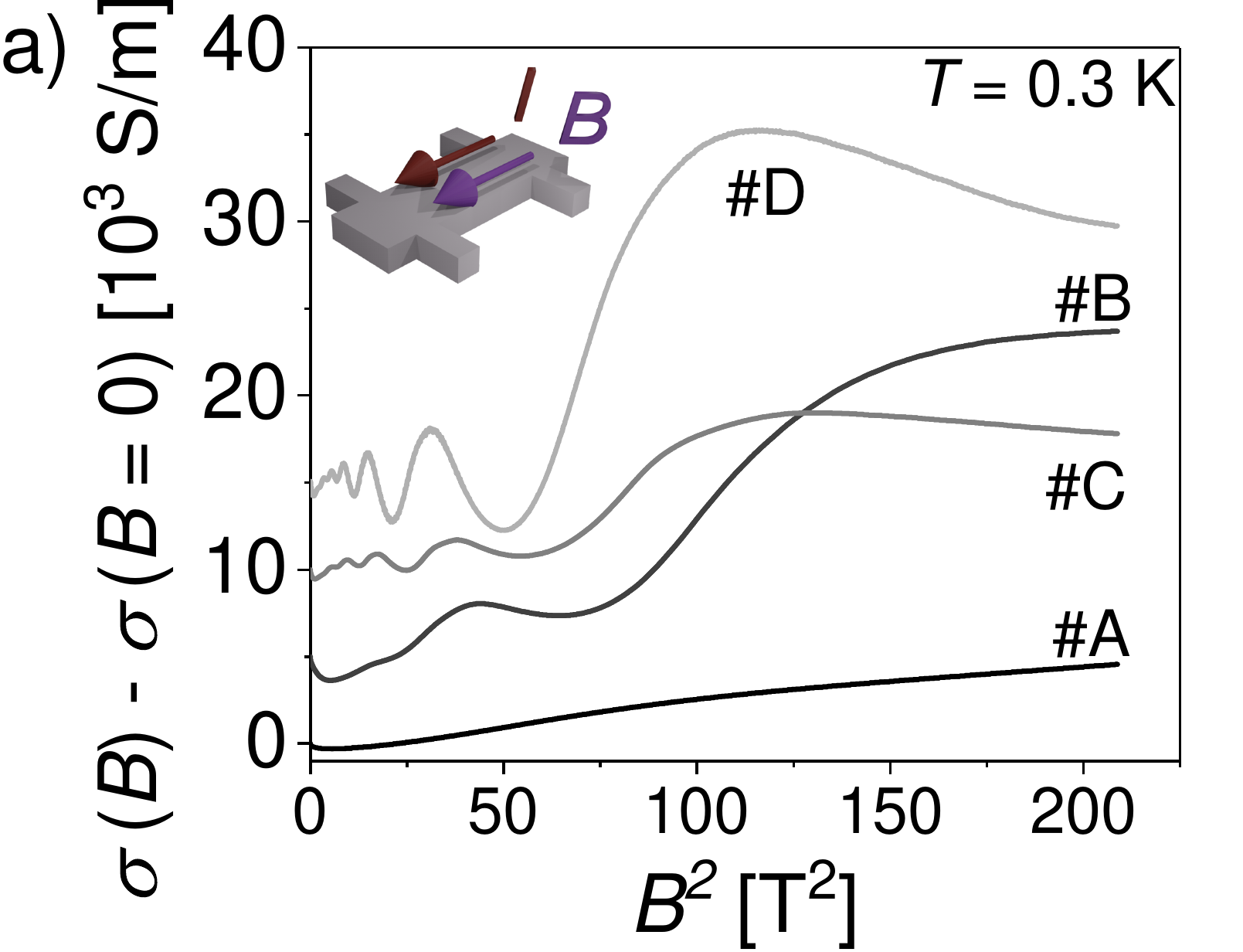}}\hfill
    \subfloat[\label{fig:WAL-test_b}]{\includegraphics[width = 0.49\textwidth]{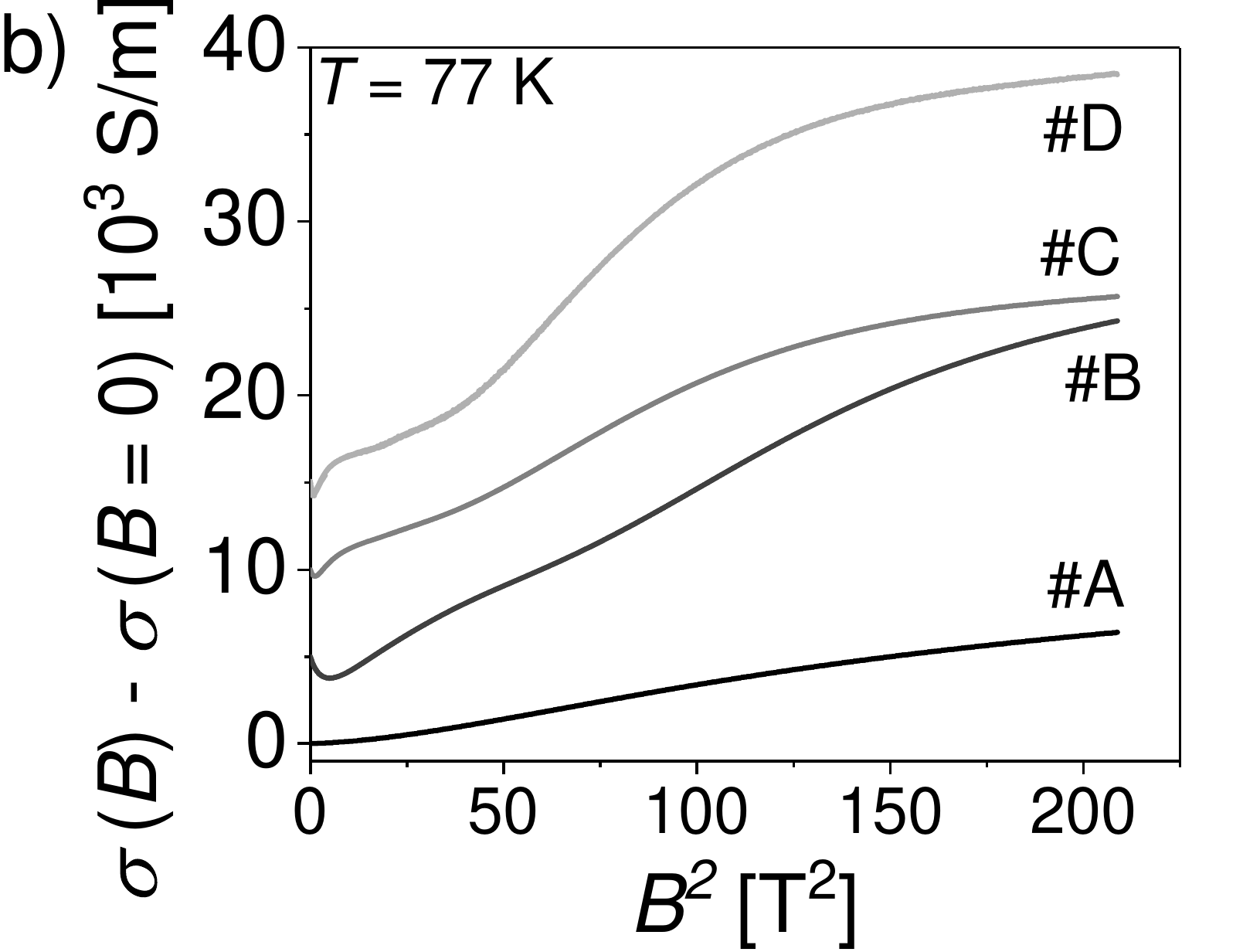}}
    
    \subfloat[\label{fig:WAL-test_c}]{\includegraphics[width = 0.49\textwidth]{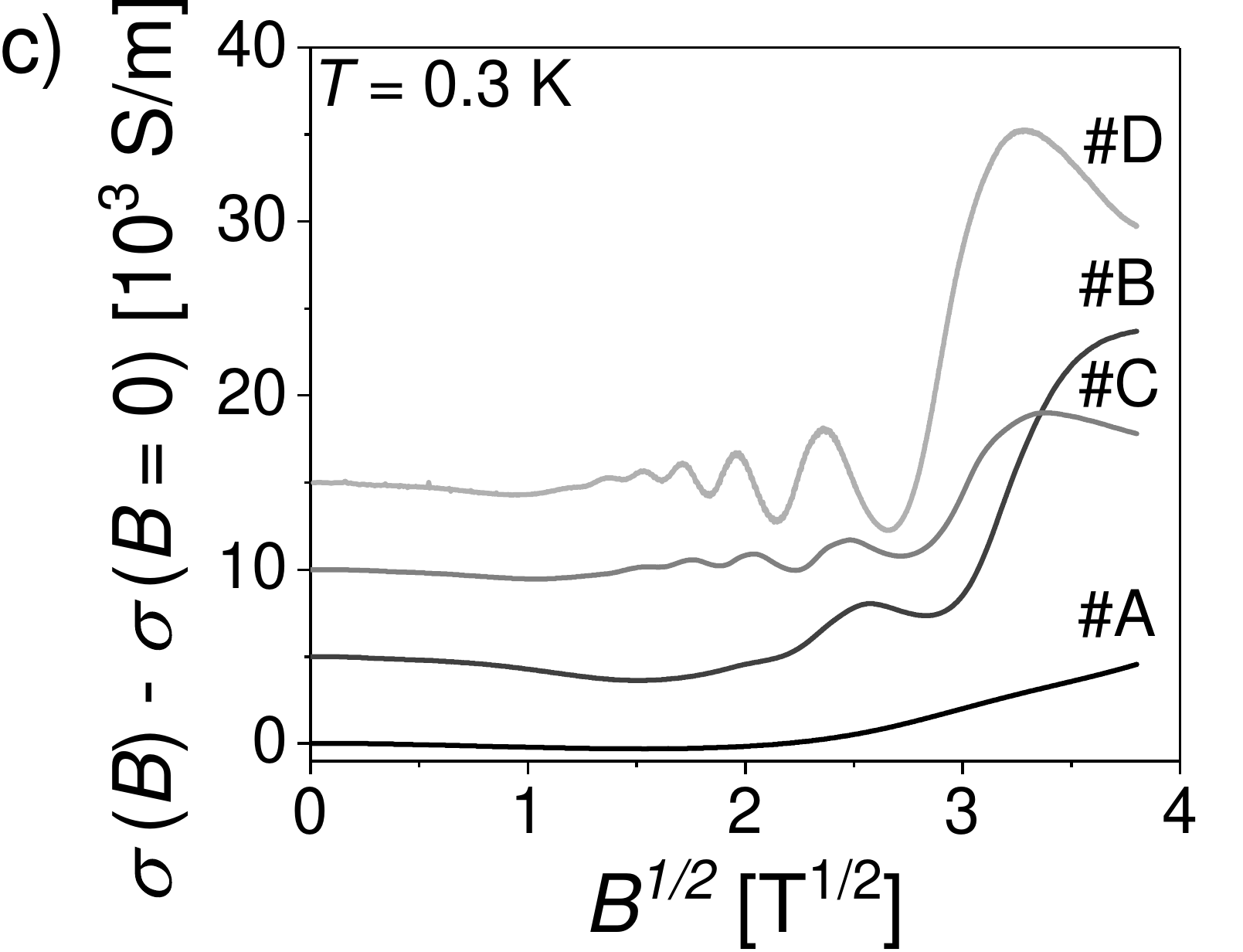}}\hfill
    \subfloat[\label{fig:WAL-test_d}]{\includegraphics[width = 0.49\textwidth]{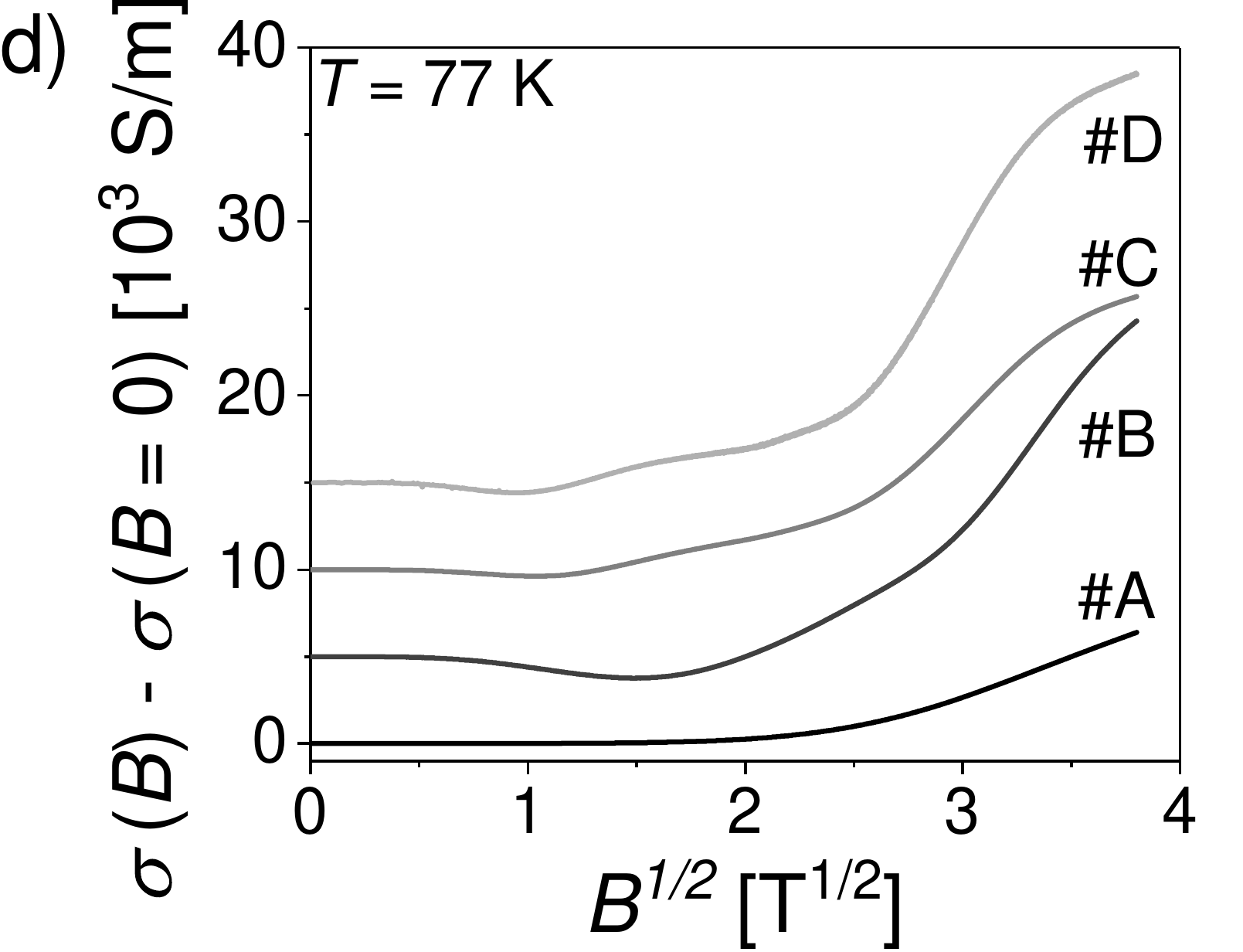}}
    \caption{Magnetoconductivity of {\aSn} films in $B_{\parallel}^{\text{in-plane}}$, plotted as a function of $B^2$ and $\sqrt{B}$ for two different temperatures. No linear dependence is present, which proves that experimental data does not agree with the weak localization theory either for low- or high-field regimes.}
    \label{fig:WAL_test}
\end{figure}

\textbf{Weak localization} (WL) is a quantum interference effect arising from inelastic scattering \cite{Kawabata_JPSJap1980, Kawabata_SolStCom1980}, which introduces positive correction to the resistance. A magnetic field changes the phase of the electron and destroys the interference, producing negative magnetoresistance. In strong magnetic fields, the WL-related conductivity is expected to be independent of the material parameters, $\Delta \sigma_{WL} \propto B^{\frac{1}{2}}$, while at low fields $\Delta \sigma_{WL} \propto \omega_c^2 $, with $ \omega_c$ - cyclotron frequency \cite{Kawabata_SolStCom1980}. In a 3D medium, we expect the scattering probability to be equal for any direction, and therefore the weak localization to be isotropic with respect to the direction of the magnetic field. As was shown in the main text and in \autoref{fig:Bparal}, it is not our case. We have also compared our results with the model for WL \cite{Kawabata_SolStCom1980} and observed neither $B^2$ nor $\sqrt{B}$ dependence of $\Delta\sigma_{xx}$, as presented in \autoref{fig:WAL_test}. However, the presence of negative MR in $B^{\text{in-plane}}_{\perp}$ at some range of fields can be a signature of non-zero contribution from WL to the overall signal.

\textbf{Mobility fluctuactions} are also known to introduce negative MR in $B^{\text{in-plane}}_{\parallel}$, which can mimic the anomaly-related field-dependence \cite{Hu_PRL2005, Xu_NatComm2019}. However, in this case, the linear MR in $B^{\text{normal}}$, starting at low magnetic fields, is also expected. We observe quasi-linear MR only in a limited range of temperatures and magnetic fields, which lets us conclude that disorder, and consequently mobility fluctuations, do not play a major role in the magneto-transport of {\aSn} epitaxial layers.

\FloatBarrier

\subsection{Semi-classical description of NLMR}

Initially, NLMR related to the chiral anomaly was proposed to occur in the ultra quantum limit in materials with linear dispersion due to the chirality of the lowest Landau level (LLL) \cite{Nielsen_PhysLett1983}. However, we observe NLMR before the ultra-quantum limit is achieved. This is evidenced by two observations. First, for sample A above $T \approx \qty{30}{\kelvin}$, MR is negative in the entire range of magnetic fields. Second, in high-mobility samples (C and D) we observe negative MR together with SdH oscillations, meaning that higher Landau levels are still occupied. We, therefore, turn to the semiclassical description of the chiral anomaly \cite{Son_PRB2013}. It was shown that in the material with the band degeneracy points, where Berry curvature is divergent, and with Fermi energy $E_F$ not coinciding with the Dirac point, the NLMR can be obtained in frames of the Boltzmann equation. Here, the anomaly-related contribution to the conductivity is given by:

\begin{equation}
    \label{eq:C_CH}
    \sigma_{CH} = \frac{e^{4} v^{3}}{4 \pi^{2} c^{2} E_{F}^{2} \hslash}\tau_{a}*B^2,
\end{equation}

with $e$ being the electron charge, $v$ the Fermi velocity, $E_{F}$ the Fermi level, and $\tau_{a}$ the intervalley (axial) scattering time. Deviations from parabolicity may appear at high magnetic fields, e.g., because of the formation of Landau levels, or reaching the LLL \cite{Son_PRB2013}. It should be also noted that for materials with field-induced Weyl nodes, such as {\aSn}, anomaly-related conductivity may exhibit non-parabolic dependence on the magnetic field \cite{Cano_PRB2017}. At the same time, a weak antilocalization (WAL) of Dirac and Weyl fermions is expected around $B = 0$ \cite{Lu_PRB2015}. Magnetoconductivity of the system then writes:
\begin{equation}
    \label{eq:NLMR_conductivity}
    \Delta \sigma_{xx} = \sigma_{xx}(B) - \sigma_{xx}(B = 0) = \sigma_{CH} + \sigma_{WAL}.
\end{equation}

\begin{figure}[htbp]
    \captionsetup[subfigure]{labelformat = empty}
    \subfloat[\label{fig:HLN_03K}]{\includegraphics[width = 0.31\textwidth]{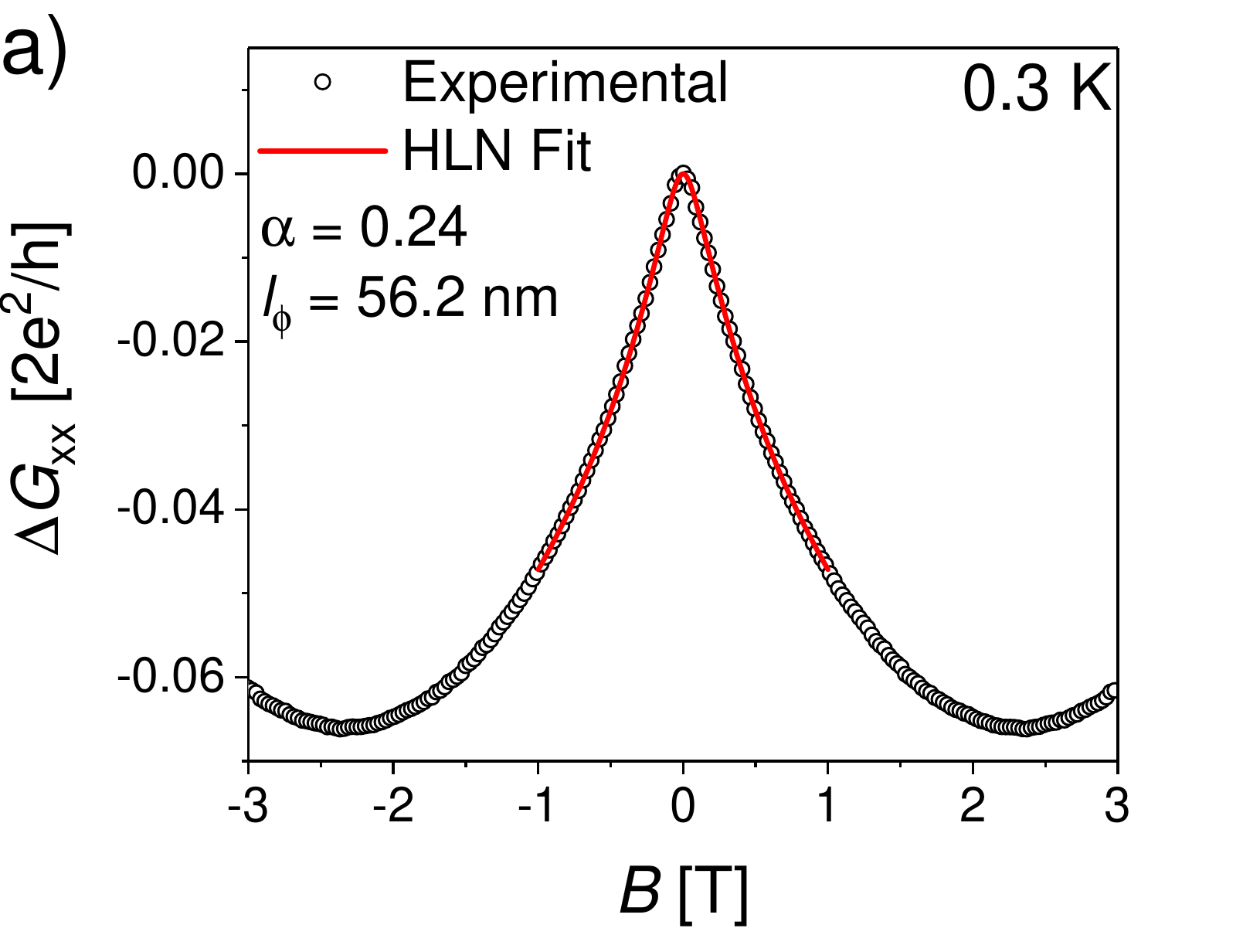}}\hfill
    \subfloat[\label{fig:HLN_4K}]{\includegraphics[width = 0.31\textwidth]{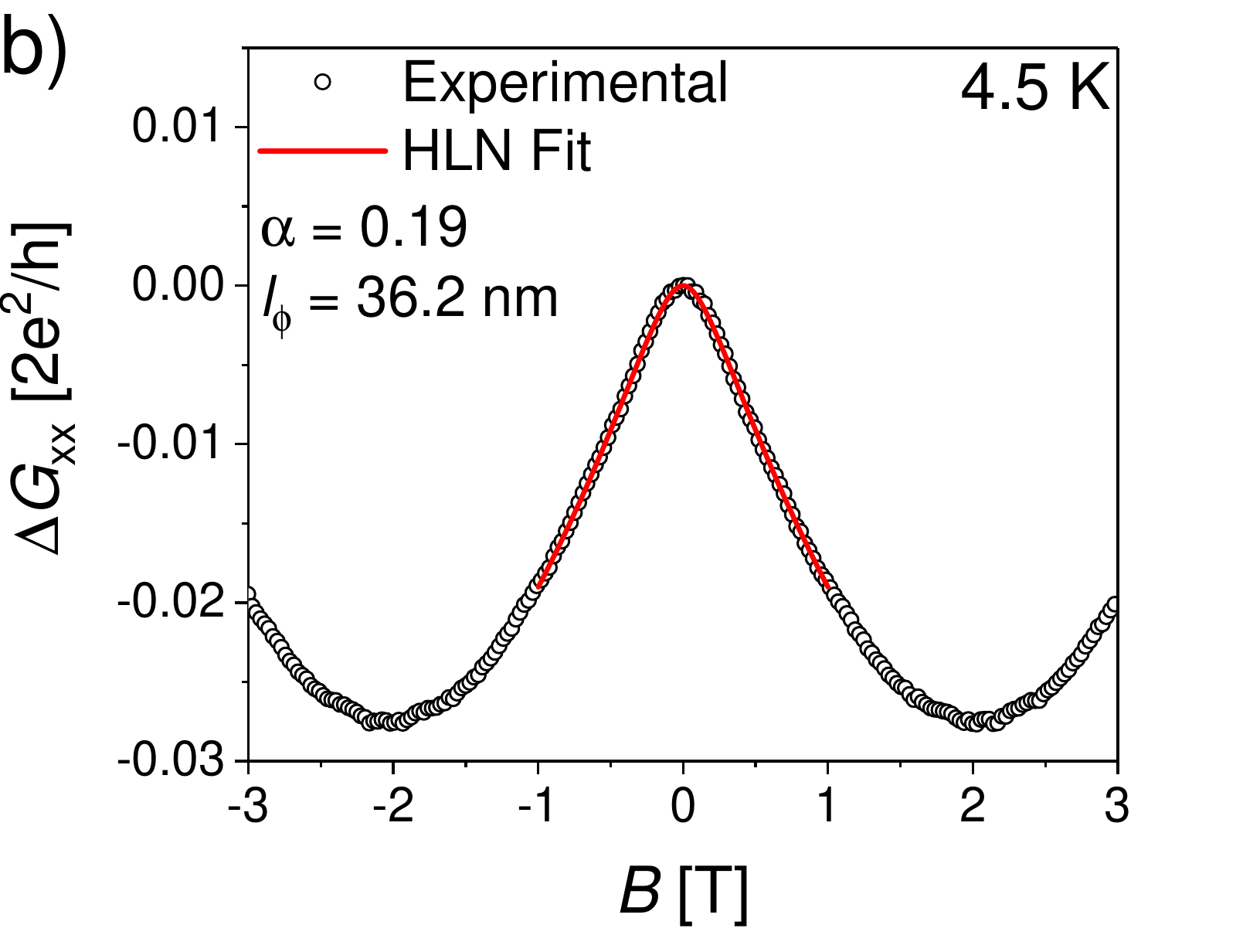}}\hfill
    \subfloat[\label{fig:HLN_10K}]{\includegraphics[width = 0.31\textwidth]{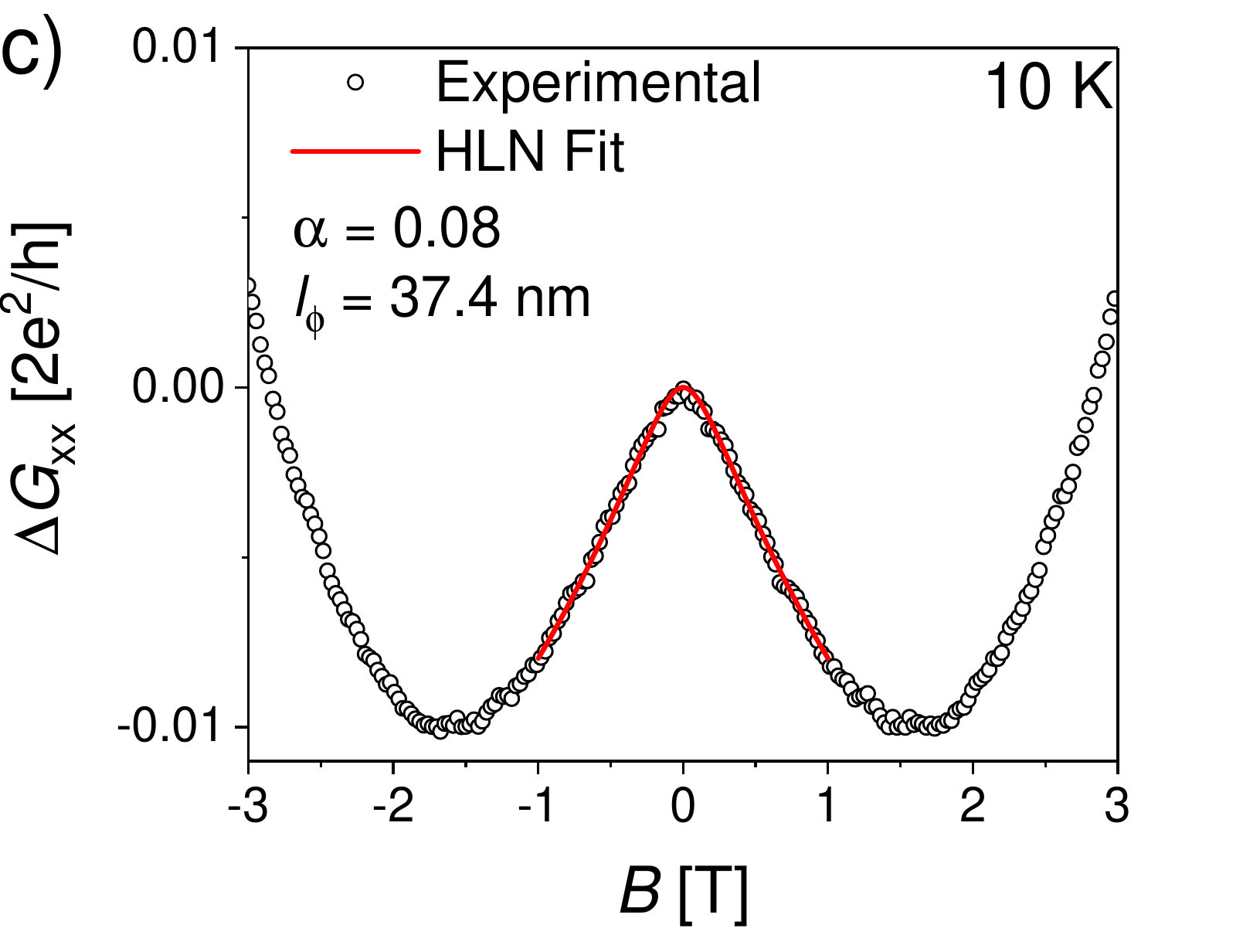}}
    
    \subfloat[\label{fig:HLN_B2_03K}]{\includegraphics[width = 0.31\textwidth]{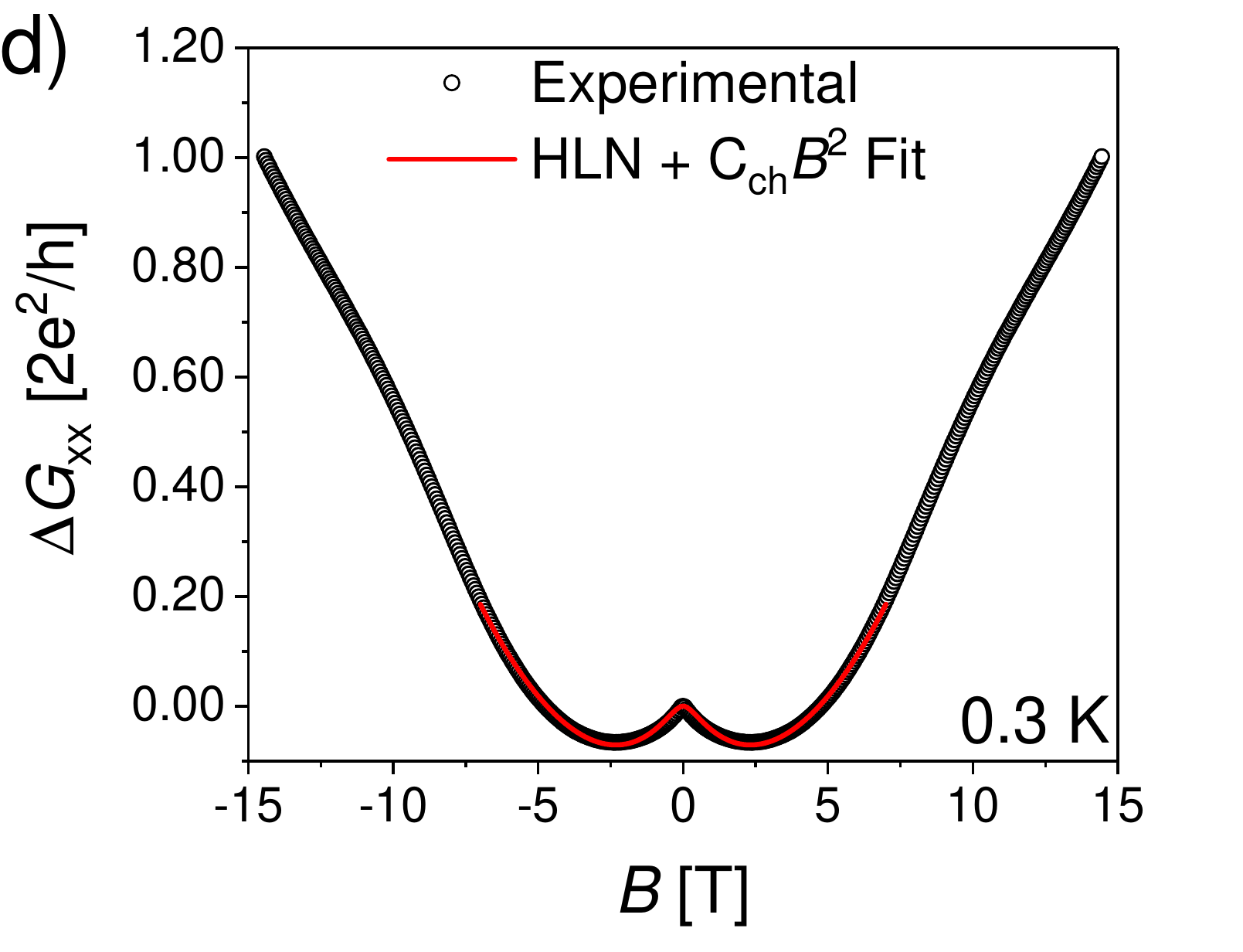}}\hfill
    \subfloat[\label{fig:HLN_B2_4K}]{\includegraphics[width = 0.31\textwidth]{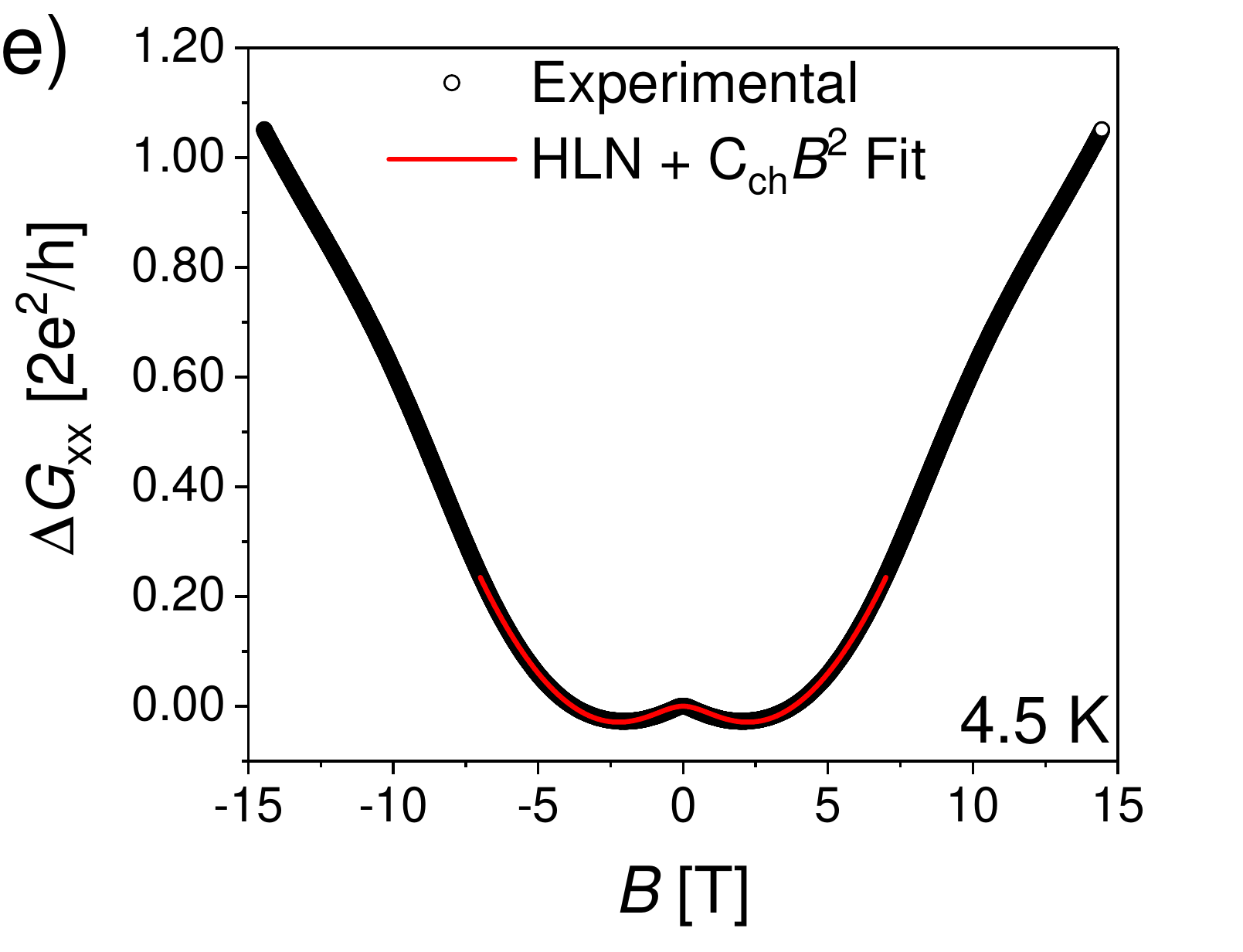}}\hfill
    \subfloat[\label{fig:HLN_B2_77K}]{\includegraphics[width = 0.31\textwidth]{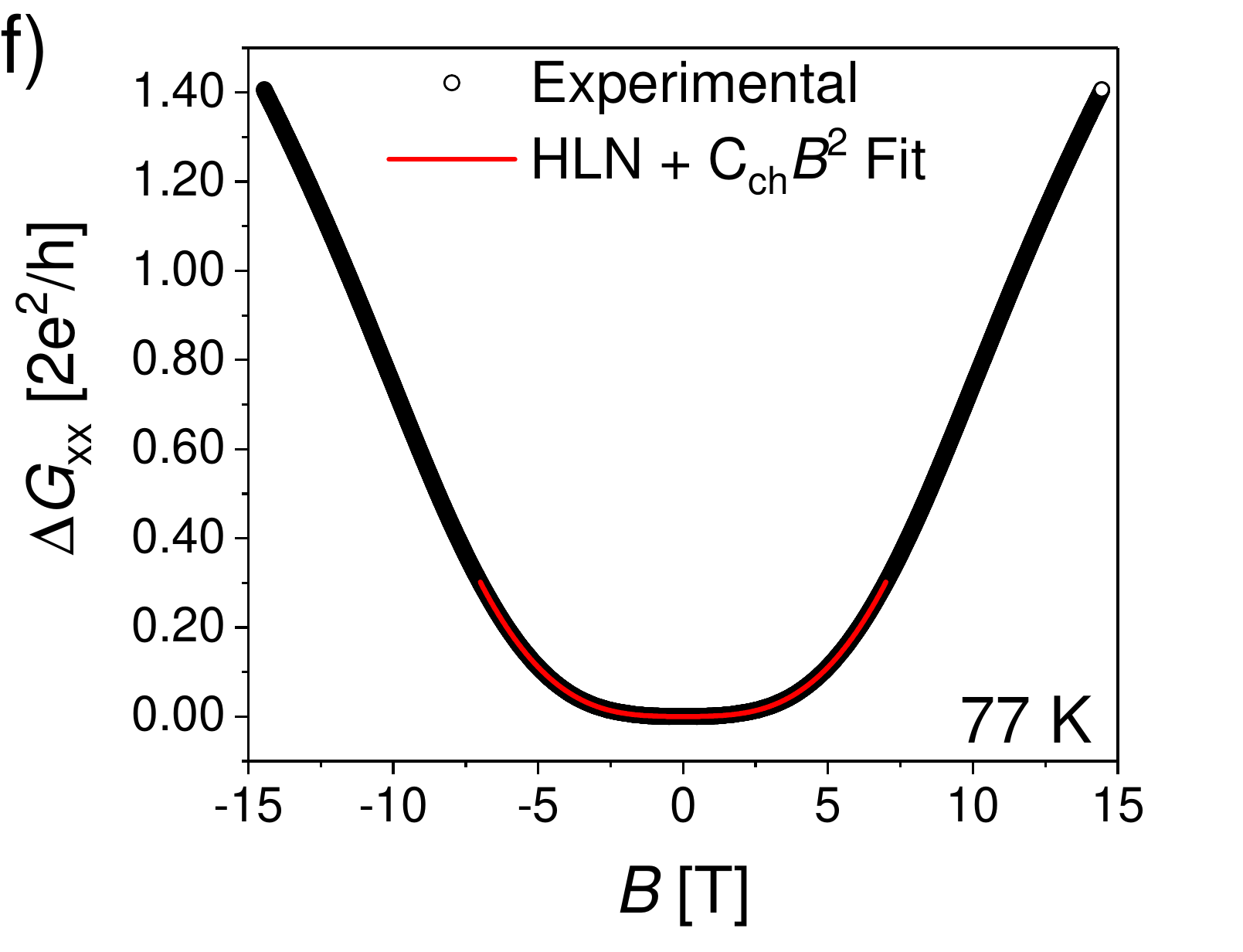}}
    \caption{Magnetoconductivity of sample A (50 nm) in $B^{\text{in-plane}}$. Low-field, low-temperature magnetoconductivity can be accurately described by the HLN formula (\autoref{eq:HLN}) with only 2 fitting parameters, as shown in panels a) to c). After including the parabolic $C_{CH} * B^2$ term, it is possible to describe the experimental data in a wider range of magnetic fields and temperatures.}
\label{fig:HLN_50nm}
\end{figure}

For the 2D system with strong spin-orbit interaction, WAL-related conductance is given by the Hikami-Larkin-Nagaoka equation \cite{Hikami_ProgTheorPhys1980}:
\begin{equation}
    \label{eq:HLN}
    \Delta \sigma_{WAL} = - \alpha \frac{e^2}{2 \pi^2 \hbar}\left[ \ln \left(\frac{B_{\phi}}{B} \right) - \psi \left(\frac{1}{2} + \frac{B_{\phi}}{B} \right) \right],
\end{equation}

where $B_{\phi} = \frac{\hbar}{4el_{\phi}^2}$, $l_{\phi}$ is the phase coherence length, and $\psi(x)$ is the digamma function. Magnetoconductivity of sample A around $B = 0$ can be described by \autoref{eq:HLN}, as shown in \autoref{fig:HLN_50nm}a-c. The phase coherence length $l_{\phi} = \qty{56.2}{\nano\meter}$ at $T = \qty{0.3}{\kelvin}$ is higher than the thickness of the layer (\qty{50}{\nano\meter}), which justifies application of the HLN formula. Adding the anomaly-related $B^2$ contribution allows to describe the magnetoconductivity of sample A in the field range $\pm \qty{7}{\tesla}$, as presented in \autoref{fig:HLN_50nm}d-f. Our experimental data obtained for this sample is therefore in agreement with the semi-classical description of chiral anomaly \cite{Son_PRB2013}. This is in contrast with MR of thicker layers (samples B, C and D). First, low-field positive MR does not exhibit the temperature dependence of WAL and is present at temperatures as high as \qty{77}{\kelvin}. Second, we do not observe parabolic scaling of magnetoconductivity with magnetic field. The explanation of this difference is beyond the scope of the current paper. Nevertheless, we expect that it can be related to the field-dependent splitting of Weyl cones in {\aSn} \cite{Cano_PRB2017} - a finite magnetic field is needed to form well-separated Weyl cones and, consequently, for NLMR to dominate the overall signal.

\newpage
\printbibliography